\documentclass{SciPost}

\hypersetup{
    colorlinks,
    linkcolor={red!50!black},
    citecolor={blue!50!black},
    urlcolor={blue!80!black}
}

\usepackage[bitstream-charter]{mathdesign}
\DeclareSymbolFont{usualmathcal}{OMS}{cmsy}{m}{n}
\DeclareSymbolFontAlphabet{\mathcal}{usualmathcal}

\fancypagestyle{SPstyle}{
\fancyhf{}
\lhead{\colorbox{scipostblue}{\bf \color{white} ~SciPost Physics }}
\rhead{{\bf \color{scipostdeepblue} ~Submission }}

\fancyfoot[C]{\textbf{\thepage}}
}

\usepackage{graphicx}
\usepackage{dcolumn}
\usepackage{bm}

\usepackage{xcolor}
\usepackage{ dsfont }
\usepackage[normalem]{ulem}
\usepackage{hyperref}

\definecolor{eggplant}{RGB}{180,33,147}

\definecolor{grey}{rgb}{.6,.6,.6}

\newcommand{\ket}[1]{|#1\rangle}
\newcommand{\bra}[1]{\langle #1|}
\newcommand{\bracket}[2]{\langle #1| #2 \rangle}

\newcommand{\mI}{\mathcal{I}}
\newcommand{\mJ}{\mathcal{J}}

\begin{document}
\pagestyle{SPstyle}

\begin{center}{\Large \textbf{\color{scipostdeepblue}{
  Who can compete with quantum computers? \\
Lecture notes on quantum inspired tensor networks computational techniques
}}}\end{center}

\begin{center}\textbf{
Xavier Waintal \textsuperscript{1$\dagger$},
Chen-How Huang \textsuperscript{2}
and Christoph W. Groth \textsuperscript{1}
}\end{center}

\begin{center}
{\bf 1} Univ. Grenoble Alpes, CEA, IRIG-PHELIQS GT, F-38000 Grenoble, France
\\[\baselineskip]
{\bf 2} Department of Physics and Nanoscience Center, University of Jyväskylä, P.O. Box 35 (YFL), FI-40014 University of Jyväskylä, Finland
\\[\baselineskip]
$\dagger$ \href{mailto:email2}{\small xavier.waintal@cea.fr}
\end{center}

\section*{\color{scipostdeepblue}{Abstract}}
\textbf{\boldmath{%
This is a set of lectures on tensor networks with a strong emphasis on
the core algorithms involving Matrix Product States (MPS) and Matrix Product Operators (MPO). Compared to other presentations, particular care has been given to disentangle aspects of tensor networks from the quantum many-body problem: MPO/MPS algorithms are presented as a way to deal with linear algebra on extremely (exponentially) large matrices and vectors, regardless of any particular application. The lectures include
well-known algorithms to find eigenvectors of MPOs (the celebrated DMRG), solve linear problems, and recent learning algorithms that allow one to map a known function into an MPS (the Tensor Cross Interpolation, or TCI, algorithm). The lectures
end with a discussion of how to represent functions and perform calculus with tensor networks using the ``quantics'' representation. They include the detailed analytical construction of important MPOs such as those for differentiation, indefinite integration, convolution, and the quantum Fourier transform.
Three concrete applications are discussed in detail: the simulation of a quantum computer (either exactly or with compression), the simulation of a quantum annealer, and techniques to solve partial differential equations (e.g.\ Poisson, diffusion, or Gross--Pitaevskii) within the ``quantics'' representation.
The lectures have been designed to be accessible to a first-year PhD student and include detailed proofs of all statements.
}}

\vspace{\baselineskip}

\noindent\textcolor{white!90!black}{%
\fbox{\parbox{0.975\linewidth}{%
\textcolor{white!40!black}{\begin{tabular}{lr}%
  \begin{minipage}{0.6\textwidth}%
    {\small Copyright attribution to authors. \newline
    This work is a submission to SciPost Physics. \newline
    License information to appear upon publication. \newline
    Publication information to appear upon publication.}
  \end{minipage} & \begin{minipage}{0.4\textwidth}
    {\small Received Date \newline Accepted Date \newline Published Date}%
  \end{minipage}
\end{tabular}}
}}
}


\vspace{10pt}
\noindent\rule{\textwidth}{1pt}
\tableofcontents
\noindent\rule{\textwidth}{1pt}
\vspace{10pt}

\section{Foreword}
\label{sec:foreword}

A naive, yet popular, statement says that quantum computers can provide an exponential speedup over classical computers because their internal states live in an exponentially large space of dimension $2^N$ for an $N$-qubit quantum computer. The number of dimensions grows so fast with $N$ that classical supercomputers cannot even hold this state in memory as soon as $N>50$; hence, classical computing is supposedly doomed to address these states.

Yet, despite this supposed impossibility, a rather large number of such exponentially large states have been calculated, sometimes with machine precision, using classical algorithms
\cite{white1992,pan2019}.
The solution of this small paradox is the same as in other successes of physics: apparently very complex phenomena have internal mathematical structures that, when revealed, allow one to make precise predictions.

This text contains the notes of a set of lectures given at the
Jyväskylä summer school (Finland) during August 2025.
At the core, this is a comprehensive introduction to tensor network techniques
\cite{schollwoeck2011,orus2014,bridgeman2017,cirac2021}, including classic material (matrix product states and operators, Density Matrix Renormalization Group (DMRG) algorithm, etc.), but also more recent topics (tensor network learning algorithms, quantics representation of functions, e.g.\ solving partial differential equations, etc.). The presentation of these topics is done in the context of quantum computing (gate-based as well as quantum annealers) instead of the more traditional many-body problem. This allows one to avoid a large fraction of the usual formalism (e.g.\ second quantization) and concentrate on the core aspects of tensor networks.

The unifying principle of all these techniques, other than all of them being obviously based on tensor networks, is that they all compete in one way or another with what quantum computers are supposed to do. These are not the only competitors, though, and we could have also included e.g.\ variational Monte Carlo \cite{becca2017} as a classical counterpart to the variational quantum eigensolver popular in quantum computing \cite{louvet2024}.

The lectures have been entirely given on the blackboard, meaning that the range of material is rather limited, but that it has been covered in enough depth for the reader to be in a position to actually write their own code and implement the different algorithms. Actually, each lecture was followed by a hands-on session where the goal was to implement (from scratch) and try out as many of the algorithms as possible. It turned out that one of the students produced some neat illustrations and is now a co-author of this text. The corresponding code repositories are described in Appendix~\ref{app:code}.

The style of this manuscript is rather informal, and it contains, in addition to the scientific material, some subjective opinions on the status of this or that aspect (e.g.\ claims of quantum supremacy, discussion of the I/O bottleneck in quantum computing, etc.). I feel that such personal views are particularly useful in the field of quantum computing, where the level of hype is rather high. Many groups and companies make many claims of various types of present or future advantages that are sometimes hard to decipher. The key information of a scientific article in this field often does not lie in what is shown but in what is missing; a secondary goal of these notes is to guide the reader to where to look.

\subsection{What's in the lectures?}
These lectures describe the following set of algorithms:
\begin{itemize}
\item An introduction to tensor networks and the associated linear algebra, including the Singular Value Decomposition (SVD), the Cross Interpolation, and the associated partial-rank-revealing LU decomposition.

\item A comprehensive set of algorithms for Matrix Product States (MPS) that are seen as representations of exponentially large vectors. We show how to sample MPS, put them in orthogonal form, compress them, add two MPS, calculate the scalar product of two MPS, etc. Note that MPS
were renamed Tensor Trains in the mathematical literature. We will use both terms interchangeably.

\item A comprehensive set of algorithms for Matrix Product Operators (MPO) that are representations of exponentially large matrices. We show how to multiply two MPOs, perform MPO-MPS matrix-vector products, solve linear problems of the form $\mathrm{MPO}\times\mathrm{MPS}=\mathrm{MPS}$ ($Ax=b$), find the lowest eigenvector of an MPO (the celebrated DMRG algorithm), etc.

\item A detailed introduction to the Tensor Cross Interpolation (TCI) learning algorithm. TCI transforms very large matrices or vectors (in the form of a function that returns the value for given indices) into an MPO or MPS.

\item An introduction to the quantics tensor-train representation, which allows one to use the above algorithms to solve partial differential equations. We discuss how the Fourier transform translates into a simple MPO-MPS product in this context and can therefore be performed exponentially faster than the regular Fast Fourier Transform.

\item An introduction to quantum computing and its link with tensor networks.
\end{itemize}
On the other hand, we do not discuss:
\begin{itemize}
\item More advanced tensor networks such as PEPS, PEPO, MERA, or tree tensor networks.
\item Any work involving fermions and bosons. The two examples shown involve only qubits and spins.
\item Usage of symmetries; advanced time-evolution techniques such as TDVP; and many other tensor network techniques such as belief propagation, iDMRG, etc.
\end{itemize}

Readers who want to proceed to more advanced topics will find some pointers on this website:
https://tensornetwork.org. This website contains a lot of resources on the topic, including a list of tensor network libraries. See also the https://tensor4all.org website for the aspects related to TCI. The lectures themselves were recorded, and the videos can be found at this address: {\color{blue}{ https://www.youtube.com/@Quantuminspiredalgorithms}}. In the hands-on sessions, most students have used Python together with the NumPy library, but other programming languages, such as Julia, are equally valid choices. While implementing the algorithms described below from scratch is very useful for pedagogical purposes, multiple implementations already exist and can be used once one understands how they work. A popular choice is the ITensor library \cite{fishman2022}.

\subsection{Structure of the lectures}

Even though the presentation of tensor networks could have been done in an
abstract way, with no relation to an actual application, we chose to link it to
two physical problems: the simulation of a quantum computer (which is presented in
section \ref{sec:qc}) and the simulation of the transverse field Ising model
(which is presented in section \ref{sec:tfi}). These two sections are independent of
tensor networks; they just state the problem to be solved.

The remainder of these notes is split into three parts:
\begin{itemize}

\item First, the general concepts of tensor networks are introduced in Sections~
\ref{sec:basics} and~\ref{sec:exact}. Section~\ref{sec:basics} just defines
the various objects and operations while section \ref{sec:exact} gives a first
set of algorithms to simulate quantum computers. These algorithms are ``exact''
as opposed to the algorithms discussed in the rest of these lectures which use a (controlled) approximation.

\item Second, the central concept of tensor networks -- low-rank compression -- is discussed in Sections~\ref{sec:mps} and~\ref{sec:hamiltonian_models}.
Section~\ref{sec:mps} introduces the necessary tools and discusses the approximate simulation of a quantum computer as a first application. We arrive at the actual DMRG algorithm to find the ground state of a many-body problem in 
Section~\ref{sec:hamiltonian_models}, which is rather late. This is the original and still the main application of tensor networks, so it was impossible not to include it. However, in these notes it plays a relatively minor role.

\item Third, we leave the realm of many-body physics and quantum computers in Sections~
\ref{sec:mpo_mps_are_large_vectors}, \ref{sec:tci}, and~\ref{sec:quantics}.
From there on, MPO and MPS are just considered as convenient representations of very large matrices and vectors, allowing one to perform linear algebra on objects that are just too big to be held in memory in their naive form.
Section~\ref{sec:mpo_mps_are_large_vectors} discusses several algorithms that complete the linear algebra toolbox of MPO/MPS. We still lack a way to turn problems into this
framework. This is solved in section~\ref{sec:tci} where we discuss the tensor cross interpolation learning algorithm. The lectures culminate with section~\ref{sec:quantics} that discusses how all the above can be used to solve partial differential equations using the quantics representation.
\end{itemize}
\section{The quantum computer: a machine for performing certain matrix-vector multiplications}
\label{sec:qc}

A quantum computer is a well-controlled out-of-equilibrium quantum many-body system
that one intends to use to perform a calculation. Such a system can be described at several levels: from the actual underlying physics (usually described in terms of its Hamiltonian, i.e.\ with time and energies) up to an abstract representation used to describe quantum algorithms (the gate-based quantum computer). In this section, we briefly present this latter model, which will serve as a reference point \cite{nielsen2000}. We will not discuss the quantum algorithms themselves; rather, we will look at what a quantum computer is supposed to do at a very general level and ask what prevents us (or not) from doing the same thing on a classical computer.

\subsection{An exponentially large internal state}
The abstract ``gate-based'' quantum computer is defined as follows. We have a set of
$N$ two-level systems, called quantum bits or qubits (for instance, the spin of an electron), that can be in the states $\ket{0}$ and $\ket{1}$. The most general state of the quantum computer has the form
\begin{equation}
\ket{\Psi} = \sum_{i_1i_2\cdots i_N} \Psi_{i_1i_2\cdots i_N} \ket{i_1i_2\cdots i_N},
\end{equation}
where the sum runs over all qubit values $i_a \in \{0,1\}$, and
$\ket{i_1i_2\cdots i_N}$ is a shorthand for the tensor product
$\ket{i_1i_2\cdots i_N} = \ket{i_1}\otimes \ket{i_2}\otimes\cdots\otimes\ket{i_N}$.
The tensor $\Psi_{i_1i_2\cdots i_N}$ can be thought of as a large vector containing $2^N$
complex values. The potential capabilities of quantum computers stem from the fact that this vector is exponentially large and, once $N\gtrsim 50$, cannot be stored in a classical computer.

\subsection{Quantum circuits}
When one operates a quantum computer, one initializes it with an initial
state $\ket{\Psi}^{(0)}$ (usually $\ket{\Psi}^{(0)} = \ket{000\cdots{}0}$),
and the state of the system evolves according
to the Schrödinger equation, which transforms $\ket{\Psi}^{(n)}$ into
$\ket{\Psi}^{(n+1)} = \hat U^{(n)} \ket{\Psi}^{(n)}$,
with the evolution operator $\hat U^{(n)}$ given by
\begin{equation}
\hat U^{(n)} = T e^{-i\int_{t_n}^{t_{n+1}} dt \hat H(t)}.
\end{equation}
where $T$ is the time-ordering operator and $\hat H(t)$ is the Hamiltonian of the system.
In physics, the system is described by $\hat H(t)$. In quantum computing, one actually starts with the evolution operators $\hat U^{(n)}$, assuming that someone else has worked out
how to engineer appropriate Hamiltonians. The different evolution operators that one will use are called ``gates'' and fall into two categories, depending on whether they act on a single qubit or on two qubits. A single-qubit gate is defined on one qubit as $\hat U\ket{i} =
\sum_j U_{ji}\ket{j}$, where the $2\times 2$ matrix $U_{ij}$ is unitary. In terms of the wavefunction $\Psi^{(n)}$, such a single-qubit gate on qubit $a$ translates into a matrix that acts as
\begin{equation}
\label{eq:1qubit_gate}
\Psi_{i_1i_2\cdots{}i_N}^{(n+1)} = \sum_{i'_a} U^{(n)}_{i_ai'_a}
\Psi_{i_1\cdots{}i_{a-1}i'_a i_{a+1}\cdots{}i_N}^{(n)}.
\end{equation}
Likewise, a two-qubit gate acting on qubits $a$ and $b$ is described by a $4\times 4$ matrix
and transforms the wavefunction into
\begin{equation}
\label{eq:2qubit_gate}
\Psi_{i_1i_2\cdots{}i_N}^{(n+1)} = \sum_{i'_a,i'_b} U^{(n)}_{i_ai_b,i'_ai'_b}
\Psi_{i_1\cdots{}i_{a-1}i'_a i_{a+1}\cdots{}i_{b-1}i'_b i_{b+1}\cdots{}i_N}^{(n)}.
\end{equation}
Depending on the quantum hardware, some gates are easier to implement than others. Typical examples include the one-qubit gates (the first three are the Pauli matrices)
\begin{eqnarray}
X &=& \begin{pmatrix} 0 & 1 \\ 1 & 0 \end{pmatrix}, \\
Y &=& \begin{pmatrix} 0 & -i \\ i & 0 \end{pmatrix}, \\
Z &=& \begin{pmatrix} 1 & 0 \\ 0 & -1 \end{pmatrix}, \\
H &=& \frac{1}{\sqrt{2}} \begin{pmatrix} 1 & 1 \\ 1 & -1 \end{pmatrix}, \\
T &=& \begin{pmatrix} e^{i\pi/8} & 0 \\ 0 & e^{-i\pi/8} \end{pmatrix},
\end{eqnarray}
and the controlled NOT (or C-NOT) two-qubit gate
\begin{equation}
\label{eq:control_not}
C_X = \begin{pmatrix} 1 & 0 & 0 & 0 \\
                      0 & 1 & 0 & 0 \\
                      0 & 0 & 0 & 1 \\
                      0 & 0 & 1 & 0 \end{pmatrix}.
\end{equation}
In the above gate, the two qubits are not equivalent: there is the control bit (c) and the target bit (t). The matrix assumes the following ordering of the two-qubit states:
$\ket{0}_\text{c}\ket{0}_\text{t}$,
$\ket{0}_\text{c}\ket{1}_\text{t}$,
$\ket{1}_\text{c}\ket{0}_\text{t}$ and
$\ket{1}_\text{c}\ket{1}_\text{t}$. Three-qubit gates are often more complicated to implement
(nature only provides two-body interactions), but can be constructed from combinations
of one- and two-qubit gates. Overall, a quantum circuit looks like this
\begin{center}
\includegraphics[scale=0.5]{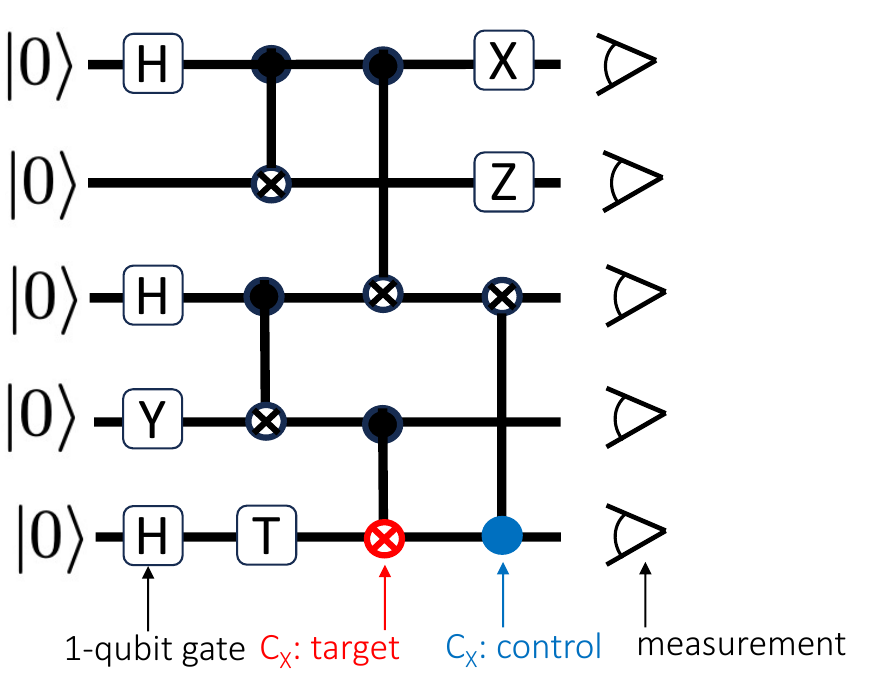}
\end{center}
They are read a little like music, with one line per qubit.

The last ingredient of the quantum computer gate model is measurement. When a qubit $a$ is
measured, it returns the value $\alpha$ ($\alpha = 0$ or 1) with probability
\begin{equation}
P_\alpha = \sum_{i_1\cdots{}i_{a-1} i_{a+1}\cdots{}i_N} \left|
\Psi_{i_1\cdots{}i_{a-1} \alpha i_{a+1}\cdots{}i_N}^{(n)} \right|^2.
\end{equation}
After measurement, the new wavefunction becomes
\begin{equation}
\Psi_{i_1\cdots{}i_N}^{(n)}\rightarrow
\delta_{i_a,\alpha} \frac{1}{\sqrt{P_\alpha}}
\Psi_{i_1\cdots{}i_{a-1} \alpha i_{a+1}\cdots{}i_N}^{(n)}.
\end{equation}
And that's essentially it: the above set of equations entirely describes what a quantum computer is supposed to do. The entire field of quantum algorithms (which we will not discuss) consists of using these rules to perform useful computations.
A good entry point to this literature is \cite{nielsen2000}.

\subsection{Summary}
So, in a nutshell, a quantum computer allows one to perform a
subset of linear algebra, namely matrix-vector multiplications, with very special
matrices (the ``gates'' that act only on certain indices and belong to a fixed set of unitary matrices) on exponentially large vectors (the wavefunction $\Psi_{i_1\cdots{}i_N}^{(n)}$). The appeal of quantum computers clearly comes from the exponentially large size of those wavefunctions. However, a very strong downside is that at the end of the calculation one does not hold the corresponding exponentially large vector of $2^N$ values, but only a much smaller set: $N$ bits of (probabilistic) information. We get a single sample of the distribution $|\Psi_{i_1\cdots{}i_N}^{(n)}|^2$. This is one of the two Achilles' heels of quantum computing, which we dub the I/O bottleneck (well, more the O bottleneck for this aspect).

It is very important to realize that the largest part of the Hilbert space of dimension
$2^N$ will remain forever inaccessible to quantum computers (and classical methods). This can be understood using a simple counting argument. Suppose that the quantum circuit consists of $D$ layers of gates ($D$ being the depth of the circuit). We also suppose that each layer is packed with as many gates as possible (meaning that all the qubits are acted upon). Lastly, each gate is parametrized by a few angles. Then the total
dimension of the subspace that can be spanned by these circuits is $O(D N)$, which is obviously much smaller than $2^N$.

Now let us put some realistic numbers. Suppose that we work with $N=100$ qubits. The total dimension of the Hilbert space is $2^{100} \approx 10^{30}$. Typical depths that can be considered with existing hardware are of the order of $D\approx 100$, but let’s suppose that this number is scaled up to $D=10^6$. The explorable subspace would still have 20 orders of magnitude fewer degrees of freedom than the full Hilbert space. So the question really is: does this subspace belong to the ``relevant'' part of the Hilbert space? And, conversely, is the ``relevant'' part of the Hilbert space amenable to classical simulations? The word relevant is defined very loosely here, but there are several scientific articles that start to give it a more precise meaning. For instance, the problem of the ``barren plateaus'' in the variational quantum eigensolver (VQE) algorithm has been traced back to the fact that most of the states in the $O(N D)$-dimensional subspace manifold are essentially chaotic, hence irrelevant \cite{larocca2025}.

In this set of lectures, we will discuss a set of classical techniques that also allow us to explore a finite subspace of the full Hilbert space. In addition to providing us with very powerful and useful techniques, this will help us put the claims of quantum computing into perspective.

\subsection{Digression on decoherence and fidelity}
We simply cannot leave the subject of quantum computing without discussing, however shortly, the phenomenon of decoherence, the other Achilles' heel of quantum computing.
Indeed, quantum entanglement is both a resource (when it is obtained in a precise way with degrees of freedom that are fully under control) and the main obstacle to building a quantum computer (when it occurs between qubits and other degrees of freedom such as a phonon or a two-level system).
In practice, the fidelity of the state $F(n) = \bra{\Psi^{(n)}} \rho^{(n)} \ket{\Psi^{(n)}}$ between the targeted state $\ket{\Psi^{(n)}}$ and the density matrix $\rho^{(n)}$ actually obtained decreases exponentially as
\begin{equation}
F(n)\sim e^{-\epsilon n},
\end{equation}
with an average error rate per gate $\epsilon$.
Actually, the error rate $\epsilon$ includes decoherence but not only: other more mundane phenomena also affect the precision of the calculation.
For instance, the energy difference of the two-qubit states may vary a little or a microwave pulse may be a little too long or slightly less intense than expected.
The bottom line is that an analog machine executes every operation with finite accuracy.
For the best current quantum hardware, $\epsilon$ lies somewhere around $10^{-3}$ (often less when used as a system for real quantum circuits as opposed to benchmarks on single qubits or pairs of qubits).
This phenomenon strongly limits the depth of the circuit that can be used in practice.
It is the main obstacle towards building a useful working quantum computer, as we explain in \cite{waintal2024}.
The hope of quantum computing is to use quantum error correction to address this problem \cite{nielsen2000}.
In a nutshell, quantum error correction uses several physical qubits to build one ``logical'' qubit of better quality than the physical ones.
For instance, one could use $\ket{000000}$ as logical $\ket{0}_\text{L}$ and $\ket{111111}$ as logical $\ket{1}_\text{L}$, while constantly measuring the parity of pairs of physical qubits to verify that they remain the same (i.e.\ as $00$ or $11$ but not $10$ or $01$, which correspond to errors) and prevent the other ``non-computational states'' (such as $\ket{101101}$) from acquiring a significant amplitude.
The exact theoretical construction is only slightly more sophisticated than that.
The practical construction, on the other hand, adds many layers of complexity to the hardware, so that it is unclear how far one will be able to go in practice.
We will not go further in that direction; the reader interested in a critical discussion can have a look at \cite{waintal2019}.

\section{Tensor networks: basic notation and operations}
\label{sec:basics}

We will now introduce the main theoretical tool used in these lectures: tensor networks
and the set of operations to manipulate them. Formally, a tensor with $N$ indices is a function from $\{0,\ldots,d_1-1\} \times \{0,\ldots,d_2-1\}\times\cdots\times\{0,\ldots,d_N-1\}$ to the field of e.g.\ real or complex numbers. It is the generalization of vectors and matrices to objects with any number of indices.

\subsection{Defining tensors}
A vector $v_i$ is a tensor with a single index (orange in the drawing below). It is represented by a circle (or another shape) with a single leg. A matrix $M_{ij}$ has two indices and is represented by a circle with two legs (green). A tensor $T_{ijk}$ with three indices has three legs (blue):
\begin{center}
\includegraphics[scale=0.3]{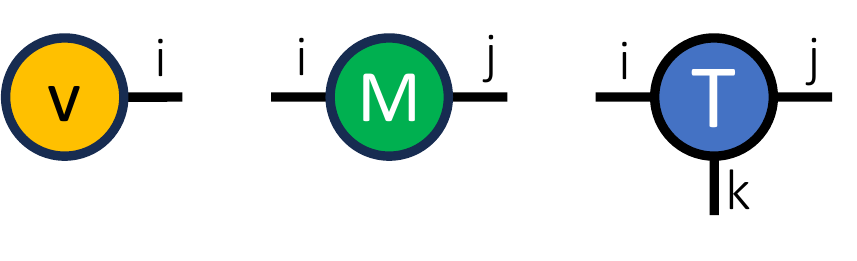}
\end{center}
More generally, a tensor $F_\mathbf{\sigma}$ with
$\mathbf{\sigma}=(\sigma_1,\sigma_2,\ldots,\sigma_{N})$ is said to be of degree $N$.  We denote by $d_l$ the dimension of $\sigma_l$, meaning that
$0\le \sigma_l<d_l$. When all the dimensions are equal, we denote them by $d=d_l$.
\begin{center}
\includegraphics[scale=0.4]{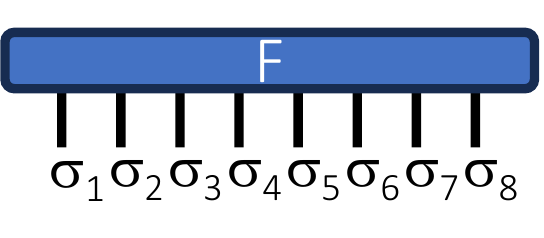}
\end{center}
To specify the value of an index, we simply draw the corresponding value next to its leg.
For instance, the vector $v_j$ defined as $v_j = T_{0j0}$ is drawn as follows:
\begin{center}
\includegraphics[scale=0.3]{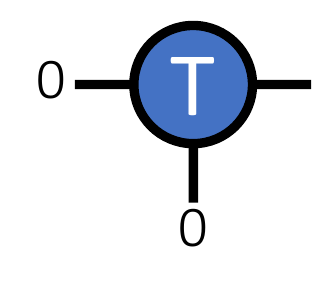}
\end{center}

Several special tensors appear frequently. An important one is the Kronecker
tensor $K$, also known as \emph{copy} and defined for any number of legs as
$K_{ijkl\ldots} = \delta_{ij} \delta_{jk} \delta_{kl}\cdots{}$ (1 if all indices are equal; 0 otherwise). It is represented by a black disk:
\begin{center}
\includegraphics[scale=0.4]{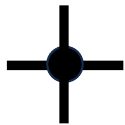}
\end{center}

Another special tensor is the flattening tensor\footnote{The flattening tensor is also sometimes called combiner.} $F_{ij\alpha}$, which we denote by a triangle. If the size of index $i$ is $d_i$, then $F_{ij\alpha}=1$ if
$\alpha = i+jd_i$, and $0$ otherwise (it follows that $d_\alpha = d_id_j$). 
\begin{center}
\includegraphics[scale=0.4]{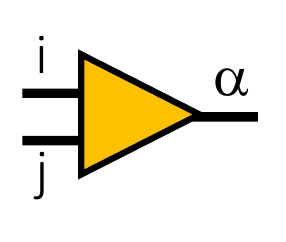}
\end{center}
Two identical flattening tensors facing each other cancel each other out:
$\sum_{ij} F_{ij\alpha} F_{ij\alpha'} = \delta_{\alpha\alpha'}$ and
$\sum_{\alpha} F_{ij\alpha} F_{i'j'\alpha} = \delta_{ii'}\delta_{jj'}$.
\begin{center}
\includegraphics[scale=0.3]{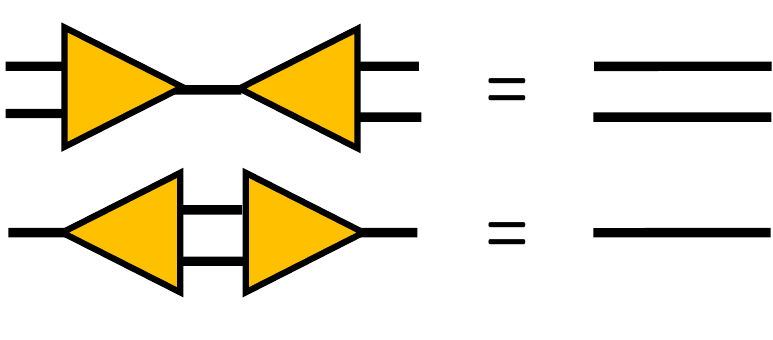}
\end{center}
The flattening
tensor can be used, for instance, to flatten a matrix into a vector where all the rows
(or columns) are placed one after the other.
\begin{center}
\includegraphics[scale=0.3]{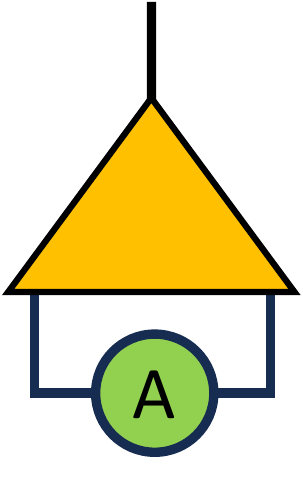}
\end{center}
The precise meaning of this drawing will be explained in the next paragraph through
a process known as ``contraction''. A note of warning for future reference: this particular contraction is never done explicitly (that would be grossly inefficient), but is performed using the index algebra explained in the factorization section.

\subsection{Tensor contraction and tensor networks}
There exist a number of basic operations that one can perform with tensors. The first one is the contraction of two tensors. Contraction is a generalization of the matrix-matrix or matrix-vector product: one identifies one index of the first tensor with another index of the second tensor (with matching dimension), and sums over the possible values of this index.
For instance, the expression
\begin{equation}
\sum_{j=0}^{d-1} T_{ijk} v_j \equiv C_{ik}
\end{equation}
is denoted graphically as
\begin{center}
\includegraphics[scale=0.3]{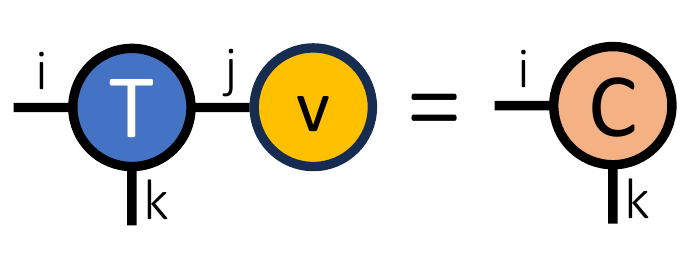}
\end{center}
Any connection between two tensors implies that the involved indices take the same values and are summed over; this is Einstein’s implicit sum notation.

With these notations, the tensor product between two vectors
$(V \otimes W)_{ij} = V_i W_j$ is simply represented by putting the tensors next to each other
(no repeated indices):
\begin{center}
\includegraphics[scale=0.3]{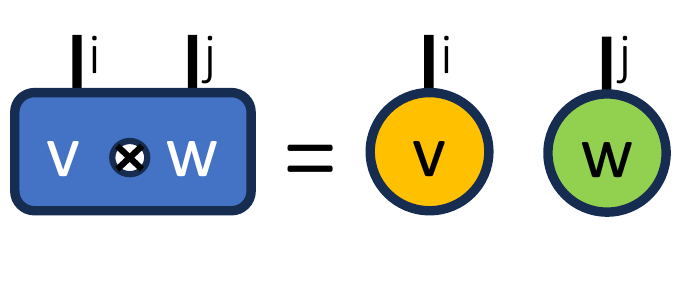}
\end{center}
The scalar product is (the star indicates complex conjugation):
\begin{center}
\includegraphics[scale=0.3]{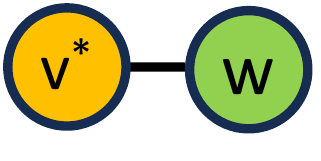}
\end{center}
The trace of a matrix reads
\begin{center}
\includegraphics[scale=0.3]{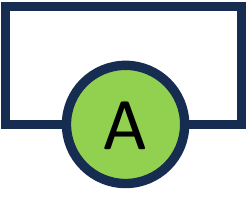}
\end{center}

A tensor network is simply a collection of tensors where some of the indices are contracted. The following tensor network, for instance, evaluates to a number. After one has performed the internal summations over all indices, a single number remains:
\begin{center}
\includegraphics[scale=0.3]{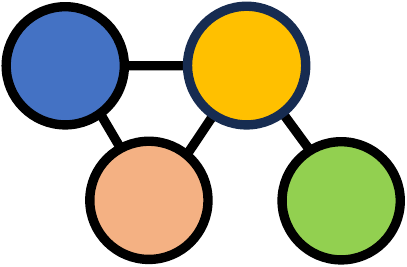}
\end{center}
The following tensor network evaluates to a matrix: it has two free indices, which are not summed over, and are also referred to as physical indices.
\begin{center}
\includegraphics[scale=0.3]{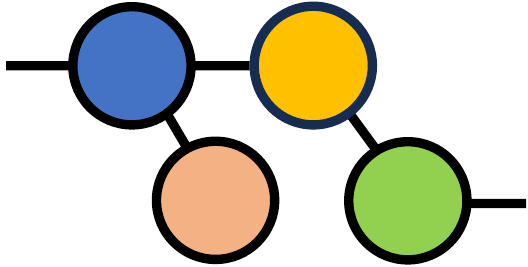}
\end{center}
More generally, a tensor network is an undirected graph where, on each node, stands a tensor
of a degree equal to the number of edges of the node plus the number of physical indices. There are many more interesting tensor networks than the above two trivial examples, and we will see some of them in the course of these lectures. Contracting a tensor network is easy in some cases (as we shall see), but can be exponentially difficult in others. Finding the best order in which to execute the various contractions is, in general, a difficult (NP-complete) problem. We will also see several examples where the order in which one performs the contractions is crucial to keep the computational complexity minimal. So, a common game when designing tensor network algorithms is to look at the algorithmic complexity of the different steps to decide on the best strategy. For instance, in the following example (we take all the bond dimensions to be equal to $D$), doing the contraction vertically first is a bad idea:
\begin{center}
\includegraphics[scale=0.3]{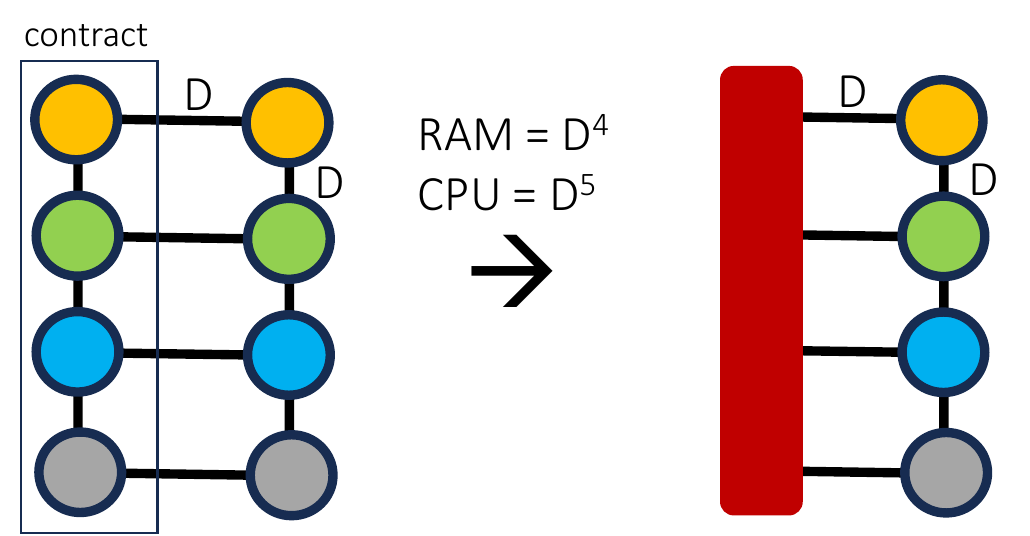}
\end{center} (the CPU scaling above is $D^5$ but it would be $D^{21}$ for an MPS with $20$ tensors).
It is much better to start from the top, contract horizontally, and proceed step by step to the bottom:
\begin{center}
\includegraphics[scale=0.3]{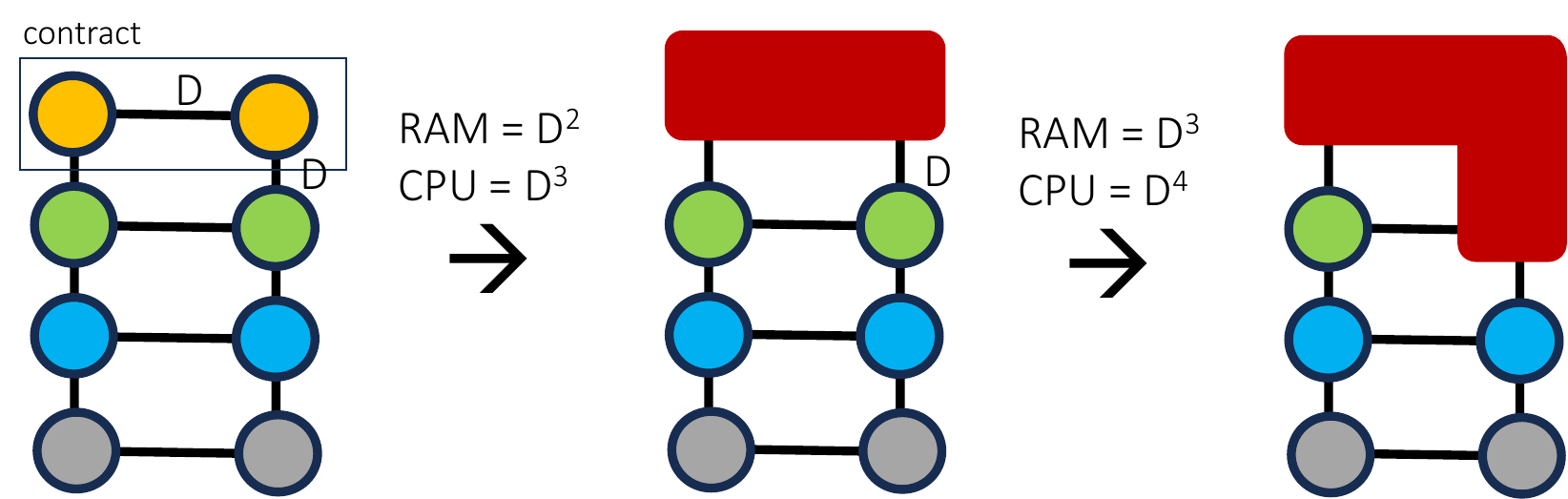}
\end{center} (the CPU scaling above is $D^4$ and would remain $D^{4}$ for an MPS with $20$ tensors).
There are two metrics to look for when contracting a tensor network: the computational
time (CPU) and the memory footprint (RAM). Sometimes one can trade some of one resource for the other \cite{ayral2023}.

\subsection{Factorization}
The second operation that one can perform with a tensor is to fuse and defuse indices.
In an actual computer, there are no matrices or three-leg tensors: memory is organized as a single, very long vector. To store a matrix $M_{ij}$, one unfolds it as a vector $M_\alpha = M_{ij}$, with $\alpha = j d_i + i$ (Fortran style, column-major), or $\alpha = i d_j + j$ (C style, row-major). The resulting index $\alpha$ is a ``composite'' index and the corresponding index operation is denoted:
$\alpha = i \otimes j$ (C style) or $\alpha = j \otimes i$ (Fortran).
\begin{center}
\includegraphics[scale=0.3]{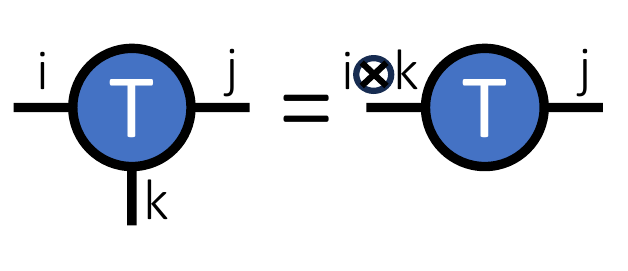}
\end{center}

In tensor network algorithms, indices are constantly fused and defused. A good tensor network library such as ITensor \cite{fishman2022} provides support for the corresponding bookkeeping of ``which indices point to what'', as these operations are simple, yet very prone to errors. The fusing and defusing of indices can also be understood in terms of the flattening tensor:
\begin{center}
\includegraphics[scale=0.3]{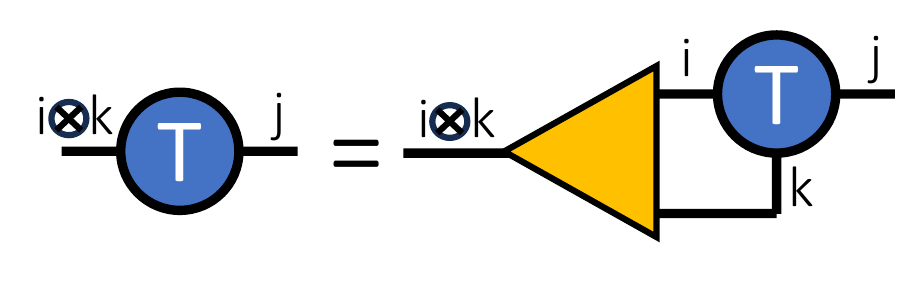}
\end{center}

The main application of index fusing is to map a three- or four- (or more-) legged tensor
onto a matrix. Indeed, once we have a matrix, we recover all the known results from linear algebra, and we can use the corresponding factorization routines. In these lectures, we will use three different types that we will explain in detail in turn:
\begin{itemize}
\item The $QR$ factorization,
\item The Singular Value Decomposition (SVD),
\item The $LU$ factorization, and in particular its partial rank-revealing version.
\end{itemize}
For the moment, let us consider the first one: any matrix $A$ may be written as the product $A=QR$ of a unitary matrix $Q$ and an upper triangular matrix $R$. The $QR$ factorization is simply the process of orthogonalizing a matrix: if we write the matrix $A$ as its set of columns stacked together, $A=(\mathbf{a}_1\;\mathbf{a}_2\cdots\mathbf{a}_P)$, then we first normalize $\mathbf{a}_1$, orthogonalize $\mathbf{a}_2$ with respect to $\mathbf{a}_1$, and continue until we have a full set of orthogonal vectors.

The overall process of fusing, factorizing the resulting matrix, and finally defusing is known as factorization. Sometimes it is also performed in an approximate way (much more about that will follow), and it is then known as factorization with compression. In the case of QR decomposition, for example, the overall factorization of a four-legged tensor $U_{ijkl}$ takes the following form:
\begin{center}
\includegraphics[scale=0.3]{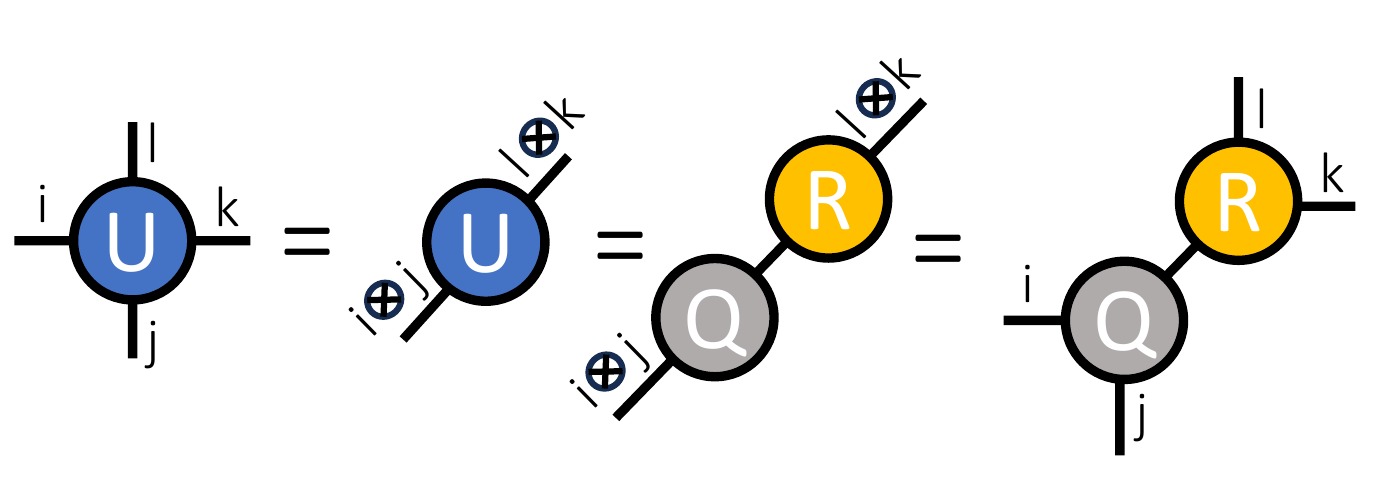}
\end{center}

Note that this factorization is far from unique. For instance, one could factorize the same tensor as follows:
\begin{center}
\includegraphics[scale=0.3]{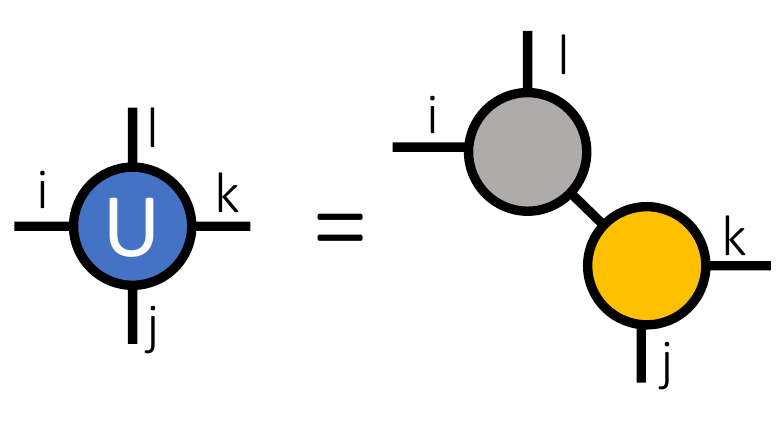}
\end{center}

\subsection{Example: Factorizing the controlled NOT gate}
Let us consider the concrete case of the controlled NOT (C-NOT) two-qubit gate shown in
Eq.~\eqref{eq:control_not}. The matrix of Eq.~\eqref{eq:control_not} was actually an example
of fused indices: the tensor $[C_X]_{c_{\rm in}t_{\rm in}c_{\rm out}t_{\rm out}}$ has four legs
(control-input, target-input, control-out, target-out), and the actual matrix that was shown
was $[C_X]_{c_{\rm in}\otimes t_{\rm in}, c_{\rm out}\otimes t_{\rm out}}$. In plain English, this tensor does nothing to the target qubit if the control qubit is in state
$\ket{0}$ and flips the target qubit if the control qubit is in state $\ket{1}$.
In terms of the Pauli matrices, the C-NOT reads
\begin{equation}
C_X = \frac{1_{\rm c} + Z_{\rm c}}{2} 1_{\rm t} +
\frac{1_{\rm c} - Z_{\rm c}}{2} X_{\rm t}.
\end{equation}
This is a rank-2 (the internal index takes two values) factorization:
\begin{equation}
[C_X]_{c_{\rm in}t_{\rm in}c_{\rm out}t_{\rm out}} =
\sum_{a=0}^{1}
A_{a,c_{\rm in}c_{\rm out}}
B_{a,t_{\rm in}t_{\rm out}},
\end{equation}
with the four tensors (seen as matrices) given by
$A_0 = (1+Z)/2$, $A_1 = (1-Z)/2$, $B_0=1$, and $B_1=X$.
It follows that the usual notation used for C-NOT,
\begin{center}
\includegraphics[scale=0.3]{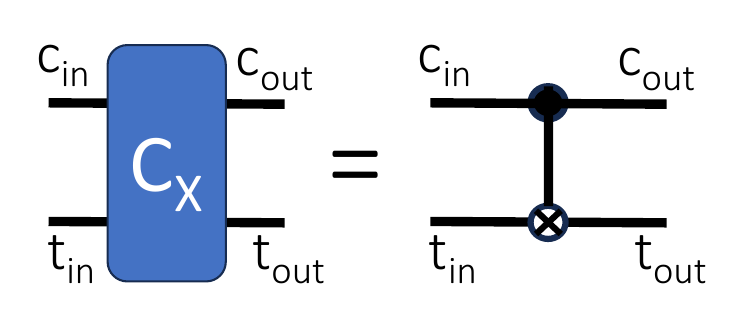},
\end{center}
is not just a convenient notation, but it also has a meaning in the tensor sense:
this four-legged $2\times 2 \times 2\times 2$ tensor factorizes as the product of two
$2\times 2 \times 2$ tensors. This is already a form of (exact) compression, since the most general two-qubit gate factorizes into a product of rank 4. As a side remark, this is the reason the Google team used a different two-qubit gate in their 2019 quantum supremacy experiment \cite{arute2019}: they wanted a gate that created as much entanglement as possible, hence full rank, in order for the corresponding experiment to be as hard to simulate as possible (but at the cost of the gate being useless for actual computations).

A very large fraction of the algorithms that we will discuss in these lectures (but not all) amount to a sequence of contractions and factorizations. The overall idea is to seek the solution of a large problem (whose solution is a large tensor) in terms of its tensor network representation. The algorithms update the small tensors that form the network one after the other until the problem is solved without \emph{ever} considering the large tensor itself. This paragraph might be a bit obscure at this stage, but will become clearer later in the lectures.

A fun fact about $C_X$: at first sight, it looks like this gate does nothing to the control
qubit; it only acts on the target qubit. However, this is just an illusion: remembering that
$Z = HXH$ and $X = HZH$ (the Hadamard gate maps the eigenstates of $Z$ onto the eigenstates of $X$), one can rewrite $C_X$ as
\begin{align}
C_X &= H_{\rm c}H_{\rm t}\left[
\frac{1_{\rm c} + X_{\rm c}}{2} 1_{\rm t} +
\frac{1_{\rm c} - X_{\rm c}}{2} Z_{\rm t}
\right] H_{\rm c}H_{\rm t} \nonumber \\
&= H_{\rm c}H_{\rm t}\left[
1_{\rm c}\frac{1_{\rm t} + Z_{\rm t}}{2}  +
X_{\rm c}\frac{1_{\rm t} - Z_{\rm t}}{2}
\right] H_{\rm c}H_{\rm t}.
\end{align}
In other words, switching basis switches the role of the control and target qubits!

\section{Basic quantum computer emulators}
\label{sec:exact}

To be truly useful, tensor networks must be used in conjunction with a compression scheme.
This aspect will be the subject of most of these lectures. However, for the present section, we limit ourselves to an ``exact mode'' of using tensor networks, which already has some interesting applications and will allow us to make use of the concepts introduced above.

\subsection{A quantum circuit is a tensor network.}
Let us consider an explicit quantum circuit that builds a GHZ (Greenberger-Horne-Zeilinger) state:
\begin{center}
\includegraphics[scale=0.3]{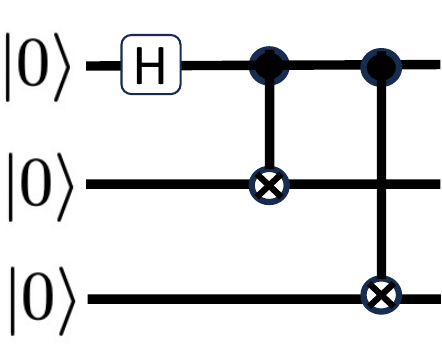}
\end{center}
It creates the state
\begin{equation}
\ket{\Psi} = \frac{1}{\sqrt{2}} \left( \ket{000} + \ket{111} \right),
\end{equation}
using a Hadamard gate and two C-NOT gates, as one can verify easily.
Now please look back at Eq.~\eqref{eq:1qubit_gate} and Eq.~\eqref{eq:2qubit_gate}, which define one- and two-qubit gates. These are actually the same as the definition of the contraction of two tensors. It follows directly that the quantum circuit (which is intended as a set of instructions that the quantum computer must run) is also a tensor network, and contracting this tensor network results in the many-qubit wavefunction:
\begin{center}
\includegraphics[scale=0.3]{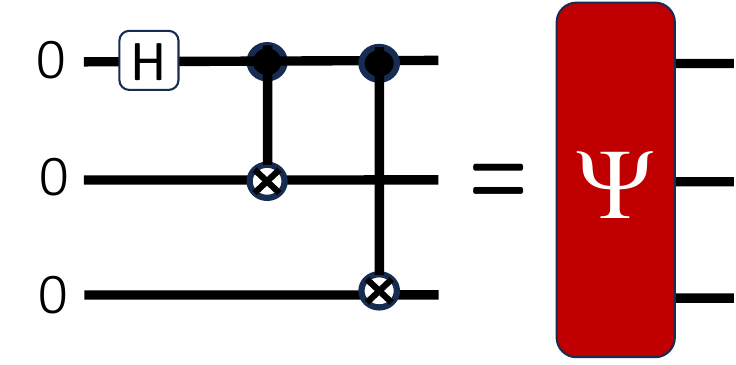}
\end{center}
The problem of emulating a quantum computer on a classical one is therefore reduced to contracting the corresponding quantum circuit. There exist various strategies for doing this.

\subsection{Long and narrow quantum circuits: the full state simulator}
Let us start with the simplest emulator,
the so-called state vector emulator of a quantum computer.
One begins by allocating a very large vector $\Psi_\alpha$ of $2^N$ complex numbers.
This is $16\times 2^N$ bytes of memory when double precision is used, so 16 kB for 10 qubits,
16 MB for 20 qubits, 16 GB for 30 qubits, and 16 TB for 40 qubits. On a laptop, one should therefore be able to simulate about 20-30 qubits in this mode. Then,
one interprets $\alpha$ as
\begin{equation}
\alpha = i_1 \otimes i_2 \otimes \cdots{} \otimes i_N.
\end{equation}
The initialization of all qubits in state $\ket{0}$ amounts to setting the first
element of the vector to 1 and all others to zero:
$\Psi^{(0)}_\alpha = \delta_{\alpha,0}$. To perform the contraction of the circuit, one
simply applies Eq.~\eqref{eq:1qubit_gate} and Eq.~\eqref{eq:2qubit_gate} one after the other, from left to right:
\begin{center}
\includegraphics[scale=0.25]{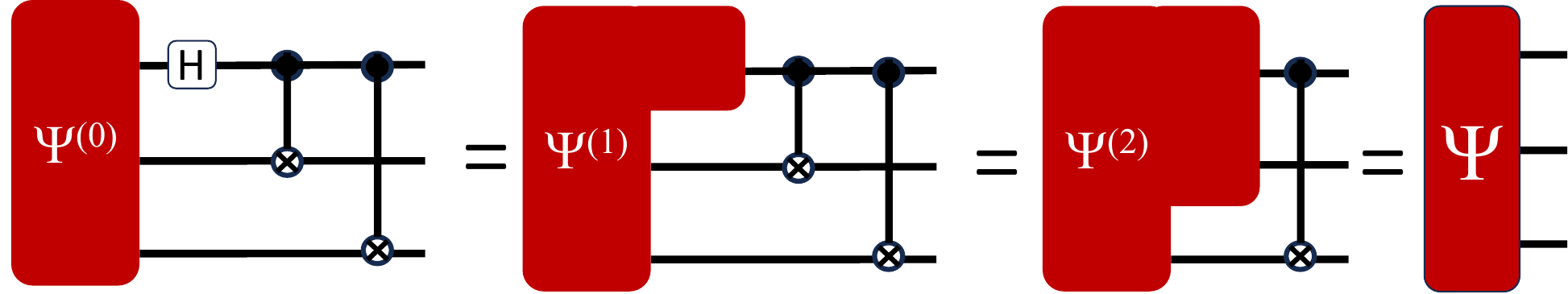}
\end{center}
This emulator has an exponential memory footprint but a run time that scales linearly with the number $N_g$ of gates: $O(N_g 2^N)$. It is therefore suitable for very deep circuits with very few qubits. But its main appeal is that it is straightforward to implement
(although efficient parallel versions may be tricky).

\subsection{Tall and skinny circuits: exact MPS simulations of a quantum computer}
We will now consider the opposite limit where there are many qubits but the depth of the circuit is rather limited. This second emulator uses a Matrix Product State (MPS), which is
the tensor network we will use most often in these lectures. 

\subsubsection{MPS definition}
An MPS is simply a linear tensor network like in this schematic:
\begin{center}
\includegraphics[scale=0.3]{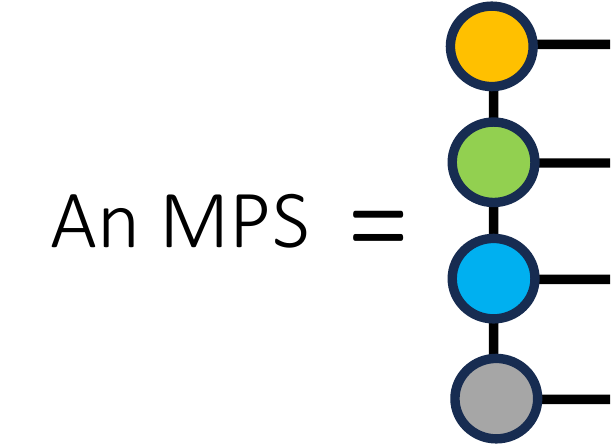}
\end{center}
It is the most common tensor network in the literature, and we will see it over and over in these lectures. We seek a decomposition of the form
\begin{center}
\includegraphics[scale=0.25]{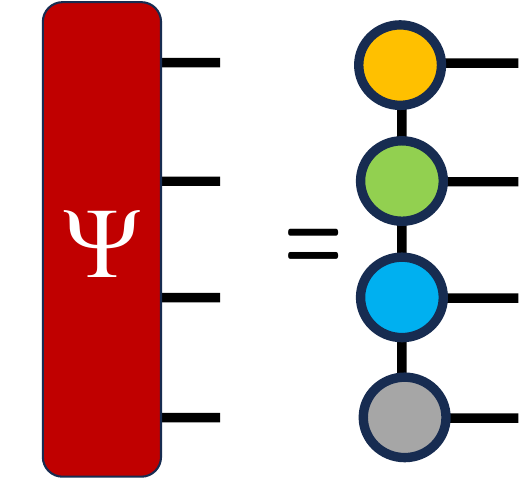}
\end{center}
Such a decomposition can always be found. An algorithm to build such a representation goes as follows. One fuses the last $n-1$ indices of the tensor and factorizes the corresponding
matrix. One repeats the procedure with the remaining tensor until one has exhausted all the physical indices. This algorithm is best explained graphically:
\begin{center}
\includegraphics[scale=0.25]{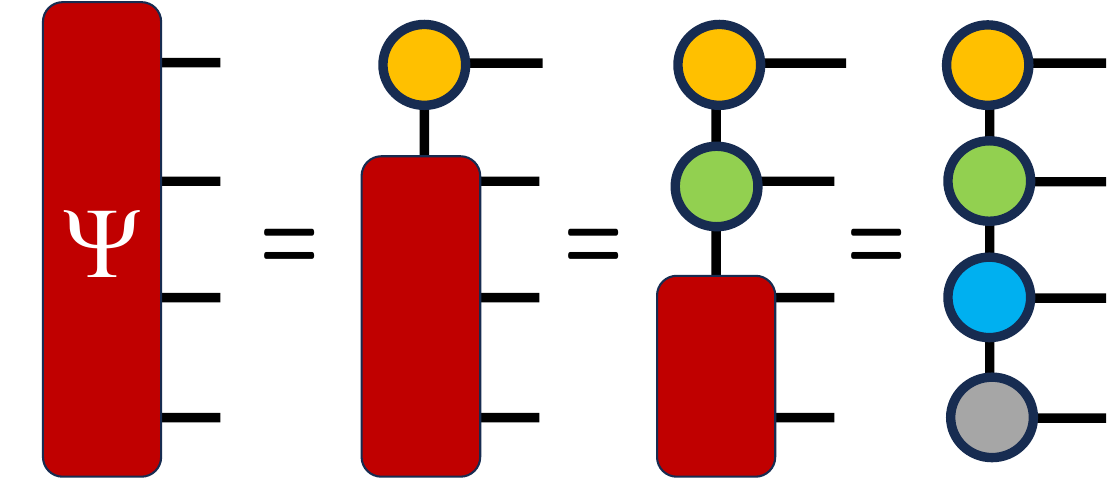}
\end{center}
It is not a very practical algorithm except for small tensors, because it requires access to all the $d^N$ elements of $\Psi$. We will see a much faster algorithm later in these lectures (Tensor Cross Interpolation) in section \ref{sec:tci}. This construction, however, already reveals a few properties of MPS.
When factorizing the $a$-th physical leg, the matrix which is factorized is of size
$d \chi(a-1)$ times $d^{N-a}$, where $\chi(a)$ is the rank of the virtual (vertical line in the drawing) index. Since the rank of a matrix is smaller than the smallest of its dimensions, it follows that the rank $\chi(a)$ grows at most as fast as $d^{a}$, with a maximum in the middle of the MPS where $\chi \le d^{N/2}$.
The cost is proportional to $d^n$, i.e.\ is exponential in $n$.

Let us get away from drawings for an instant.  The explicit form of the
MPS of $\Psi$ is
\begin{equation}
\label{eq:mps}
\Psi_{i_1\cdots{}i_N}= \sum_{\alpha_1\cdots{}\alpha_{N-1}} M^1_{\alpha_1}(i_1)
M^2_{\alpha_1\alpha_2}(i_2)M^3_{\alpha_2\alpha_3}(i_3)\cdots{}
M^{N}_{\alpha_{N-1}}(i_{N}),
\end{equation}
where the matrix $M^{a}(i_a)$ is actually a three-index tensor that we treat as
a matrix that depends on the last (physical) index. In other words,
\begin{equation}
\Psi_{i_1\cdots{}i_N}=  M^1(i_1) \times M^2(i_2) \times M^3(i_3)\times\cdots\times
M^{N}(i_{N})
\end{equation}
is just a product of (physical index dependent) matrices, hence the name ``MPS''.
(To be precise, the first element $M^1(i_1)$ and the last element $M^{N}(i_{N})$ are, respectively, a row and a column vector.)

Note that the MPS decomposition is by no means unique; there is what is known as ``gauge freedom'': Consider an invertible matrix $U$, then replacing $M^1(i_1)$ with
$M^1(i_1) U$ and $M^2(i_2)$ with $U^{-1} M^2(i_2)$ leaves the MPS unchanged. We will later use this freedom to work with a very convenient gauge known as the ``canonical form''.

\subsubsection{MPS exact emulator: nearest neighbor gates}
We now have all the tools to build a basic MPS-based quantum computer emulator.
We start with all qubits in state 0:
\begin{equation}
\Psi_{i_1\cdots{}i_N}=  \delta_{i_1,0}\cdots{}\delta_{i_N,0}
\end{equation}
(but any other product state would work as well). This state is obviously a (trivial)
MPS with rank $\chi=1$ everywhere. Now we want to apply a gate to this MPS. The goal is to put the resulting state back into MPS form so that we can proceed with the rest of the circuit. If the gate is a single-qubit gate, then we can trivially contract the gate with the corresponding MPS tensor as follows:
\begin{center}
\includegraphics[scale=0.25]{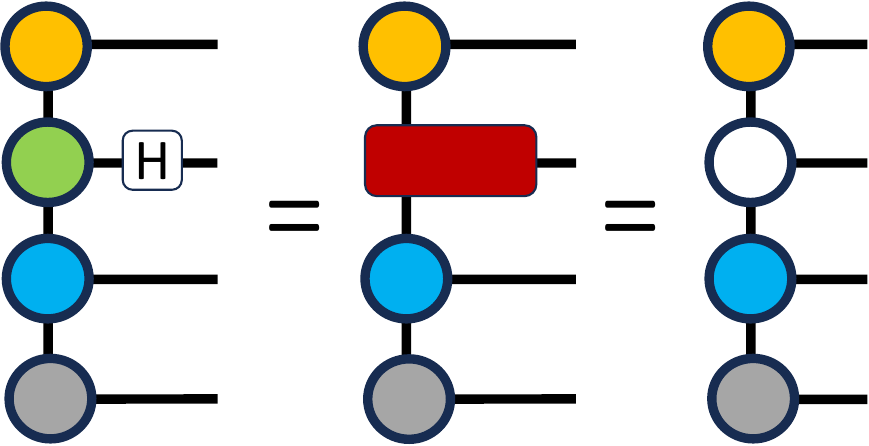}
\end{center}
(in the last step, we did nothing: we redrew the square as a circle to highlight the fact that the state was already in MPS form). Importantly, the rank $\chi$ does not increase when we apply 1-qubit gates. Now, if the gate is a two-qubit gate between neighboring qubits (say, a C-NOT), then we need an extra step, since after contraction we don't have an MPS anymore. We simply factorize the resulting tensor using e.g.\ QR or SVD:
\begin{center}
\includegraphics[scale=0.25]{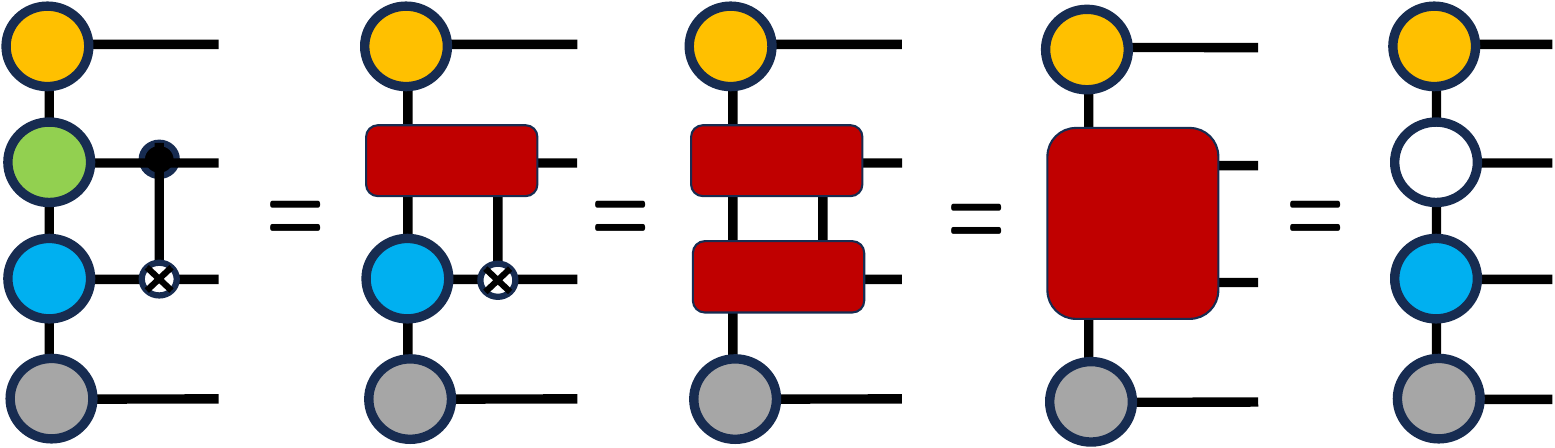}
\end{center}
The factorization corresponds to the last step above.

The rank $\chi$ now may increase by a factor of 2 (up to 4 for the most general two-qubit gate). This is obvious from the middle step above, which is actually already in MPS form if we fuse the two vertical indices connecting the red squares. It follows that the computational complexity of this algorithm is exponential in depth but only linear in the number of qubits $N$. This contrasts with the previous state-vector algorithm that was exponential in $N$ even when there was actually no entanglement in the state. This is a very general statement about algorithm complexity: we have exploited an additional piece of information (the fact that the initial state of a quantum computer is a product state) and this results in reduced computational complexity. Note that in the application of the C-NOT gate above, we have used the factorization of the $C_X$ gates discussed before. If the gate has not been factorized, we may either factorize it or use the following sequence directly:
\begin{center}
\includegraphics[scale=0.25]{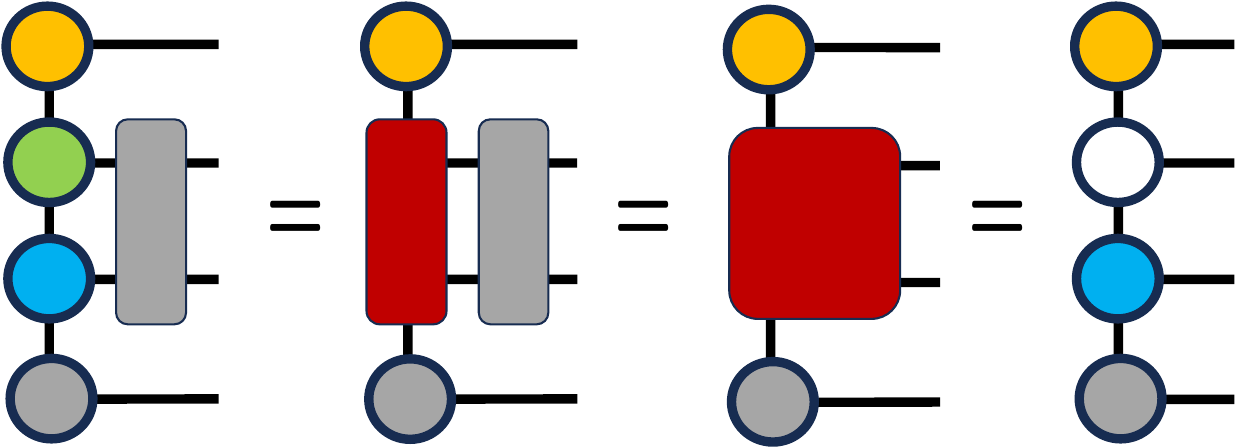}
\end{center}

\subsection{Extension to arbitrary gates: introducing the MPO-MPS product}
The last missing piece is being able to treat two-qubit gates acting on qubits that are not neighbors.
The proper way to do that is to introduce so-called Matrix Product Operators (MPO), which are to matrices what MPS are to vectors. They look like this:
\begin{center}
\includegraphics[scale=0.3]{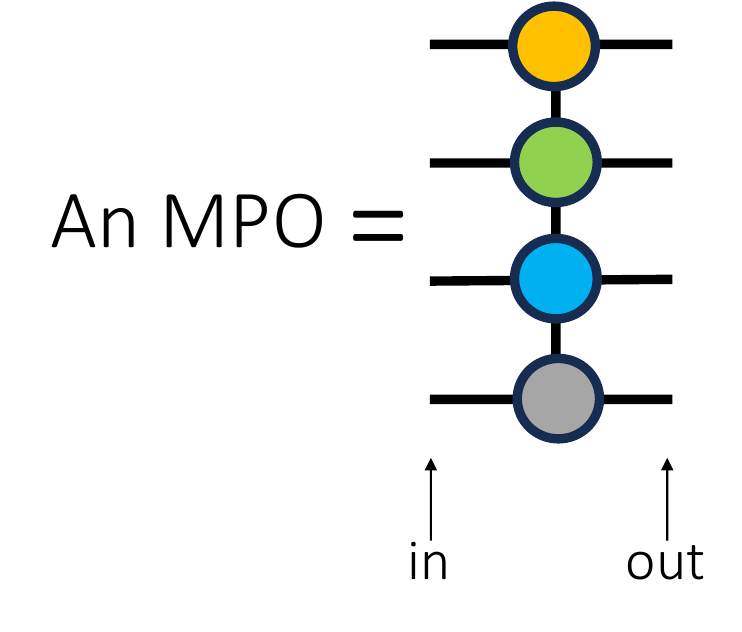}
\end{center}
We will see them almost as often as MPS. 
Note that an MPO can be put into MPS form by fusing the output and input indices together for each tensor. 
This may come in handy when applying some MPS
algorithms to them.
In our case, building the MPO amounts to introducing the
4-index tensor $I_{ii'\alpha\alpha'} = \delta_{ii'}\delta_{\alpha\alpha'}$, where
$i$ and $i'$ are the input and output physical indices while $\alpha$ and $\alpha'$ are the virtual indices. Denoting this tensor by an orange square graphically, the MPO for a C-NOT between the first and the last qubit looks like:
\begin{center}
\includegraphics[scale=0.25]{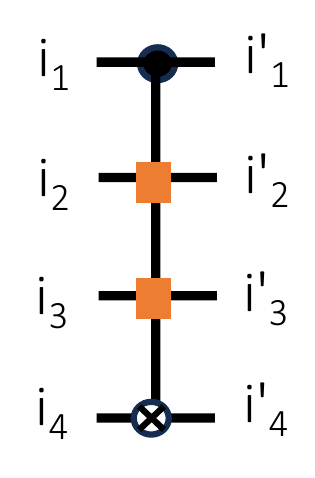}
\end{center}
Now, to apply the gate, we need to perform an MPO-MPS product, which is the tensor-network version of a matrix-vector product. This can be done in several ways. Here we present the so-called ``zip-up'' algorithm:
\begin{center}
\includegraphics[scale=0.25]{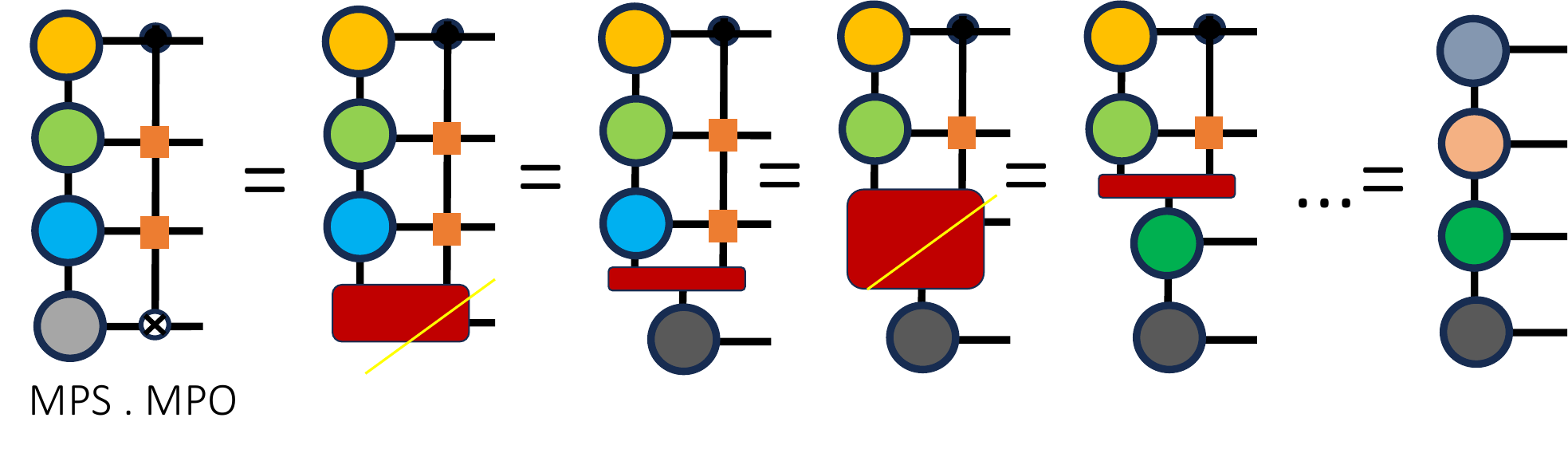}
\end{center}
One starts with contracting the bottom two tensors. Then we split (say with SVD or QR decomposition) the resulting tensor across the yellow line. This step already provides the bottom tensor of the resulting MPS.
Then one contracts the next two tensors (one after the other), splits across the yellow line, and repeats until the full MPS is obtained.

An alternative to introducing the above MPO is to stick to neighboring gates using the so-called SWAP 2-qubit gate. The SWAP gate does exactly what its name suggests and can be constructed with three C-NOT gates:
\begin{center}
\includegraphics[scale=0.25]{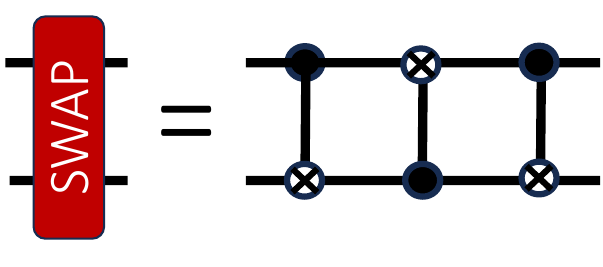}
\end{center}
One can easily verify that SWAP $\ket{01} = \ket{10}$, i.e.\ that it permutes indices, transforming $\Psi_{i_1 i_2}$ into $\Psi_{i_2 i_1}$. To apply a two-qubit gate to distant qubits, one simply brings them together, applies the gate, and then (optionally) returns them to their original positions using the following sequence:
\begin{center}
\includegraphics[scale=0.25]{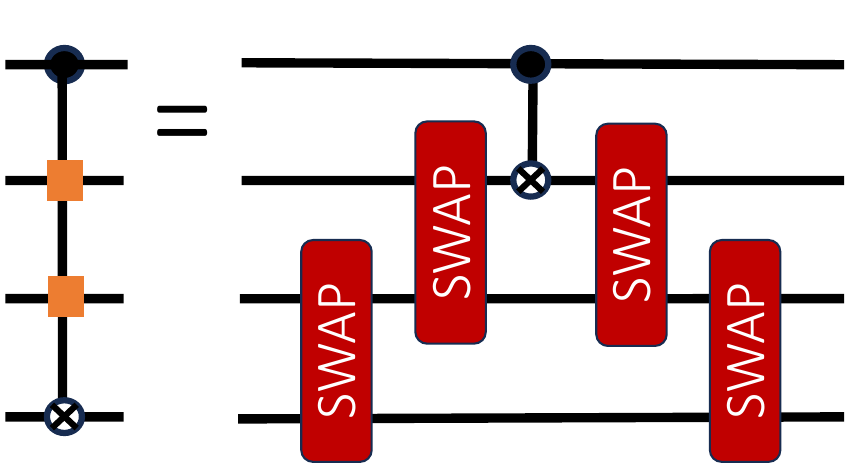}
\end{center}
Note that the SWAP gate can generate entanglement. This method is typically significantly more costly than using the MPO approach.

\begin{figure}
  \begin{center}
    \includegraphics[width=10cm]{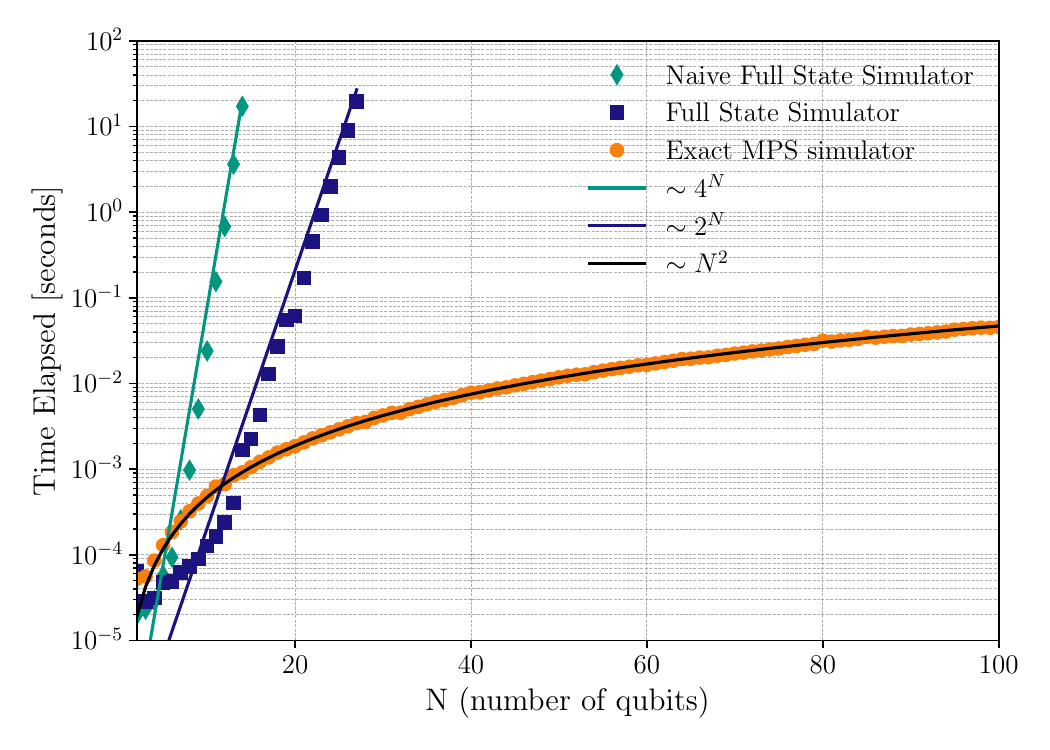}
    \caption{ \label{fig:ghz} Time needed for the construction of a GHZ state
      $[\ket{00\cdots{}0}+\ket{11\cdots{}1}]/\sqrt{2}$ with $N$ qubits using the full state simulator (exponential scaling) and the exact MPO.MPS simulator (linear scaling per gate, here $\chi=2$ is exact). Contributed by Chen-How Huang.}
  \end{center}
\end{figure}

During the hands-on sessions, the students began with the implementation of the
above two algorithms: the full-state simulator and the exact MPS simulator
for an arbitrary quantum circuit. Fig.~\ref{fig:ghz} contains a comparison between the run time of these two algorithms in a case which is particularly favorable to
the MPS approach: the simple circuit with one Hadamard gate and $N-1$ C-NOT
gates that builds the GHZ state. The full state simulator has a run time which
scales as $\sim N 2^N$, while the MPS approach scales exponentially faster as $\sim N^2$ because the GHZ state is a simple rank-$2$ MPS (this statement can be proved using the addition of two MPS explained in section \ref{sec:adding}). The figure also shows a very naive algorithm where one explicitly builds the dense matrix representing the action of the quantum circuit before applying it to the initial state ($\sim N 4^N$).

\subsection{Calculating observables from an MPS}
\subsubsection{The tensor network route to observables}
We said that we have a full-fledged MPS-based emulator, and that's true because we hold the entire state of the system. Furthermore, the MPS structure allows us to easily calculate observables, such as correlation functions. Suppose we want to calculate $\bra{\Psi} Z_j Z_k \ket{\Psi}$. Then we need to construct the tensor network
\begin{center}
\includegraphics[scale=0.3]{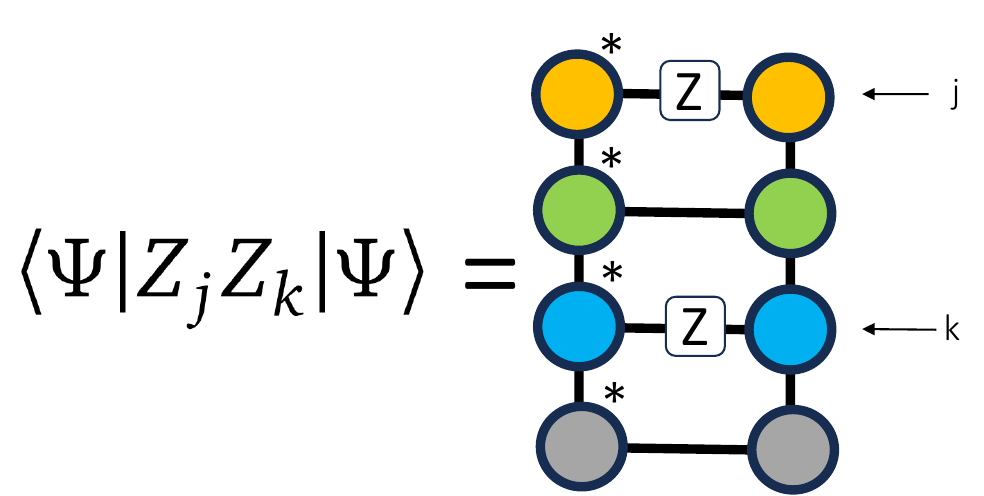}
\end{center}
and then contract it. The contraction is performed vertically as follows:
\begin{center}
\includegraphics[scale=0.25]{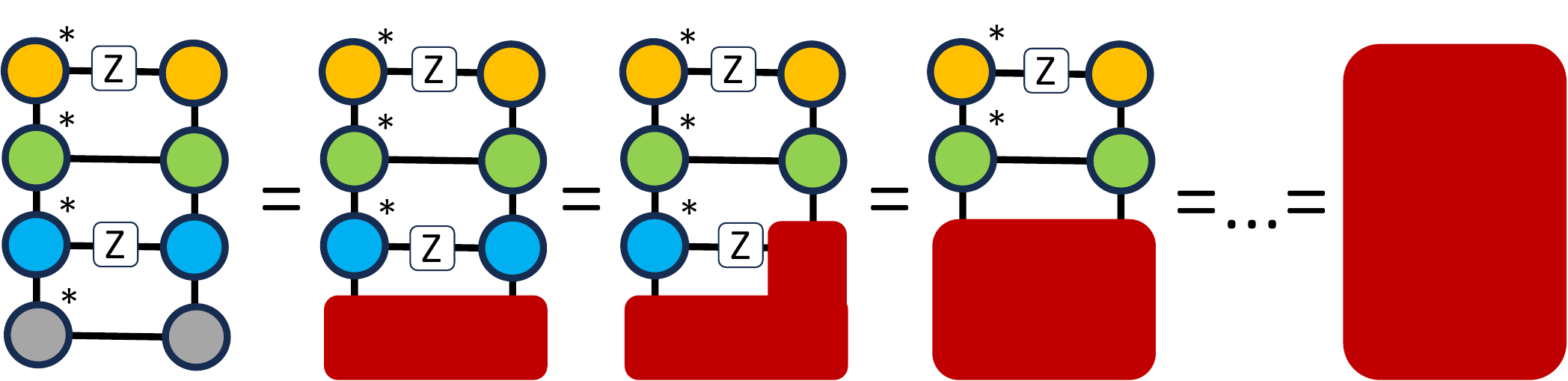}
\end{center}

\subsubsection{Direct sampling of an MPS}
Another route to calculate observables is to get samples from the MPS, i.e.\ we want to obtain $M$
bitstrings $\mathbf{i} =i_1\cdots{}i_N$ that are distributed according to $P(\mathbf{i})=P(i_1\cdots{}i_N)=|\Psi_{i_1\cdots{}i_N}|^2$. Observables such as correlation functions are then measured
by averaging over these configurations $\mathbf{i}(1)\cdots{}\mathbf{i}(M)$:
\begin{equation}
\bra{\Psi} Z_j Z_k \ket{\Psi} = \sum_\mathbf{i} (-1)^{i_j}(-1)^{i_k} P(\mathbf{i})
\approx \frac{1}{M}\sum_{\alpha=1}^M  (-1)^{\mathbf{i}(\alpha)_j}(-1)^{\mathbf{i}(\alpha)_k}.
\end{equation}
Honestly, this is not a very nice route, since using the algorithm of the previous section is far quicker and more accurate. Here we are limited by the law of large numbers; hence our accuracy will only improve as $1/\sqrt{M}$. Indeed, $1/\sqrt{M}$ means that each additional digit in accuracy corresponds to a factor of 100 increase in computing time. Since many applications require at least three to four digits, this can quickly become problematic. However, this is what one would do on an actual quantum computer (where one has no other choice). Hence, if we want to claim that we can emulate a quantum computer, we must show that we can sample an MPS.

Fortunately, an important property of an MPS is that it can be sampled exactly. Suppose that we have $\Psi$ in the form of an MPS and we want to sample a single element $i_1^*\cdots{}i_N^*$ from the distribution $P(i_1\cdots{}i_N)=|\Psi_{i_1\cdots{}i_N}|^2$. In the most general case, one has to resort to Markov chain Monte Carlo (e.g.\ Metropolis algorithm) for this task, which has some limitations (ergodicity, thermalization, correlations). For MPS, however, we have a simple specific algorithm for this task. We will use the Bayesian chain rule
\begin{equation}
P(i_1\cdots{}i_N)=  P(i_N|i_1\cdots{}i_{N-1})\cdots{}P(i_3|i_1i_2)P(i_2|i_1)P(i_1),
\end{equation}
where $P(A)$ denotes the probability to have $A$ and $P(A|B)$ is the probability of $A$ given $B$. So we will start with sampling $i_1$, then we will sample $i_2$ knowing the sample we got of $i_1$, and continue until we have obtained the entire bitstring.

To implement this scheme we need to calculate first $P(i_1)$, which is given by
\begin{equation}
P(i_1) = \sum_{i_2\cdots{}i_N} P(i_1\cdots{}i_N).
\end{equation}
Graphically, $P(i_1)$ is the following tensor network:
\begin{center}
\includegraphics[scale=0.25]{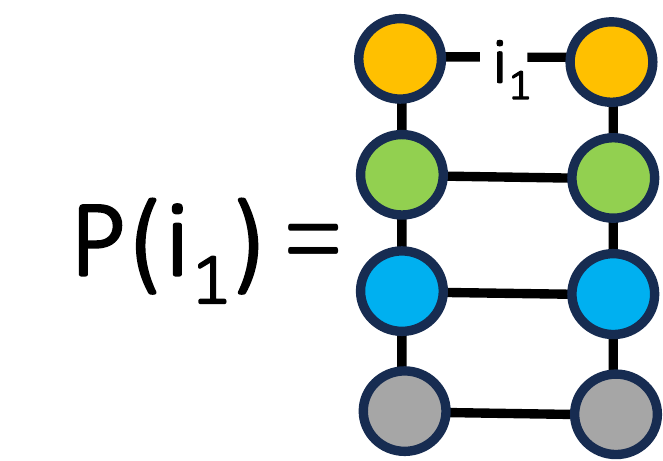}
\end{center}
Now, to calculate this number, we need to contract this tensor network. The strategy will
be to start from the bottom and slowly contract our way up. The sequence of contraction is:
\begin{center}
\includegraphics[scale=0.25]{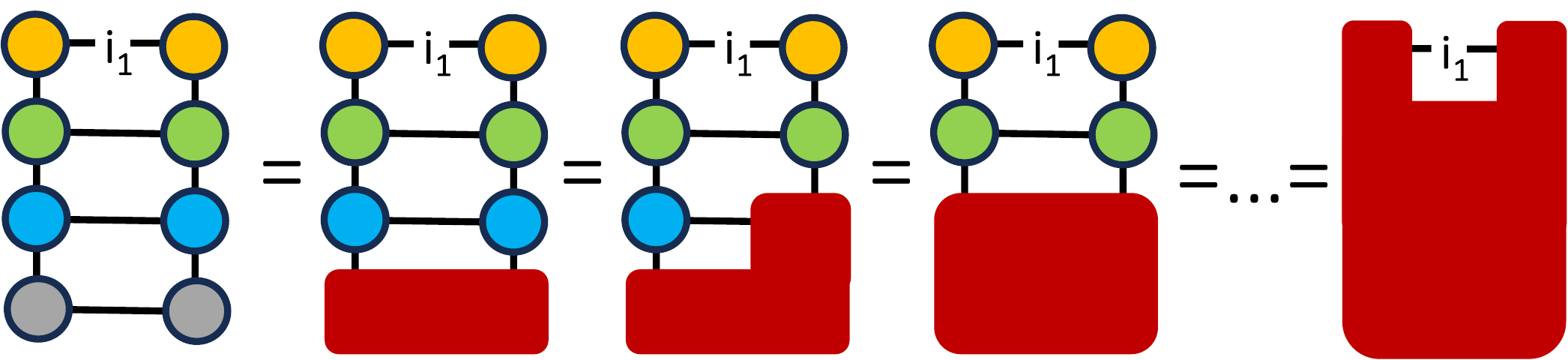}
\end{center}
In the above algorithm, the memory needed to keep the red tensor in memory is $\chi^2 d$, and the computing time scales as $N \chi^3 d$. Very importantly, it is linear in $N$, while the original tensor is exponential in $N$. We will see later that using the canonical form of the MPS can further simplify this calculation.

Note that there are many different contracting sequences that can be chosen. Picking the wrong one will lead to the correct result (which does not depend on the order of contraction) but can be catastrophic in terms of computing time and memory footprint. For instance, contracting all the vertical lines first leads to an intermediate step where there are two instances of $\Psi$, hence two objects of size $d^N$.

Once we have calculated $P(i_1)$, we draw a random number uniformly distributed inside $[0,1]$. If this number is smaller than $P(i_1=0)$, then $i_1^*=0$, otherwise $i_1^*=1$. To proceed, we compute $P(i_1^*,i_2)$, which is given by
\begin{center}
\includegraphics[scale=0.25]{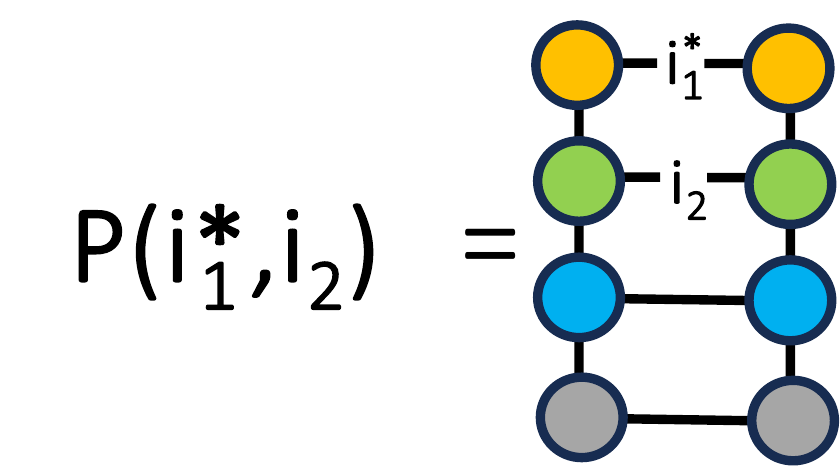}
\end{center}
Then we sample $i_2$ using $P(i_2|i_1=i_1^*) = P(i_1=i_1^*,i_2)/P(i_1=i_1^*)$ to get $i_2^*$. The algorithm continues until we have sampled all qubits.

\subsection{Amplitude simulations of a quantum computer}
The above simulations are not fair to classical computers because the results of the calculation consist of the entire wavefunction $\Psi$ for all possible indices. An actual quantum computer does not yield this exponentially large piece of information -- only a single sample ($N$ bits with randomness).
The question is therefore whether there are ways to obtain samples without calculating the entire distribution. Indeed, this can be done by calculating only a few amplitudes.

The algorithm to do so was proposed in \cite{bravyi2022} and is very simple. One starts with an empty circuit which has the trivial distribution
$P^{(0)}_{i_1\cdots{}i_N} = |\Psi^{(0)}_{i_1\cdots{}i_N}|^2 = \prod_\alpha \delta_{i_\alpha,0}$.
We draw a sample from this distribution, which is trivial, and get $\ket{00\cdots{}0}$.
Next, we add the first gate. Let's suppose it is a two-qubit gate between qubit 1 and
qubit 2. The trick is to note that this gate is only going to affect these two qubits,
not the other ones. Hence, we can re-use the end of our sample, and we only need to resample the first two qubits. More formally, suppose that we have a sample $\ket{i_1^{(n)}\cdots{}i_N^{(n)}}$
distributed according to  $P^{(n)}_{i_1\cdots{}i_N} = |\Psi^{(n)}_{i_1\cdots{}i_N}|^2$.
We only need to compute the four amplitudes (two for 1-qubit gates)
\begin{equation}
q_{i_1i_2} = \Psi^{(n+1)}_{i_1i_2i_3^{(n)}\cdots{}i_N^{(n)}}.
\end{equation}
This allows us to draw $(i_1^{(n+1)},i_2^{(n+1)})$ from the distribution
\begin{equation}
P^{(n+1)}_{i_1i_2} = \frac{|q_{i_1i_2}|^2}{\sum_{i_1i_2} |q_{i_1i_2}|^2},
\end{equation}
and use $i_\alpha^{(n+1)}= i_\alpha^{(n)}$ for the other bits.
The correctness of this algorithm stems from the fact that
\begin{equation}
P^{(n+1)}_{i_3i_4\cdots{}i_N} \equiv \sum_{i_1i_2} P^{(n+1)}_{i_1\cdots{}i_N} = \sum_{i_1i_2} P^{(n)}_{i_1\cdots{}i_N} \equiv P^{(n)}_{i_3i_4\cdots{}i_N},
\end{equation}
which is trivial once one remembers that the gates are unitary.
Then we simply use Bayes' rule to calculate the probability of $(i_1,i_2)$ given the rest of the sample. This proves that being able to calculate amplitudes (sometimes called a ``strong'' simulation because we obtain knowledge of the full state) is more difficult than just being able to sample (correspondingly called a ``weak'' simulation). Indeed, there are quantum computing algorithms for computing amplitudes, but they are significantly more challenging than just measuring the qubits at the end (try googling ``Hadamard test'').

So, in order to produce samples, we are left with the calculation of a ``few'' (of the order of the number of gates in the circuit) amplitudes of the form
\begin{center}
\includegraphics[scale=0.25]{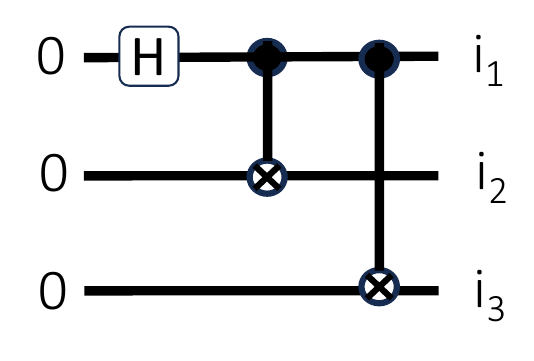}
\end{center}
The above tensor network is \emph{just a number} (for fixed values of $i_1\cdots{}i_N$), not an
exponentially large object. Calculating it can be (exponentially) difficult, though. Yet it is always simpler than the exact MPS approach. There exist multiple strategies to compute these numbers.

One simple possibility is to use the exact MPS approach discussed above for \emph{half} of the depth of the circuit, and then treat the remaining half with another MPS backwards, starting from the \emph{end} of the circuit (we can do this, because at the end, we're back to a simple product state). To get the amplitude, we simply calculate the scalar product between the two obtained MPS (calculating a scalar product is essentially the same as the algorithm used to calculate the partial sums to sample an MPS) \cite{ayral2023}. If the rank scales as $\chi \sim e^{\alpha D}$, where
$\alpha$ is a constant that depends on the type of circuit, and $D$ is the depth of the circuit, then using the fact that calculating a scalar product requires $O(N \chi^3)$ operations, the total computing time scales as $e^{\alpha 3D/2}$, which is significantly smaller than the $e^{\alpha 2D}$ operations needed to perform the full exact MPS evolution and calculate the amplitudes at the end.

There are, however, better strategies to contract a tensor network \cite{pan2019}. Generally speaking, they are based on an analysis of the underlying graph structure of the tensor network to determine the best order to execute the contraction. This best order usually cannot be determined exactly (it is an exponentially hard problem \cite{pfeifer2014}), but good heuristic approaches are known. Practical implementations must also consider how to split the work in such a way that it may be performed on multiple
CPUs or GPUs. A common approach is to \emph{slice} certain indices. Slicing consists of fixing the value of a certain index inside a process and distributing the different values of this index over multiple processes.
This way, each different process has a simpler tensor network to contract. The results of all the processes are added together at the end of the calculation.

\subsection{Some remarks on ``quantum supremacy''}
One last remark before we move on: the difficulty of simulating a quantum state exactly (essentially the parameter $\alpha$ above) depends very strongly on the type of quantum circuit. Empirically, it seems that
useful circuits, i.e.\ those that have a lot of internal structure, are much easier to simulate than circuits that have been purposely designed to be somewhat random (by using e.g.\ random angles for some of the rotations). An extreme example of the latter are the so-called quantum supremacy experiments, designed explicitly to be as hard to simulate as possible \cite{arute2019}. The corresponding quantum circuit essentially consists of applying gates with random angles to all the qubits at once, the two-qubit gates being designed such that the entanglement grows as fast as possible (before decoherence sets in!). An example of a random circuit for nearest-neighbor two-qubit gates in one dimension is shown below:
\begin{center}
\includegraphics[scale=0.5]{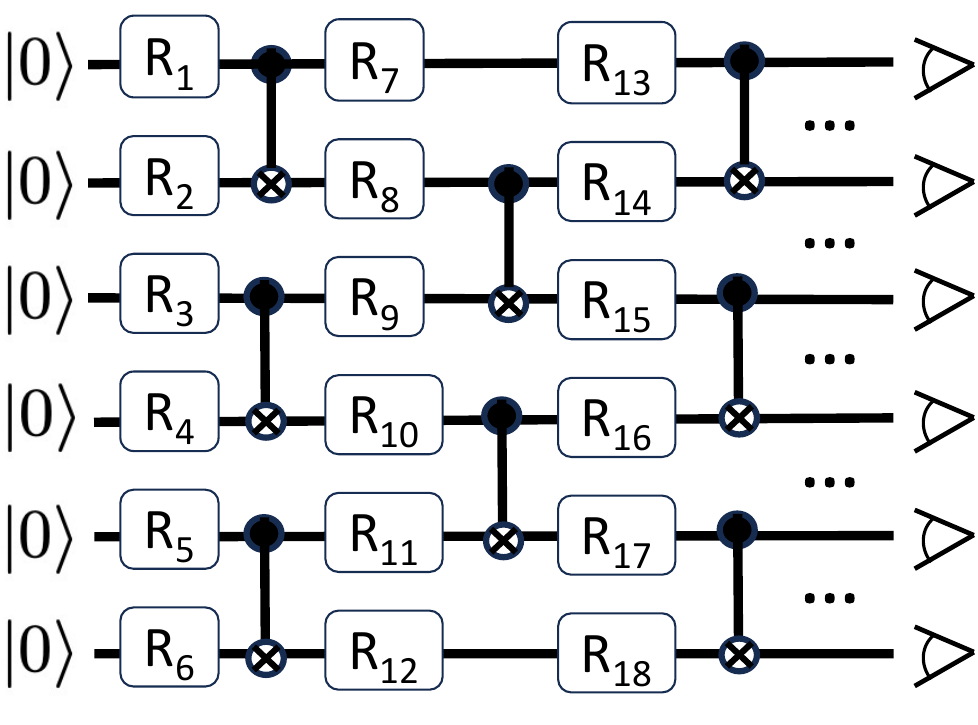}
\end{center}
where the $R_i$ gates are different rotations around an arbitrary axis of an arbitrary angle. The actual quantum supremacy experiments correspond to a 2D version of this circuit.

Besides the fact that the authors of these
experiments greatly overestimated the difficulty of performing the associated simulations -- the initial estimate of 10\,000 years on the largest supercomputer on Earth had to be re-examined down to 6 seconds \cite{morvan2024} (see Table I) -- they produce a totally chaotic state. Such a state has no structure, with all the amplitudes $\Psi_{i_1\cdots{}i_N}\sim 1/\sqrt{2^N}$ being almost equally probable. Almost, but not exactly: some qubit configurations are slightly more probable than others, which is the only remaining quantumness present in these trivial states. In other words, if one cannot simulate these experiments, there is absolutely no way to distinguish their output from the output of a perfect random generator. The supremacy tag might be a little exaggerated, in our humble opinion. These are merely experiments that are difficult to simulate, but the list of things that are difficult to simulate is very long. For instance, try to predict the shapes of the pieces of a glass that you break by throwing it against a wall. It's a very hard task, even if the experiment is perfectly controlled. Yet it's not really worth calling ``classical supremacy''.
It remains, however, that in general the exact simulation techniques discussed in this section do have an exponential cost in the number $N$ of qubits or the depth $D$ of the circuit. In general, but not always.

Before leaving the topic of exact quantum computer emulators, we would like to briefly mention that the story does not end here. For instance, very fast emulators exist for
quantum circuits that consist only of a restricted set of gates (the so-called Clifford gates) or mostly of such gates \cite{aaronson2004}. Even though these states may be highly entangled, systems with thousands of qubits can be easily simulated. A very different class of algorithms is quantum Monte Carlo, which is the classical alternative to the variational quantum eigensolver (VQE) algorithm that has been proposed for quantum computers.

We have now introduced many common tools and algorithms of tensor networks. It is time to introduce the central idea behind most practical and useful applications of tensor networks: compression.

\section{Compressing many-body states with matrix product states}
\label{sec:mps}

In this section, we describe \emph{approximate} algorithms to simulate quantum computers.
Note that in doing so, we're turning history upside down, since these algorithms came out
much later than the corresponding algorithms for e.g.\ finding the ground state of a many-body Hamiltonian.
For pedagogical purposes, however, discussing the case of the quantum computer is significantly
simpler, so we'll start with that.

At the core of the algorithms below, and essentially all the other tensor network algorithms that will be discussed in these lectures, is the notion of \emph{low rank compression}, which, as we shall see, is intimately linked to the concept of entanglement.

\subsection{Low-rank matrices and the singular value decomposition}
We begin with some basic concepts of linear algebra that are sometimes not as well known as they should be.
Consider a $P\times Q$ matrix $A$ that can be written in terms of $Q$ juxtaposed vectors $\mathbf{a}_i$:
\begin{equation}
A = \left(\mathbf{a}_1 \mathbf{a}_2 \cdots \mathbf{a}_Q\right).
\end{equation}
Suppose the matrix has rank $\chi<\min(P,Q)$.
This means, by definition, that only $\chi$ of the vectors $\mathbf{a}_i$ (say, the first $\chi$ ones for concreteness) are independent.
In other words, there exists a $\chi\times Q$ matrix $C$ such that
\begin{equation}
\forall j,\quad \mathbf{a}_j = \sum_{k=1}^\chi \mathbf{a}_k C_{kj},
\end{equation}
with $C_{kj}=\delta_{kj}$ for $j\le\chi$.
Now, defining the $P\times \chi$ matrix $B$ by $B_{ik}=[\mathbf{a}_k]_i$, we arrive at
\begin{equation}
A = BC.
\end{equation}
In other words, we have compressed the matrix $A$ (which contains $PQ$ numbers) into a product of two (potentially much smaller) matrices, containing a total of $(P+Q-\chi)\chi<PQ$ numbers.
The question is, of course, how to find these two matrices $B$ and $C$!

\subsubsection{A glimpse at the cross interpolation formula}
In the case where the matrix is exactly of rank $\chi$, this construction is fairly simple.
Since $B$ is full rank, it contains a $\chi\times\chi$ submatrix of full rank. Let us denote this block by $A_{11}$, so that $\det A_{11}\ne 0$.
The matrix $A$ can be written in block form as
\begin{equation}
A =
\begin{pmatrix}
A_{11} & A_{12} \\
A_{21} & A_{22}
\end{pmatrix}.
\end{equation}
Setting
\begin{equation}
B =
\begin{pmatrix}
A_{11} \\
A_{21}
\end{pmatrix},
\end{equation}
we can obtain the blocks of
\begin{equation}
C =
\begin{pmatrix}
C_{11} & C_{12}
\end{pmatrix}
\end{equation}
from the relation $A=BC$: $C_{11}=1$ and $C_{12}=A_{11}^{-1}A_{12}$.
More explicitly, we have
\begin{equation}
A =
\begin{pmatrix}
A_{11} \\
A_{21}
\end{pmatrix}
A_{11}^{-1}
\begin{pmatrix}
A_{11} & A_{12}
\end{pmatrix}.
\end{equation}
In other words, we have obtained the explicit form of the compressed matrix in terms of its first
$\chi$ rows and $\chi$ columns. The right-hand side of the above equation is, as we shall see later, the cross interpolation. It is exact here, but can serve as an approximation
when $A$ is only approximately low rank. We now need to discuss what we mean
by ``approximately low rank''.

\subsubsection{The Singular Value Decomposition}
The answer lies in the Singular Value Decomposition (SVD), also known as principal component
analysis in some contexts. Let's suppose that $Q<P$ (otherwise, we consider the transpose of $A$). The $Q\times Q$ matrix $A^\dagger A$ is Hermitian; therefore it can be diagonalized. It is also positive semi-definite; therefore, all its eigenvalues are positive.
We write $A^\dagger A = V^\dagger \Lambda^2 V$, where $V$ is a unitary and $\Lambda$ a diagonal
matrix such that $\Lambda_{ij} = \delta_{ij} \lambda_i$. We assume for convenience that the so-called ``singular values'' $\lambda_i$ are sorted in decreasing order
(for a reason that will become clear at the end of this subsection).

Next we consider $\bar A =  A V^\dagger \Lambda^{-1}$, where $\Lambda^{-1}$ is the pseudo-inverse of
$\Lambda$ ($\Lambda_{ij}^{-1} = \delta_{ij} \lambda_i^{-1}$ when $\lambda_i>0$, and zero otherwise). $\bar A$ can be considered to consist of $Q'\le Q$ juxtaposed vectors
$\mathbf{\bar a}_i$ followed by $Q-Q'$ null vectors:
\begin{equation}
\bar A = \left(\mathbf{\bar a}_1 \mathbf{\bar a}_2\cdots{}\mathbf{\bar a}_{Q'} 0 0 0\right).
\end{equation}
The diagonalization implies that
\begin{equation}
\mathbf{\bar a}_i\cdot \mathbf{\bar a}_j =\delta_{ij},
\end{equation}
i.e.\ these vectors are normalized and orthogonal. This is the beginning of a basis
that we can complete to obtain a full $P\times P$ unitary matrix,
\begin{equation}
U = \left(\mathbf{\bar a}_1 \mathbf{\bar a}_2\cdots{}\mathbf{\bar a}_{Q'} \mathbf{\bar a}_{Q'+1}
\cdots{}\mathbf{\bar a}_{P}
\right).
\end{equation}
We finally arrive at
\begin{equation}
A = U \Lambda V,
\end{equation}
which is the singular value decomposition.

The crucial importance of the SVD stems from the following theorem:
Finding the best rank-$\chi$ approximation of a matrix $A$ amounts to building the
truncated matrix $\tilde \Lambda$ from its $\chi$ largest singular values,
and approximating $A$ as $A\approx U \tilde\Lambda V$.

Let's prove this statement. We use the Frobenius norm
$\| A \|^2 = \text{Tr} A^\dagger A = \sum_{ij} |A_{ij}|^2$. For any unitary matrix $V$,
we have $\| VA \| = \| A \|$. We are seeking a matrix $B$ that minimizes
\begin{equation}
\| A - B\|^2 = \| \Lambda - \tilde B \|^2  = \sum_{i\ne j} |\tilde B_{ij}|^2
+ \sum_{i} |\tilde B_{ii} -\lambda_i |^2,
\end{equation}
with $\tilde B = U^\dagger B V^\dagger$. 
It is very tempting to minimize the off-diagonal and diagonal part
separately, i.e. we set $\tilde B_{ij}=0$ for $i\ne j$ and the diagonal part $\tilde B_{ii}$ is given by the first $\chi$ largest singular values. This gives rise to the remaining error
\begin{equation}
\| A - B\|^2 = \sum_{i=\chi+1}^Q \lambda_i^2
\end{equation}
which is minimal. There is, however, a loophole in the above proof: we cannot minimize
the off-diagonal and diagonal parts of $\tilde B$ separately because the matrix must remain of rank $\chi$. To complete the proof, we must therefore show that the optimum that we have found is indeed a global minimum, i.e. that
$\| A - B\|^2 \ge \sum_{i=\chi+1}^Q \lambda_i^2$ for all matrices $\tilde B$ of rank $\chi$. This is achieved using the von Neumann trace inequality whose statement and proof can be found in \cite{mirsky1975}.

The importance of this theorem cannot be overstated; it is central to almost everything
that is performed with tensor networks. If $\sum_{i=\chi+1}^Q \lambda_i^2$ is tiny
with respect to $\sum_{i=1}^\chi \lambda_i^2$ then $A$ is equal to $B$ up to a tiny error.

\subsection{Entanglement entropy, area law and volume law}
It is time to connect the concept of SVD to the concept of quantum entanglement. Let's consider a bipartite system that consists of the tensor product of two subsystems $X$ and $Y$. Let $\ket{X_i}$ and $\ket{Y_j}$ be an orthonormal basis of the respective two subsystems so that a general state of the total system takes the form
\begin{equation}
\ket{\Psi} = \sum_{ij} \Psi_{ij} \ket{X_i} \otimes \ket{Y_j}.
\end{equation}
Now, let's perform the SVD of the matrix $\Psi = U\Lambda V$, and introduce two new basis sets for $X$ and $Y$ as follows:
\begin{eqnarray}
\ket{\tilde X_\alpha} &= \sum_i U_{i\alpha} \ket{X_i}, \\
\ket{\tilde Y_\alpha} &= \sum_j V_{\alpha j} \ket{Y_j}. 
\end{eqnarray}
We thus obtain
\begin{equation}
\ket{\Psi}  = \sum_{\alpha} \lambda_\alpha \ket{\tilde X_\alpha} \otimes \ket{\tilde Y_\alpha}.
\end{equation}
This is the Schmidt decomposition, with $\bracket{\Psi}{\Psi}=1$ implying that $\sum_\alpha \lambda_\alpha^2=1$. We can represent it schematically as follows
\begin{center}
\includegraphics[scale=0.25]{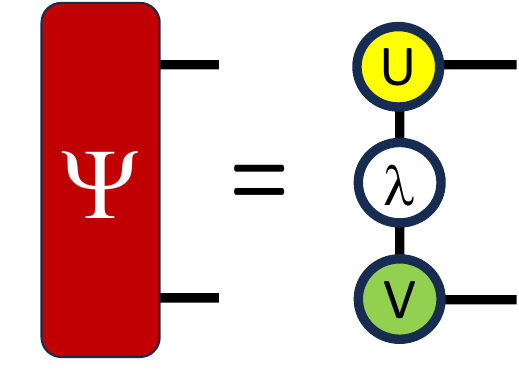}
\end{center}
It follows trivially from this decomposition that the state $\ket{\Psi}$ is a product state if and only if $\Psi_{ij}$ has a unique non-zero singular value $\lambda_1=1$.

To quantify entanglement more precisely, we introduce reduced density matrices with respect to subsystems $X$ and $Y$:
\begin{eqnarray}
\rho_X &= \text{Tr}_Y \ket{\Psi}\bra{\Psi} = \sum_{ij} [\Psi\Psi^\dagger]_{ij}\ket{X_i}\bra{X_j}, \\
\rho_Y &= \text{Tr}_X \ket{\Psi}\bra{\Psi} = \sum_{ij} [\Psi^\dagger\Psi]_{ij}\ket{Y_i}\bra{Y_j}.
\end{eqnarray}
These contain all the information necessary for calculating observables within the respective subsystem:
The average of an observable $O_X$ ($O_Y$) acting on the $X$ ($Y$) subsystem is given by $\bra{\Psi} O_X \ket{\Psi} = \text{Tr}_X [\rho_X O_X]$ (or, respectively, $\bra{\Psi} O_Y \ket{\Psi} = \text{Tr}_Y [\rho_Y O_Y]$). If the system is not entangled, both $\rho_X$ and $\rho_Y$ correspond to pure states. Otherwise, $\rho_X$ corresponds to the mixed state
\begin{equation}
\rho_X = \sum_\alpha \lambda_\alpha^2 \ket{\tilde X_\alpha}\bra{\tilde X_\alpha},
\end{equation}
and $\rho_Y$ has a similar form. Schematically, this reads
\begin{center}
\includegraphics[scale=0.25]{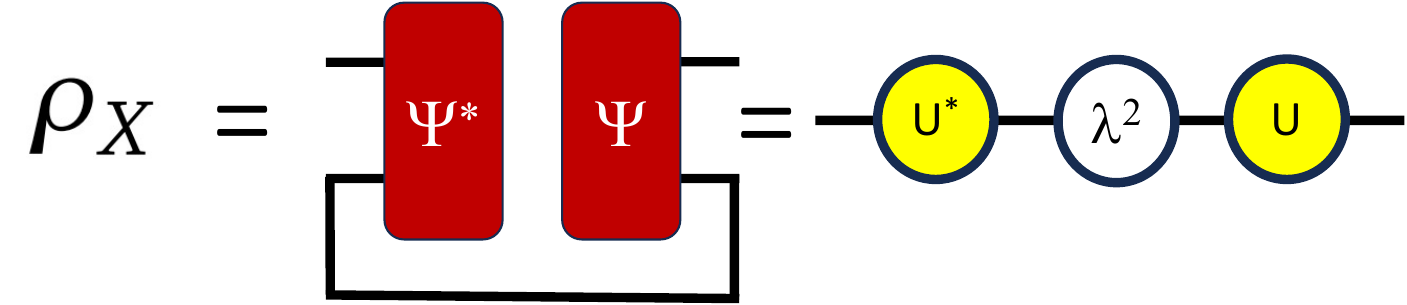}
\end{center}
We quantify the entanglement by computing the entropy $S$ associated with these reduced density matrices:
\begin{equation}
S = - \text{Tr}_X \rho_X \log \rho_X = -\sum_\alpha \lambda_\alpha^2\log\lambda_\alpha^2
= - \text{Tr}_Y \rho_Y \log \rho_Y.
\end{equation}
We interpret the values $\lambda_\alpha^2$ as the probabilities to be in the state $\ket{\tilde X_\alpha}$. In the worst-case scenario (maximal entanglement), where all the singular values are equal, $S=\log\chi$. In other words, the level of entanglement directly controls the size of the matrices we will have to deal with when working with the corresponding state in MPS form. It is therefore important to understand how $S$ scales with $N$, since it will determine the difficulty of performing the corresponding simulation.

There is an extensive literature on this subject, and we will not attempt to do it justice; see \cite{eisert2010} for a review. Most of the understanding is developed in one dimension, where rigorous theorems show that if the Hamiltonian is local and gapped, then the ground state entanglement entropy saturates at large $N$ to a finite value \cite{hastings2007}. This is the situation where most of the original successes of MPS were obtained; see the reviews \cite{schollwoeck2011,orus2014}.
Conversely, given an MPS, one can always construct a local 1D Hamiltonian of which it is the ground state \cite{fannes1992,perez-garcia2007}. More generally, for a system in $d$ dimensions with $N=L^d$ sites, the system is said to obey an ``area law'' if $S\sim L^{d-1}$, and a ``volume law'' if $S\sim L^d$. Volume-law states are widely believed to be particularly challenging for tensor network approaches. Quantum computers should target applications in which the internal state obeys a volume law; otherwise, they risk being overtaken by tensor-network approaches.

\subsection{Canonical form}
We now explain a central concept of MPS, the canonical form. As we shall see, the canonical form is what allows us to compress an individual tensor (using SVD) and, by doing so, obtain the \emph{best} compression of the overall MPS. One cannot stress too much how central the canonical form is.
We will also see another canonical form, equally important but in a different context, in the TCI section \ref{sec:tci}.

An MPS possesses a ``gauge invariance'', meaning that many different MPS represent the same state. Indeed, in the MPS expression of Eq.~\eqref{eq:mps}, which in matrix form reads
\begin{equation}
\Psi_{i_1\cdots{}i_N}=  M^1(i_1)
M^2(i_2)M^3(i_3)\cdots{}
M^{N}(i_{N})
\end{equation}
one could replace
\begin{align}
M^i(i_i) &\rightarrow M^i(i_i) U, \nonumber \\
M^{i+1}(i_{i+1}) &\rightarrow U^{-1} M^{i+1}(i_{i+1})
\end{align}
for any invertible matrix $U$ and obtain another equivalent MPS representation of $\Psi_{i_1\cdots{}i_N}$. Out of the different possibilities to ``fix the gauge'', a particularly useful one is known as the canonical form.  It is realized when all the tensors $M^i$ are unitary matrices (in a sense that will be explained below), except the central tensor $M^{i^*}$ at the ``orthogonal center'' $i^*$. The canonical form is extremely useful and almost all algorithms use it for one purpose or another. Note that in the literature there are subtle differences in the definitions of the canonical form, which we will not dwell on.

The procedure to obtain this canonical form is to start from an end of the MPS and iteratively use the $QR$ decomposition by flattening the physical leg with one of the virtual ones until one reaches $i^*$, then repeat from the other end. It is best explained graphically:
\begin{center}
\includegraphics[scale=0.35]{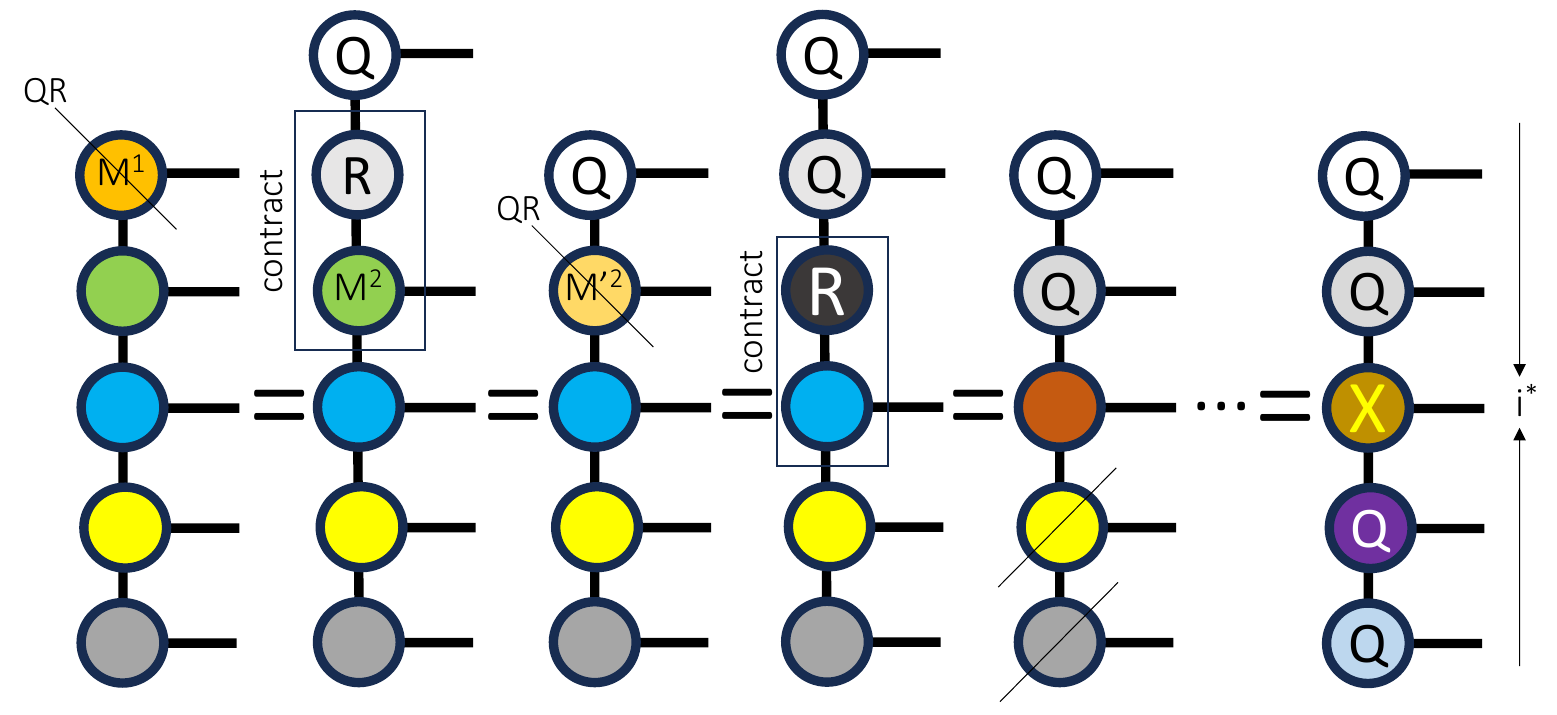}
\end{center}
In the above algorithm, one should pay attention to the flattening of the indices in the step $M^{\prime 2} = QR$ (as indicated by a thin diagonal line that partitions the tensor), which is different for the indices above and below $i^*$.

The canonical form has several interesting properties. The first is that the tensors are now nicely conditioned since all matrices except the central one are isometries. In other words, all the singular values are now positioned in a single place, in the tensor $M(i^*)$ (the $X$ in the drawing). It is typically this tensor that will be optimized in e.g.\ DMRG.

Another property is the direct consequence of the fact that the $Q$ matrices are isometries and can be represented graphically as
\begin{center}
\includegraphics[scale=0.35]{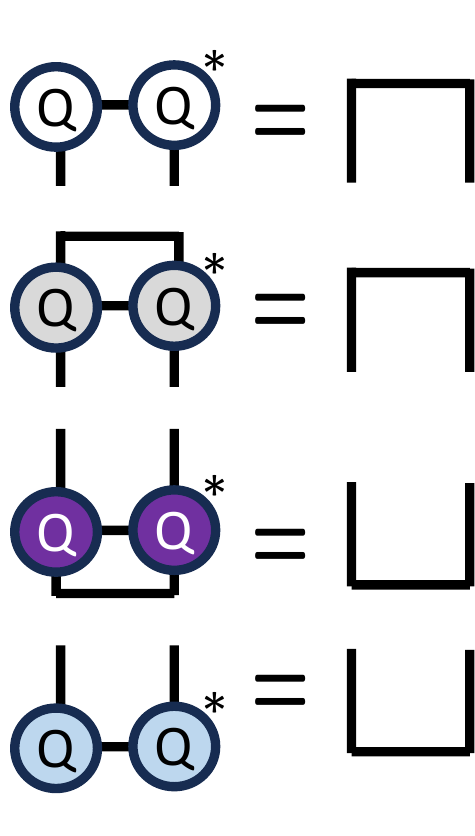}
\end{center}
In the above diagram the asterisks indicate that we have taken the complex conjugate of the tensor. It follows that the norm of the tensor is entirely given by the $M(i^*)$ tensor:
\begin{center}
\includegraphics[scale=0.35]{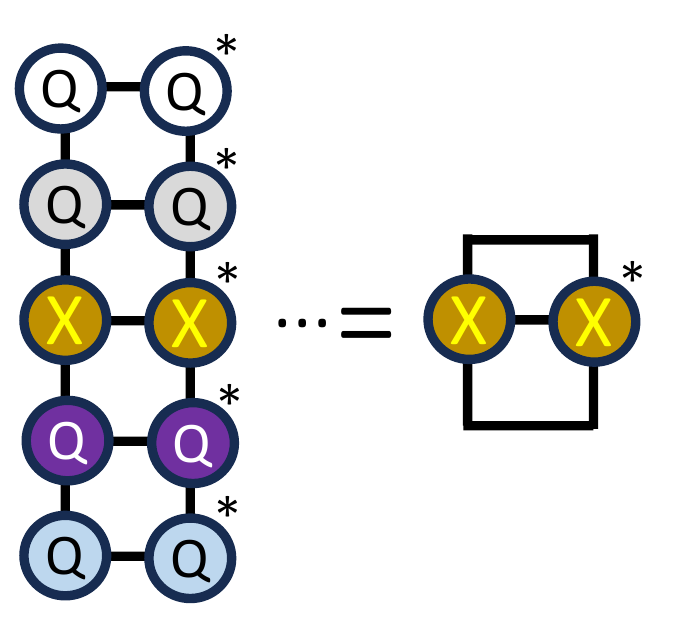}
\end{center}
or in equation form:
\begin{equation}
\sum_{i_1\cdots{}i_N} |\Psi_{i_1\cdots{}i_N}|^2 = \sum_{\alpha i \alpha'} |M^{i^*}_{\alpha\alpha'}(i)|^2.
\end{equation}
A similar telescopic simplification occurs when one calculates an observable $A$ that acts only on site $i^*$. Schematically, we have
\begin{center}
\includegraphics[scale=0.35]{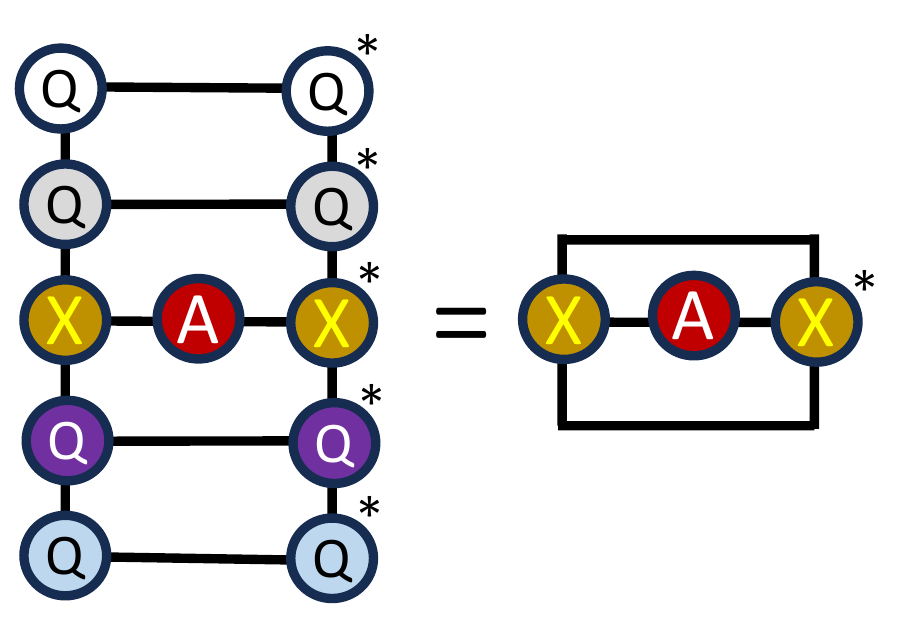}
\end{center}

The canonical form also gives us directly the reduced density matrix after taking the trace over part of the system. For instance, in our above example, if system $X$ contains the upper two qubits and system $Y$ the lower three, then $\rho_Y$ is given by:
\begin{center}
\includegraphics[scale=0.35]{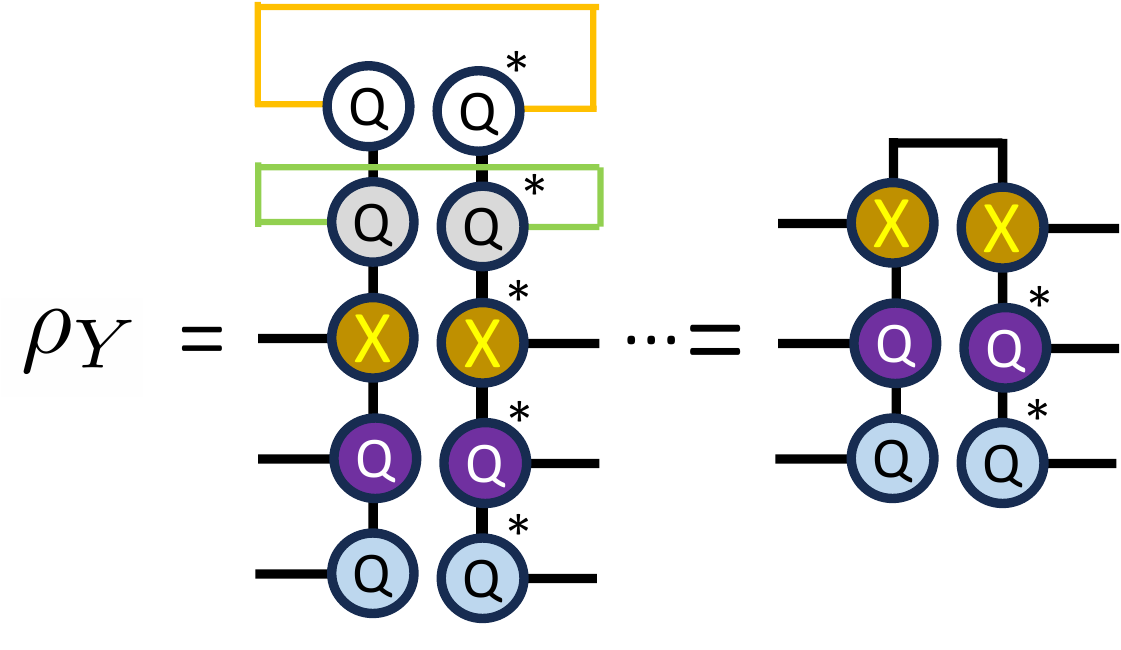}
\end{center}
It follows that the entanglement entropy can be directly obtained by performing an SVD of $M(i^*) = U \Lambda V$ (the choice of the virtual index with which the physical one is flattened determines to which subsystem $i^*$ belongs), and we get the familiar-looking formula, now extended to MPS states: $S = -\sum_\alpha \lambda_\alpha^2 \log\lambda_\alpha^2$. Hence, in order to calculate the entanglement entropy, we do not need to SVD an exponentially large matrix; doing it for the $\chi\times 2\times \chi$ tensor $M(i^*)$ is sufficient. In return, this means that the maximum level of entanglement possible with a rank-$\chi$ MPS is $S = \log\chi$, so we will not be able to describe exactly systems that are more entangled.

\subsubsection{SVD compression of an MPS}
We are now ready to start compressing some states. Suppose that we are given an MPS and we want to compute a low-rank approximation of it. If we first put it in canonical form and perform an SVD of the $M(i^*)$ tensor, then we obtain an SVD of the \emph{entire} MPS, seen as
a large matrix of size $2^{i^*} \times 2^{N-i^*}$:
\begin{center}
\includegraphics[scale=0.35]{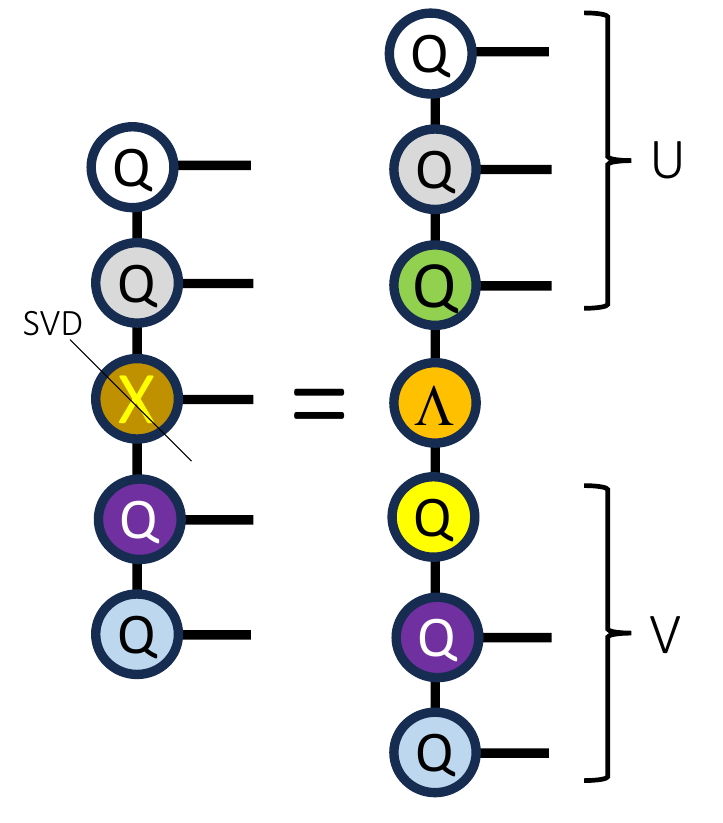}
\end{center}
Hence, the best low-rank approximation is obtained by keeping only the largest singular values! This process is repeated for all values of $i^*$. In practice, one usually performs the QR decomposition from top to bottom, then truncates the SVD from bottom to top. Note that it is only when the MPS is in canonical form that optimal truncation using the SVD can be performed. A common error is to apply SVD compression to the raw MPS without first putting it into canonical form.

\subsection{Approximate TEBD for quantum circuits}
Let's go back to our quantum computer application. Our first algorithm is a variant
of the exact MPS quantum computer simulator that we have seen already. The only difference is that we are going to apply the gates \emph{approximately}. In the context of Hamiltonian dynamics (which we will see later), this is known as the Time Evolution Bond Decimation (TEBD) algorithm \cite{vidal2004} and we shall keep the same name for quantum circuits. Graphically, the algorithm reads
\begin{center}
\includegraphics[scale=0.25]{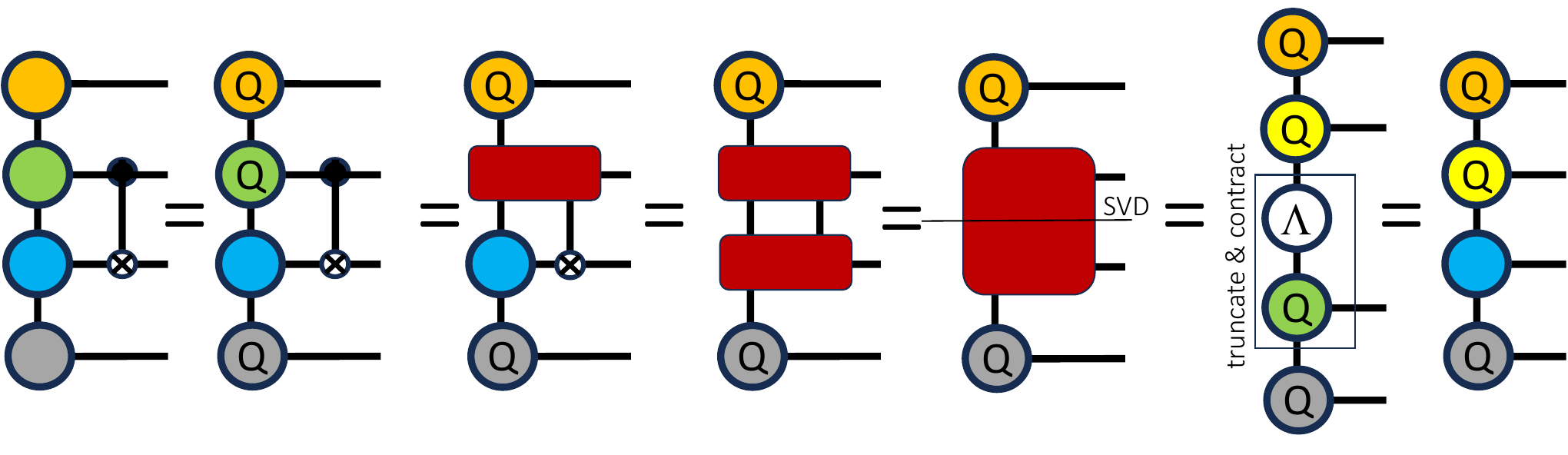}
\end{center}

The first step consists in placing the orthogonal center on one of the two tensors where the two-qubit gate will be applied (strictly speaking, one does not need to perform the $QR$ on the green tensor). Then one performs the contractions as we did in the exact case. To restore the MPS form, one performs an SVD and \emph{truncates} the singular values. This can be done by fixing the target bond dimension $\chi$ or a tolerance $\tau$ for the total weight of the discarded singular values. If the number of non-zero singular values before truncation is
$\chi'$, then we seek $\chi$ such that
\begin{equation}
\sum_{\alpha=\chi+1}^{\chi'} \lambda_\alpha^2 < \tau.
\end{equation}

A nice feature of this TEBD scheme is that one can calculate the fidelity of the calculation. Starting from a state $\ket{\tilde \Psi}^{(n)}$, if the exact application of the gate gives
$\ket{\tilde\Psi'}=\hat U^{(n)} \ket{\tilde\Psi}^{(n)}$ and $\ket{\tilde\Psi}^{(n+1)}$ is the approximate solution after truncation, then the fidelity $f^{(n)}$ of applying the gate reads
\begin{equation}
f^{(n)} = |\bra{\tilde\Psi'} \tilde\Psi\rangle^{(n+1)} |^2
= \sum_{\alpha=1}^{\chi} \lambda_\alpha^2,
\end{equation}
so that $|1-f^{(n)}|<\tau$. Conversely, $\tau$ bounds the squared norm of the discarded state. When one applies multiple gates, the fidelity is typically multiplicative so that the overall fidelity $F$ between the exact and approximate simulation,
\begin{equation}
\label{eq:fid}
F(n) = |\bra{\Psi^{(n)}} \tilde\Psi\rangle^{(n)} |^2 \approx \prod_{p=1}^n f^{(p)}
\approx (1-\tau)^n,
\end{equation}
decreases exponentially with an error rate controlled by the bond dimension. This is also the law observed in actual quantum computers due to decoherence.

\begin{figure}
  \begin{center}
    \includegraphics[width=10cm]{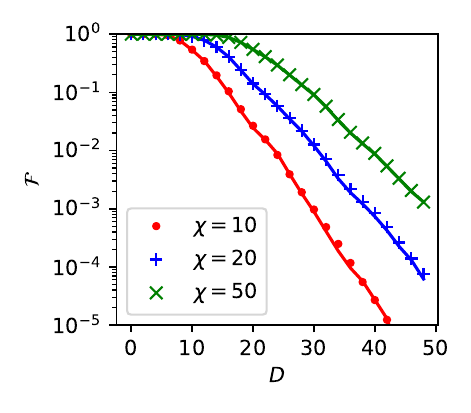}
    \caption{ \label{fig:fidelity}
      Fidelity $F$ versus depth $D$ for $N=20$ qubits, a random quantum circuit and various values of $\chi= 10,20,50$. The symbols correspond to an exact calculation of $F$ (possible for this small system), and the lines correspond to the right-hand side of Eq.~\eqref{eq:fid}. Adapted from \cite{zhou2020}.}
  \end{center}
\end{figure}

An example of the application of this TEBD algorithm is shown in Fig.~\ref{fig:fidelity} for a random circuit close to those used in the Supremacy experiment \cite{arute2019}. We note three things in this figure: first, the fidelity stays at $F=1$ for the first few layers of the circuit. This is expected because until the bond dimension has reached the cap value that we have decided, the algorithm is exact. Second, we observe that the product of the fidelities per gate is indeed a good measure of the overall fidelity, similarly to what is observed in actual quantum computers (the lines fall on the symbols). A direct consequence of that fact is that the fidelity decreases exponentially with the number of two-qubit gates, as in actual quantum computers. However, it does not decrease with the number of one-qubit gates, in contrast to actual quantum computers, because those do not affect the entanglement and can be done exactly. Last, we observe that, as one increases the bond dimension, the slope of the exponential decrease gets flatter, indicating that we are able to change our effective error rate $\tau$. Overall, this type of simulation tells us that one should not judge a quantum computer only by the number of qubits it contains, because if its error rate is not small enough, it can be simulated using tensor networks, even for a large number of qubits.

\subsection{DMRG for quantum circuits (1 site)}
The previous algorithm is easy to implement and fast. However, since a truncation is performed after each two-qubit gate, errors may accumulate rather rapidly. We will now turn
to a better algorithm where several gates can be applied before any approximation is done.
This algorithm is the first of several that belong to the ``DMRG'' class. The Density Matrix Renormalization Group (DMRG) algorithm is the grandfather of all tensor network algorithms
and was originally derived for finding the ground state of a Hamiltonian (we will discuss this later). As we shall see, the present ``DMRG for quantum circuits'' shares many features with it.

Starting from an MPS (or a product state), our goal is to approximate the state after the application of a small quantum circuit with an MPS:
\begin{center}
\includegraphics[scale=0.3]{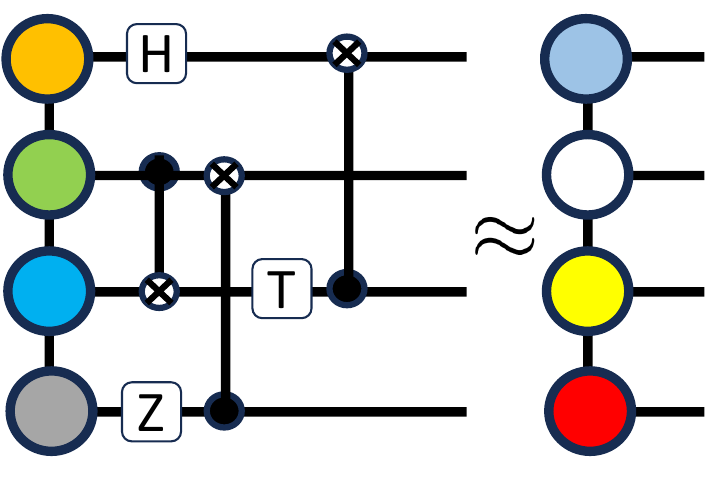}
\end{center}
We're going to sweep back and forth on the different tensors on the right-hand side of the
above (graphical) equation and optimize the corresponding tensor (while all the other tensors are considered frozen). The starting point of such an optimization is typically the
result of the above TEBD algorithm which we aim to improve.

The first step is, as often, to put the MPS in canonical form. Let's say we want to optimize the yellow tensor. The quantity to optimize is the fidelity $f^{(n)}$ defined above. Here it takes the form
\begin{center}
\includegraphics[scale=0.3]{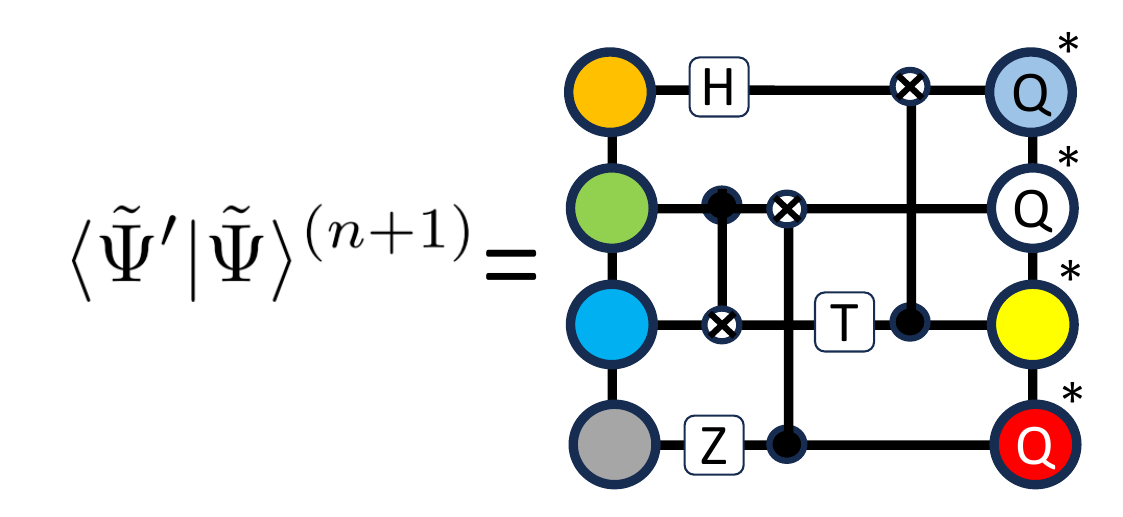}
\end{center}
Since we want to optimize only the yellow tensor, we can contract all the other indices.
Note that the order in which to perform such a contraction is not necessarily trivial.
We will not discuss it in detail here.
We arrive at
\begin{center}
\includegraphics[scale=0.3]{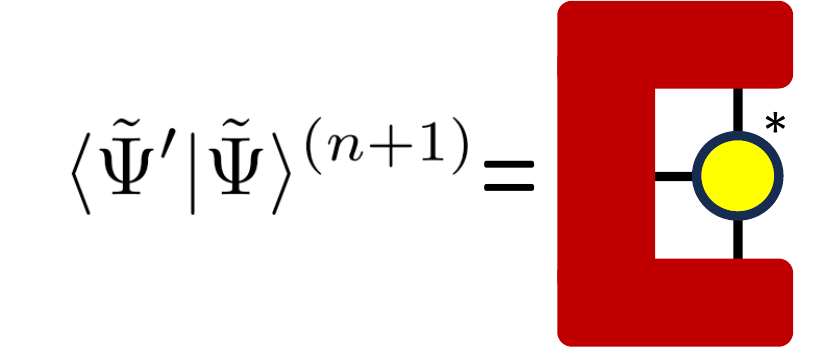}
\end{center}
What is interesting here is that this is simply a linear form in the yellow tensor. We want to optimize it subject to the constraint that the norm of the MPS is unity, which translates into the norm of the yellow tensor being unity (because of the canonical form). Introducing the corresponding Lagrange multiplier $\lambda$, we end up minimizing the simple quadratic form
\begin{align}
\mathcal{C} &= \| \ \  \ket{\Psi'} - \ket{\Psi}^{(n+1)} \|^2 -\lambda [\| \ \ \ket{\Psi}^{(n+1)} \|^2
-1 ]  \\
&= - \bra{\Psi'} \Psi\rangle^{(n+1)} - \bra{\Psi^{(n+1)}} \tilde\Psi'\rangle
  + (1 - \lambda) \bra{\Psi^{(n+1)}} \Psi\rangle^{(n+1)} + 1 + \lambda,
\end{align}
so that we need to minimize
\begin{equation}
\mathcal{C}[M_{\alpha i\alpha'}] = \sum_{\alpha i\alpha'}
-T_{\alpha i\alpha'}^* M_{\alpha i\alpha'}
-T_{\alpha i\alpha'} M_{\alpha i\alpha'}^*
+(1- \lambda) M_{\alpha i\alpha'} M_{\alpha i\alpha'}^*
+1 + \lambda,
\end{equation}
and the optimum is found by taking the derivative $\partial \mathcal{C}/\partial M_{\alpha i\alpha'}^* =0$. We arrive at $M_{\alpha i\alpha'} = \alpha T_{\alpha i\alpha'}$. The constant $\alpha$ is found by ensuring the normalization of the state
\begin{equation}
\frac{1}{\alpha^2} = \sum_{\alpha i\alpha'} |T_{\alpha i\alpha'}|^2.
\end{equation}
In other words, up to normalization
\begin{center}
\includegraphics[scale=0.3]{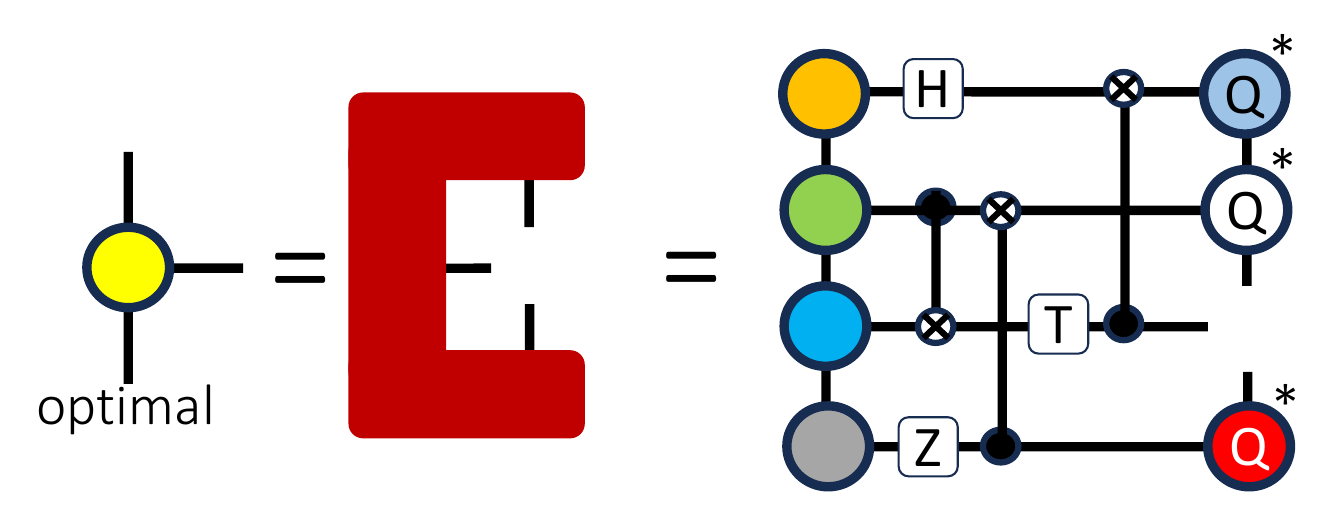}
\end{center}
Together with the optimum tensor, we also get the current fidelity of the MPS,
that is just the square of the norm of the tensor $T_{\alpha i\alpha'}$:
\begin{equation}
f^{(n)} = \bigl(\sum_{\alpha i\alpha'} T_{\alpha i\alpha'} 
M_{\alpha i\alpha'}^*\bigr)^2 = \frac{1}{\alpha^2}.
\end{equation}

A trivial variant of this algorithm is used to perform MPO–MPS products: one first uses the
zip-up algorithm described earlier, using truncated SVD instead of $QR$ to compress the output MPS. Then, in a second step, one uses the above DMRG ``fitting'' algorithm, where the quantum circuit is replaced by the MPO.

\subsection{DMRG for quantum circuits (2 sites)}
All DMRG algorithms come in two flavors: either ``single-site'', or ``two-site''.
The above algorithm is called a single-site DMRG because a single physical index is
optimized at each step.
The problem of single-site DMRG is that one cannot increase the rank $\chi$ of the tensor; it is fixed.
One would like to build a good low-rank approximation and then slowly crank up $\chi$ until the desired accuracy is reached.
Also, single-site DMRG can get trapped in local minima.
Ways to overcome these problems
within single-site DMRG exist and are called ``enrichment''.
One way to do enrichment is to slightly alter the way one gets into the canonical form: when doing the last $QR$ for the tensor just above (or below) the one to be optimized, one enriches the $2\chi \times \chi$ matrix
$Q$ with $\chi$ additional vectors (orthogonal to the previous ones) to build a
$2\chi \times 2\chi$ matrix $\bar Q$ that is now unitary.
The corresponding $R$ is full of zeros, so that the result is unchanged.
However, upon optimizing the tensor of the orthogonal center, we now have a rank $2\chi$ tensor instead of $\chi$.
Graphically, the canonicalization now reads
\begin{center}
\includegraphics[scale=0.3]{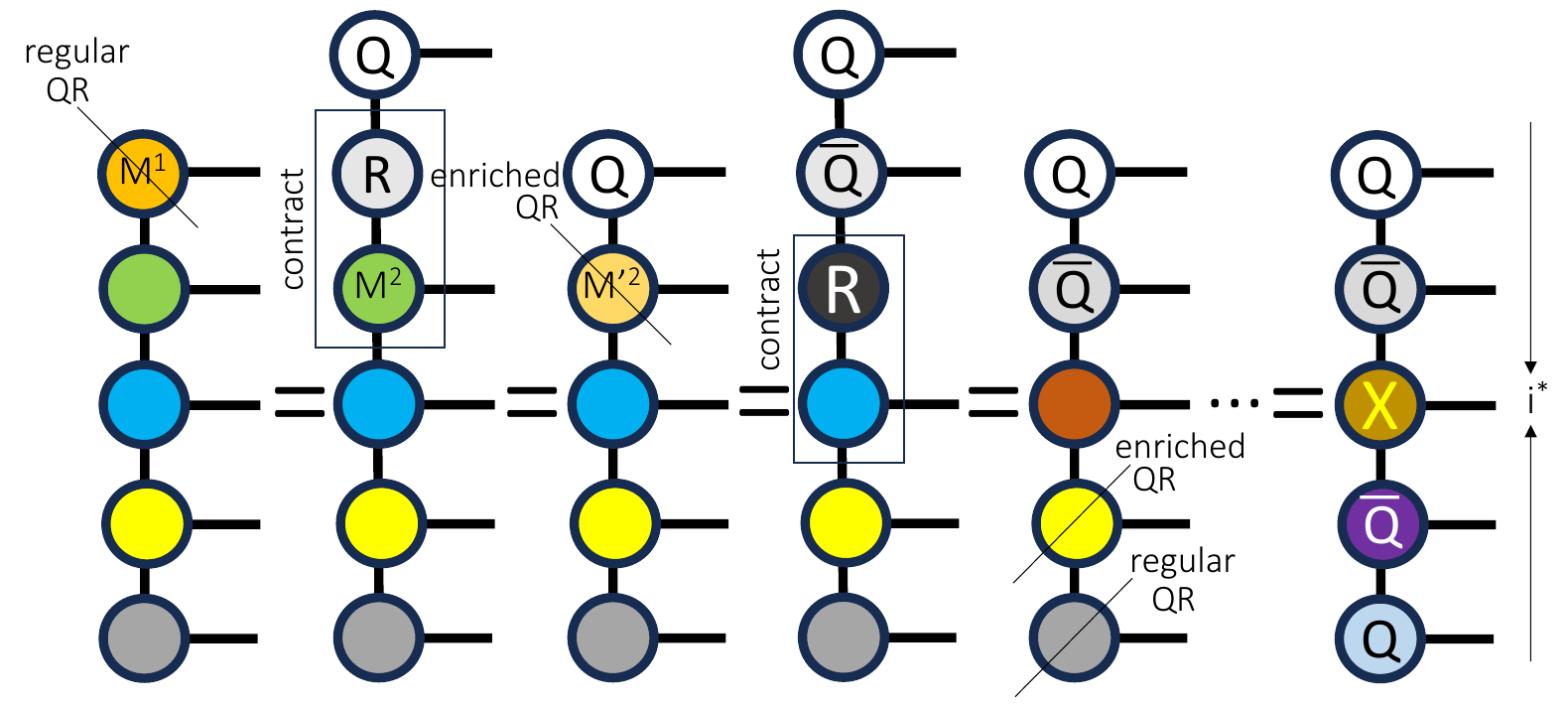}
\end{center}

However, the most common way to allow the rank to grow is to use two-site DMRG.
Two-site DMRG is a very simple variant of single-site DMRG:
One simply considers the tensors to be optimized two at a time.
One first fuses two neighboring tensors, then optimizes the resulting two-site tensors (using the exact same formula as above, except that there are now two holes instead of one), then splits the result using SVD and proceeds.
The rank is controlled during the (truncated) SVD step.
Here is a graphical representation of the procedure:
\begin{center}
\includegraphics[scale=0.3]{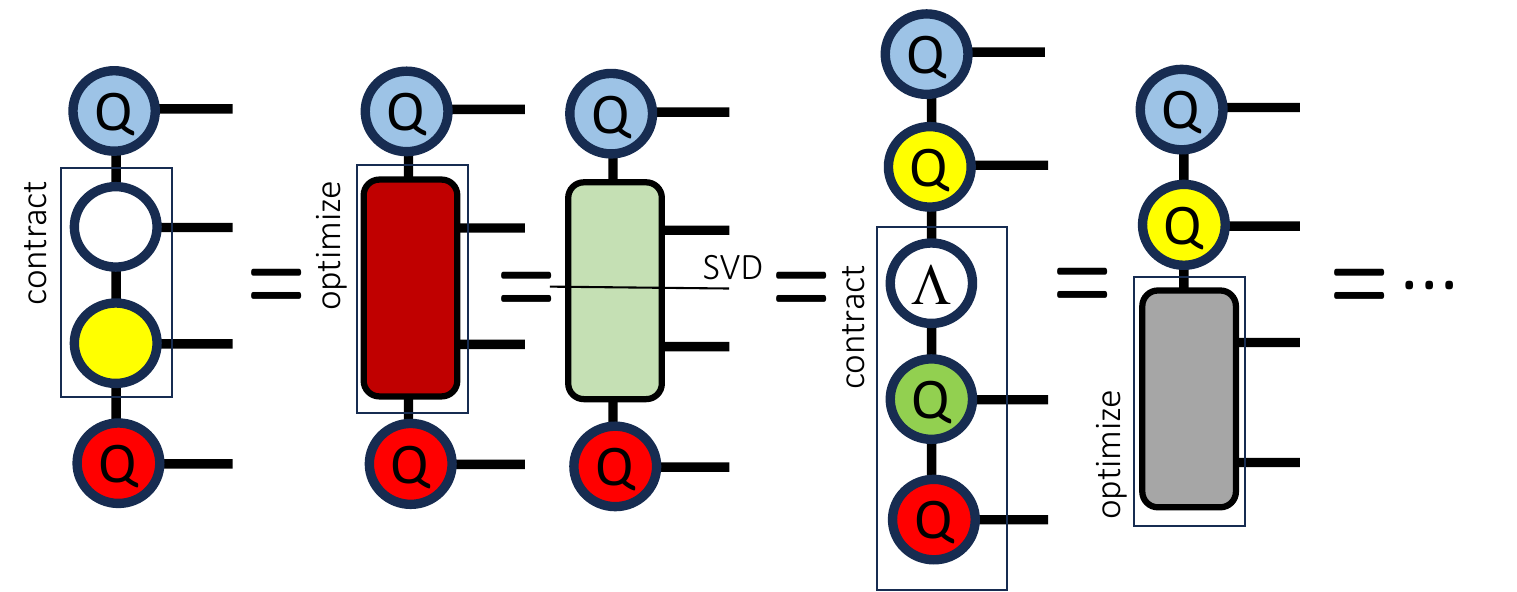}
\end{center}

We have now covered most of the standard MPO/MPS toolbox. To continue, we will introduce a
Hamiltonian so that we can make contact with the traditional many-body literature.

\section{The transverse field Ising model}
\label{sec:tfi}

In this section, we introduce the second problem that we will consider in these lectures to illustrate the various algorithms we will examine: the transverse field Ising (TFI) model \cite{pfeuty1970}. Given a set
of $N$ spins, the model is defined by its Hamiltonian
\begin{equation}
\hat H = \sum_{ij} J_{ij} Z_i Z_j - h_Z \sum_i Z_i - h_X \sum_i X_i.
\end{equation}
Here, $J_{ij}$ is the coupling matrix between spins, $h_Z$ is the magnetic field along
the $Z$ direction, and $h_X$ is the field along the $X$ direction. We will focus on two problems: finding the ground state of this type of model and its dynamics starting from a given initial state. For $h_X=0$, the model reduces to the (classical) Ising model.

The choice of using this model was motivated by the following considerations. First, it is
a genuine quantum many-body model which naively requires holding a vector of size $2^N$ in memory in order to solve it by brute force numerical diagonalization. Yet, at least in its simple form, it is one of the most tractable many-body problems. In 1D with nearest-neighbor interactions, it maps to free fermions and can therefore be solved exactly. For more complex interactions in 1D or quasi-1D (e.g.\ a ladder), the DMRG algorithm converges to essentially the exact ground state. For all dimensions, the problem is ``sign problem free,'' meaning that we can use a variety of quantum Monte Carlo techniques. Another motivation is that it is written in the same language as the quantum computing example without the added complexity of dealing with fermionic creation and destruction operators.

Despite its relative mildness, the TFI model is still the subject of active research
and can show a rich phase diagram.
For a sufficiently frustrated matrix $J_{ij}$, finding the ground state of the classical
Ising model is actually a non-trivial (NP-complete) task, and one can map pretty much all discrete optimization problems onto finding the ground state of a classical Hamiltonian. Hence, some qubit platforms have been proposed to use TFI to solve such complex optimization problems by slowly reducing $h_X$, starting from a very high value, a process known as quantum annealing. Indeed, if the field is decreased adiabatically (with respect to the avoided crossing with the rest of the spectrum), one should be able to follow the ground state at large $h_X$ ($|++\cdots{}+\rangle$ with $\ket{+} = [\ket{0}+\ket{1}]/\sqrt{2}$) down to $h_X=0$. This is perhaps the foggiest corner of quantum computing where even the theoretical existence of a quantum advantage is, to say the least, under debate. Indeed, the typical spectrum of a TFI problem looks like this:
\begin{center}
\includegraphics[scale=0.4]{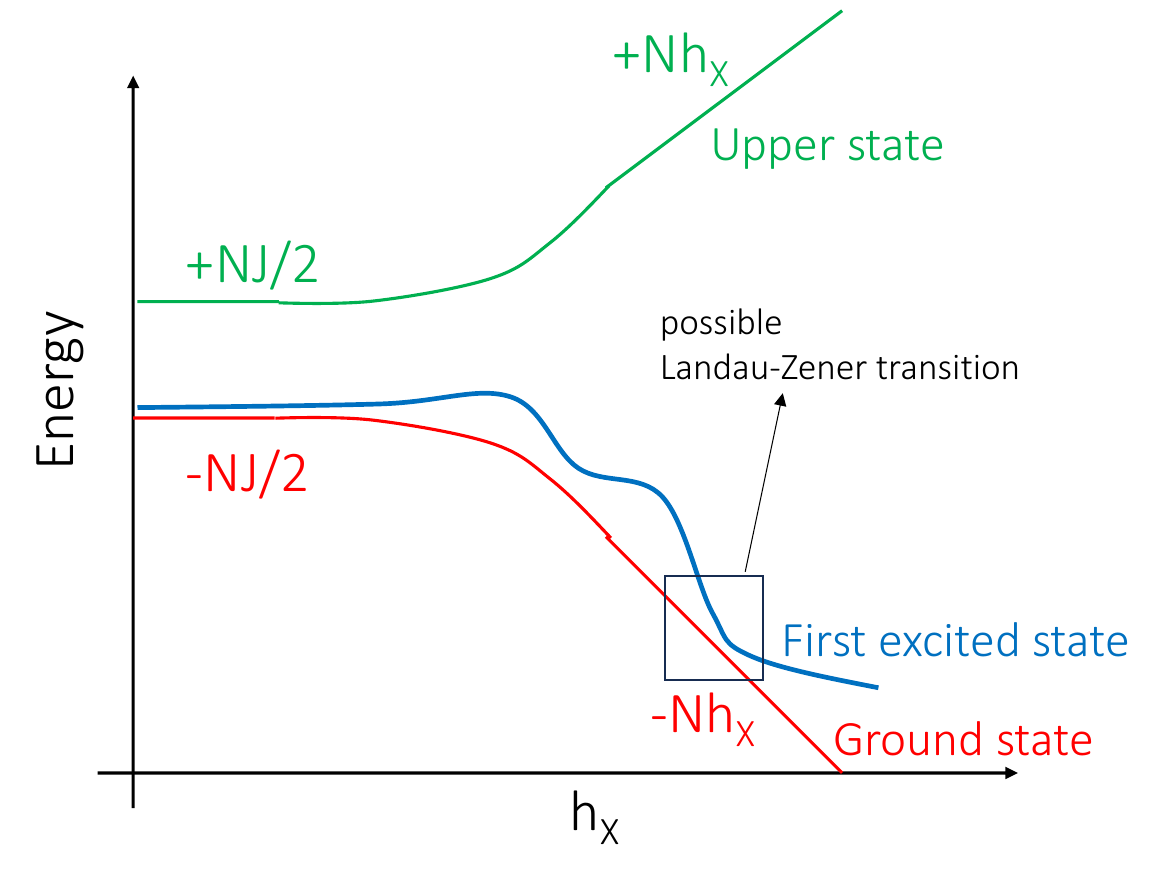}
\end{center}
If we call $\Delta$ the smallest gap between the ground state and the first excited state as
one varies $h_X$, the question is therefore whether the speed at which one decreases $h_X$, $|dh_X/dt|$, is small or large with respect to $\Delta^2$.
\begin{center}
\includegraphics[scale=0.4]{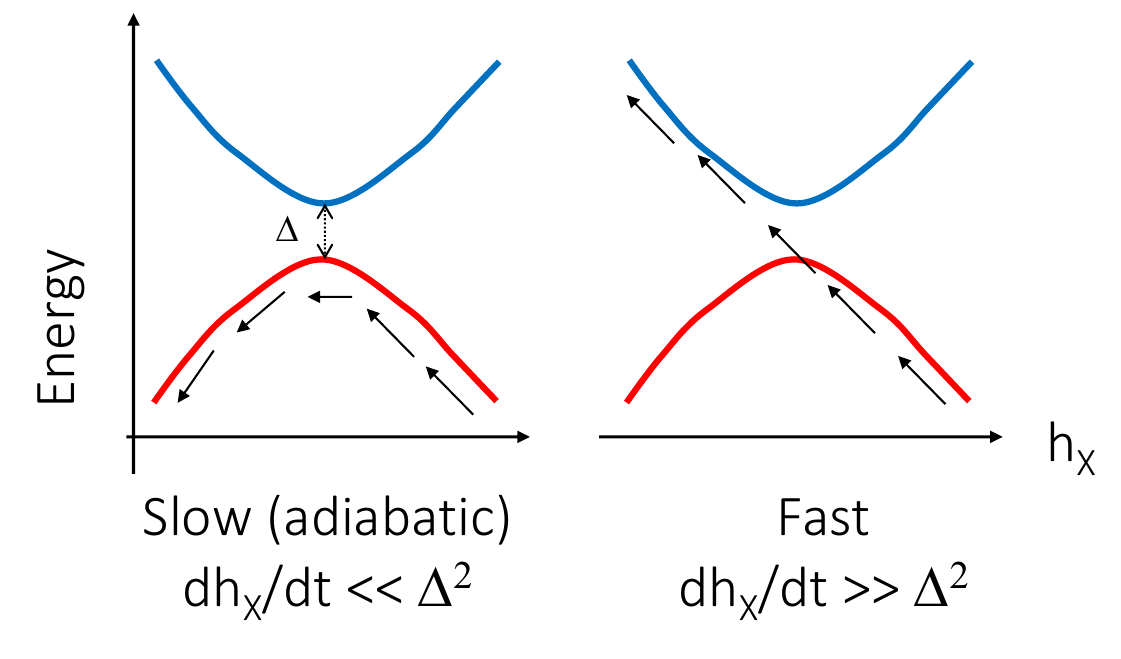}
\end{center}
This is known as the Landau-Zener transition. It can be understood already at a qualitative level by looking at the case where just two energy levels with an avoided crossing are present. The problem is that for interesting classical Ising models (i.e.\ spin-glass models), $\Delta$ has the crucial tendency to decrease exponentially with $N$, making the quantum annealing process exponentially long.

TFI also models a particular qubit platform where the state
$\ket{0}$ corresponds to the ground state of an atom (say Rubidium) and the state $\ket{1}$
to a highly excited Rydberg state (typically $n\sim 100$). The interaction in this case decays rather rapidly with the distance $r_{ij}$ between the two atoms: the coupling is antiferromagnetic and scales as $J_{ij}\sim1/r_{ij}^6$.

\section{Solving Hamiltonian models}
\label{sec:hamiltonian_models}

Let's leave quantum computing and turn to our transverse field Ising model, i.e.\ adapt
the algorithms that we have developed for unitary operators $\hat U$ to
Hermitian matrices $\hat H$. Once again, historically, it happened the other way around. We shall see that we essentially have all the ingredients
to address this new class of problems, so there is not much to do.

\subsection{Direct construction of the MPO}
The first step is to construct an MPO that represents $\hat H$.  In general, this is not at all a trivial task.  We could, of course, construct the matrix
explicitly, then transform it into an MPS (by flattening the input and output indices and then factoring these one at a time as we saw in the naive algorithm), but that would be exponentially costly.  In the next section, we will discuss an automatic algorithm (TCI) that could build the MPO for us in almost optimal time.  In the present case, however, the MPO can be constructed analytically, so we shall do so for the case where the coupling $J_{ij}$ is nearest-neighbor ($J_{ij} = J \delta_{i+1,j}$).

We write the explicit expression for the MPO of $\hat H$ as
\begin{equation}
\hat H_{i_1\cdots{}i_N;i_1'\cdots{}i_N'} = \sum_{\alpha_1\cdots{}\alpha_{N-1}} M^1_{\alpha_1}(i_1,i_1')
M^2_{\alpha_1\alpha_2}(i_2,i_2')\cdots{}
M^N_{\alpha_{N-1}}(i_N,i_N')\,,
\end{equation}
which we express in more compact notations as
\begin{equation}
\hat H = \sum_{\alpha_1\cdots{}\alpha_{N-1}} \hat M^1_{\alpha_1}
\hat M^2_{\alpha_1\alpha_2}\cdots{}
\hat M^N_{\alpha_{N-1}}\,,
\end{equation}
where $\hat M^i_{\alpha\alpha'}$ is a matrix that acts on qubit $i$ with elements
$[\hat M^i_{\alpha\alpha'}]_{ii'} = M^i_{\alpha\alpha'}(i,i')$ (using implicitly the tensor product).  We start with $\hat M^1$, which we write as
\begin{equation}
\hat M^1 =
\begin{pmatrix}
1, & Z_1, & -h_Z Z_1 -h_X X_1
\end{pmatrix}.
\end{equation}
Then we introduce $\hat M^2$ and construct the products $\hat M^1 \hat M^2$, $\hat M^1 \hat M^2 \hat M^3$, $\dots$, until we reach $\hat H$.  In this iterative construction,
the role of the three elements in the above equation is, respectively, as follows:
\begin{itemize}
\item The first element remains as ``1'' and is used to introduce the new operators of $\hat M^i$ that need not be multiplied by previous ones.
\item The second element is used to remember the previous operator $Z_{i-1}$ that needs to be multiplied by $Z_i$.
\item In the last element, we accumulate the Hamiltonian $\hat H(i)$ with $i$ spins,
$\hat H = \hat H(N)$ being our target.
\end{itemize}
In short, we want
\begin{equation}
\label{eq:explicitmpo}
\hat M^1 \hat M^2 \cdots{}\hat M^i =
\begin{pmatrix}
1, & Z_i, & \hat H(i)
\end{pmatrix}.
\end{equation}
For more complex Hamiltonians where more things need to be ``remembered'', we need to add
more elements, and the rank of the MPO increases.  So we build $\hat M^i$ for $i\in\{2\cdots{}N-1\}$ as follows:
\begin{equation}
\hat M^i =
\begin{pmatrix}
1 &  Z_i  &  -h_Z Z_i -h_X X_i  \\
0 &  0    &  J Z_i              \\
0 &  0    &  1
\end{pmatrix}\,,
\end{equation}
and we can explicitly check iteratively that it satisfies Eq.~\eqref{eq:explicitmpo}.  In particular, one has
\begin{equation}
\hat H(i) = \hat H(i-1) + J Z_{i-1}Z_i -h_Z Z_i -h_X X_i.
\end{equation}
The last vector is just made out of the third column of $\hat M^i$:
\begin{equation}
\hat M^N =
\begin{pmatrix}
  -h_Z Z_i -h_X X_i \\
  J Z_i  \\
  1
\end{pmatrix}\,.
\end{equation}
That's it, really: we now know how to construct MPOs by hand as sums of local operators or products of operators acting on nearby qubits. 
This process is known informally as the ``auto-MPO'' construction. We will see more analytical constructions of MPOs in the quantics section~\ref{sec:quantics}, but the principle is always very similar and can be generalized easily. One looks at the different terms that must be constructed (say $X_i X_{i-1}$); for each of those, one uses a different column of the row vector $\hat M^1 \hat M^2 \cdots{}\hat M^i$ to store the information that will be needed (say $X_{i-1}$); one uses the last column to accumulate the result. In practical applications, one often needs to combine this approach with a compression of the MPO \cite{hubig2017}. Note that TCI is also an alternative solution for this construction \cite{nunez2025}.

\subsection{DMRG as the diagonalization of an MPO.}
So now that we have the Hamiltonian $\hat H$ in MPO form,
we need an algorithm to find the ground state of an MPO.
This is exactly what (the actual) DMRG does, which we shall now explain.
Note that what follows is in no way tied to the many-body
problem; it could be used to diagonalize any MPO.
There are plenty of modern applications that are not tied at all to many-body physics (more on that later), or that are tied to many-body physics in a convoluted way (the MPS is something other than the wave function, e.g.\ a Feynman diagram\dots).

Getting the ground state of $\hat H$ amounts to minimizing the functional
\begin{equation}
E[\ket{\Psi}] = \bra{\Psi} \hat H \ket{\Psi},
\end{equation}
with the constraint that $\bra{\Psi} \Psi\rangle = 1$. The assumption here is that
we can choose $\ket{\Psi}$ to be an MPS with moderate bond dimensions as a variational ansatz. In other words, we want to minimize
 \begin{center}
\includegraphics[scale=0.3]{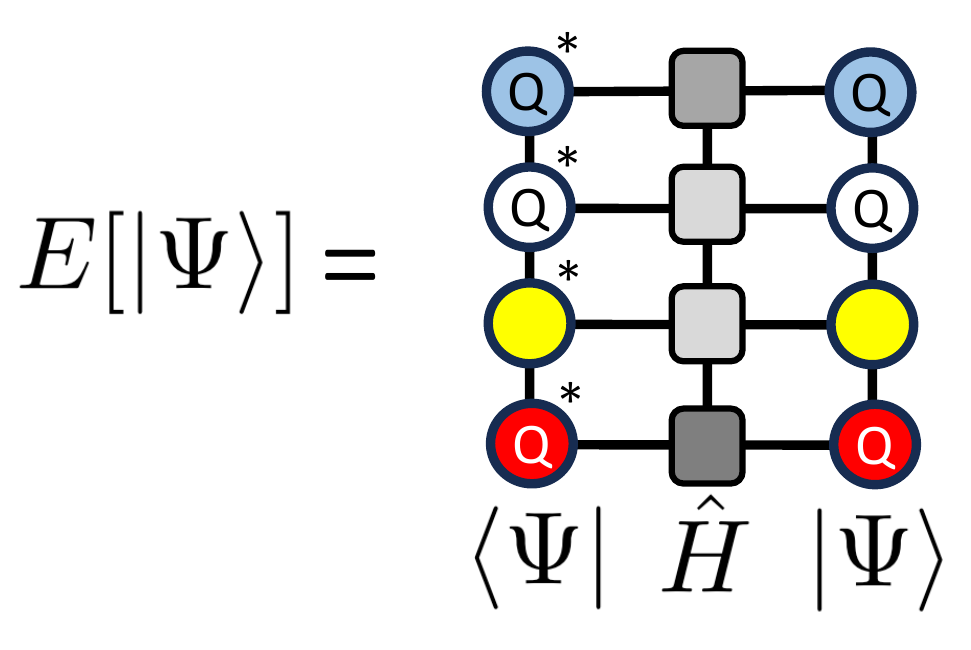}
\end{center}
As in all the DMRG algorithms, we sweep over the different tensors until convergence and optimize a single (or two) tensor at a time (at the position of the orthogonal center). We will only discuss the single-site version here; the procedure for going from single to two sites is exactly the same as that discussed in the ``DMRG for quantum circuits'' section.
One iteratively contracts the upper and lower environment to arrive at:
\begin{center}
\includegraphics[scale=0.25]{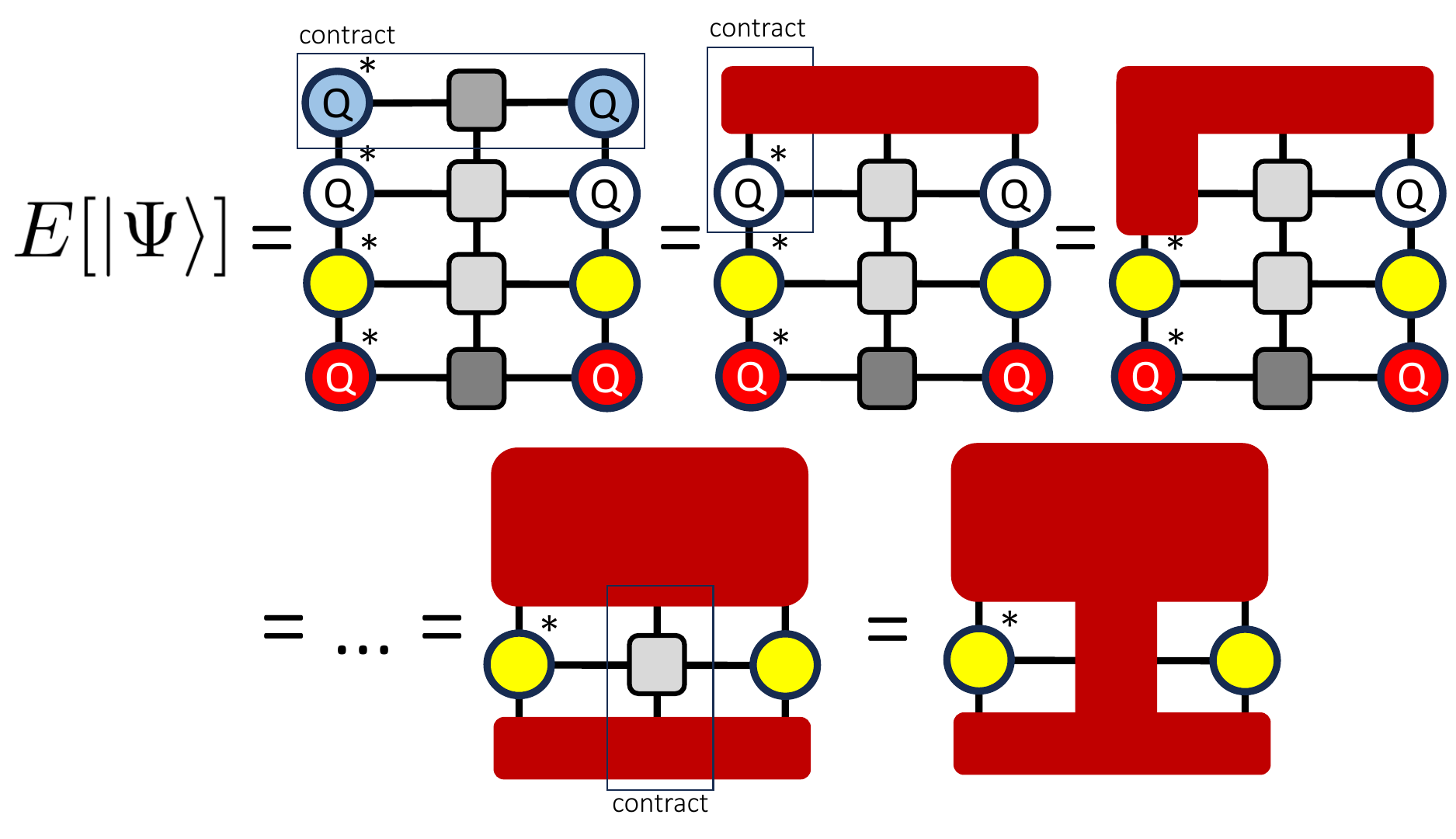}
\end{center}
(in practice, one caches the environment of different layers along the way so that one does not need to recalculate everything each time). We therefore end up with a quadratic form in terms of the yellow tensor \emph{only}. To get the optimum yellow tensor, we therefore
need to diagonalize the environment matrix
\begin{center}
\includegraphics[scale=0.25]{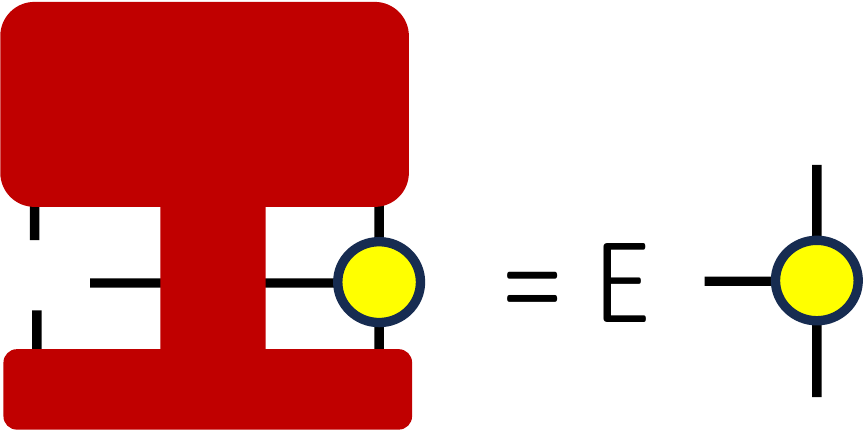}
\end{center}
and pick the lowest eigenvector. One may use a standard LAPACK routine for the diagonalization, or use a Krylov method such as Lanczos (e.g.\ from the ARPACK library). In fact, an efficient function to perform the matrix–vector product can be constructed that takes advantage of the structure of the matrix (in terms of an upper and lower environment). This is advantageous for Krylov methods since we are only interested in the ground state.
Again, we cannot even begin to do justice to all the applications found in the literature that have shaped DMRG methods by using them to solve a very large number of many-body problems in magnetism, correlated electronic systems, and more.

\subsection{Quantum dynamics with TEBD}
The TEBD algorithm that we discussed in the context of quantum computing also has
a Hamiltonian counterpart (developed before \cite{vidal2004}). The goal now is to solve the dynamics
\begin{equation}
i\frac{\partial}{\partial t} \ket{\Psi} = \hat H \ket{\Psi},
\end{equation}
or its imaginary version with $t = -i\tau$. Versatile tools exist for this purpose that are similar to DMRG, in particular the TDVP approach \cite{lubich2015,haegeman2016} seems to be the one that performs the best, at least in common situations. For such a simple model as TFI, however, we can build a TEBD
method that does not require using the MPO of the Hamiltonian. One simply uses a Trotter decomposition of $e^{-i\hat H \eta}$ (where $\eta$ is a small time step) in terms of its longitudinal part $U_{ZZ} = e^{-iJ Z_iZ_{i+1}\eta}$, $U_Z = e^{+ih_Z Z_i\eta}$, and transverse part $U_X = e^{+ih_X X_i\eta}$ such that
\begin{equation}
e^{-i\hat H \eta} \approx
\prod_{i=1}^{N} e^{+ih_X X_i\eta}
\prod_{i=1}^{N-1} e^{-iJ Z_iZ_{i+1}\eta}
\prod_{i=1}^{N} e^{+ih_Z Z_i\eta},
\end{equation}
which is applied $t/\eta$ times.
This formula is only valid to first order. Notice that in this formula, many terms commute with each other; only terms involving an $X_i$ do not commute with those involving a $Z_i$ on the same qubit. In practice, one should use the second-order version of the Trotter formula, which has small corrections for the first and last layer
\begin{equation}
e^{-i\hat H \eta} \approx
\prod_{i=1}^{N} e^{+ih_X X_i\eta/2}
\prod_{i=1}^{N-1} e^{-iJ Z_iZ_{i+1}\eta}
\prod_{i=1}^{N} e^{+ih_Z Z_i\eta}
\prod_{i=1}^{N} e^{+ih_X X_i\eta/2}.
\end{equation}

We are now back to a quantum circuit and can apply the corresponding
quantum circuit TEBD method (and/or the DMRG quantum circuit method):
\begin{center}
\includegraphics[scale=0.25]{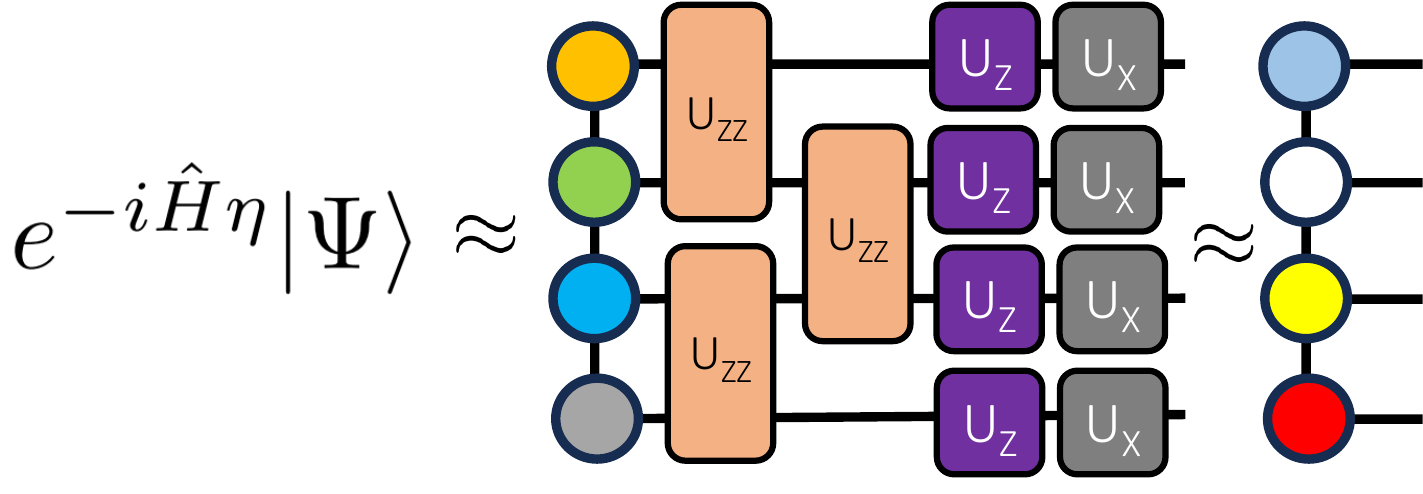}
\end{center}
Note that we need to compute the exponential of matrices. For $Z_i$ and $Z_i Z_{i+1}$,
this is trivial because these matrices are diagonal. For $X_i$, we may remember that
$X_i = H Z_i H$ so that $e^{+i X_i h_X \eta} = H  e^{+i Z_i h_X \eta}  H$.

\begin{figure}
  \begin{center}
    \includegraphics[width=10cm]{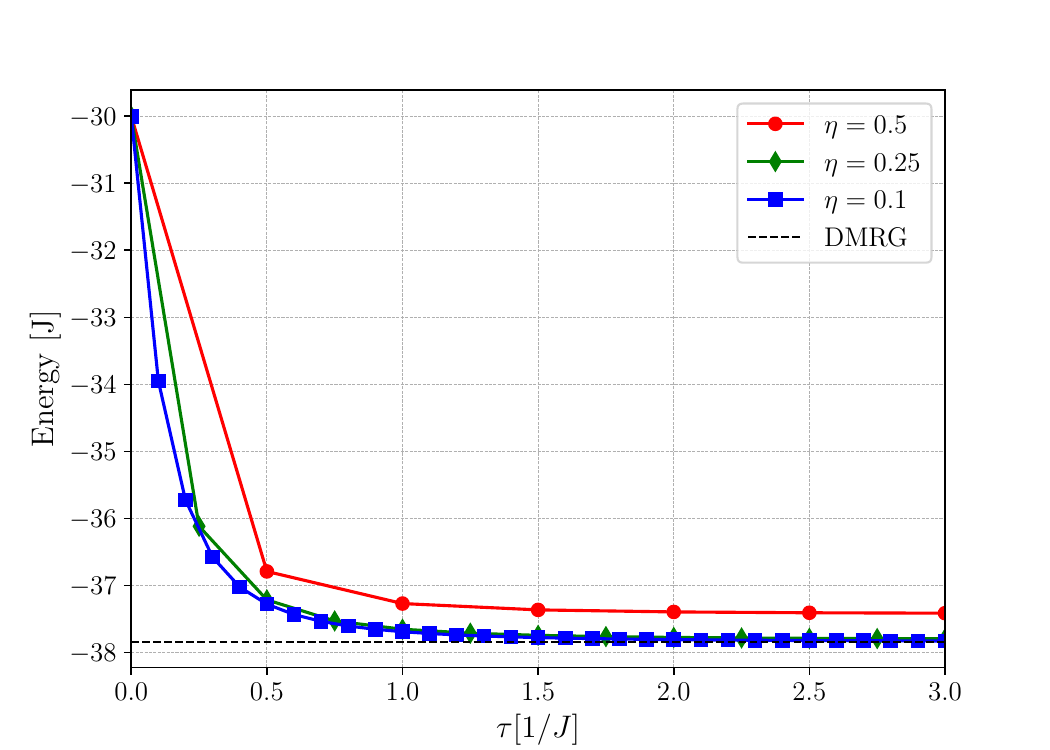}
    \caption{ \label{fig:tebd} Energy versus imaginary time $\tau$ for the TFI model
      using the imaginary-time TEBD algorithm. The energy converges to the ground state in the long-time limit. Calculations were performed with $N=30$ spins for three values of the time step $\eta$. Parameters are $h_Z=0$ and $h_X=1$. The maximum bond dimension used here is $\chi=40$ (relative precision better than $10^{-3}$), but $\chi=10$ is indistinguishable at the scale of the figure. The reference energy was obtained with the DMRG algorithm implemented using the TeNPy package \cite{hauschild2018}. (Contributed by Chen-How Huang).}
  \end{center}
\end{figure}

We may use TEBD in imaginary time to find the ground state of the TFI model, and this was one of the things done in the hands-on sessions. The result is presented in Fig.~\ref{fig:tebd}: the algorithm indeed performed as promised and found the ground state quickly. To converge to the ground state, one has to use a small value of $\eta \ll 1/J$. A too low $\eta$, however, leads to increased error due to the accumulation of $\tau/\eta$ compression steps.  One may counteract this effect by increasing the bond dimension. In the example of Fig.~\ref{fig:tebd}, the optimum is reached at $\eta\approx 0.1$ if one takes into account both time and bond dimension.

\section{MPO and MPS as large matrices and vectors}
\label{sec:mpo_mps_are_large_vectors}

So far, we have been doing quantum many-body physics. However, it should be clear by now
that most of what we have seen is far more general. Essentially, an MPS should be seen as an efficient compressed representation of a gigantic vector, and an MPO is a representation of a
gigantic matrix. Since linear algebra is pretty much everywhere in applied mathematics, chances are that these methods could be useful outside of many-body physics as well.
So far, we have seen algorithms to
\begin{itemize}
\item directly sample an MPS,
\item multiply an MPO with an MPS (or two MPOs together),
\item find the lowest eigenvalue of an MPO,
\item compress an MPS or an MPO.
\end{itemize}
This is an almost complete toolbox for linear algebra on ultra-large matrices and vectors.  Of course, we are not guaranteed that these algorithms will work: that depends on an internal structure of the solution of the problem (it must be of low rank) that may or may not be present. The biggest missing ingredient is how to construct these MPOs and MPSs for actual data. We have seen a few particular examples in the context of quantum circuits and the TFI Hamiltonian, but these approaches will not generalize to problems that are not naturally formulated in terms of tensor networks. This will be the subject of the next section, which introduces the TCI algorithm. In the remainder of this section, we complete the toolbox by adding three functionalities that are still missing.

\subsection{Element-wise multiplication between two MPS}
Suppose that we have two MPS $\Psi$ and $\Phi$ with the same physical indices.
We would like to construct an MPS for $z$ defined as $z_i = \Psi_i \Phi_i$, the element-wise product. This product can be simply written using the copy tensor:
\begin{center}
\includegraphics[scale=0.25]{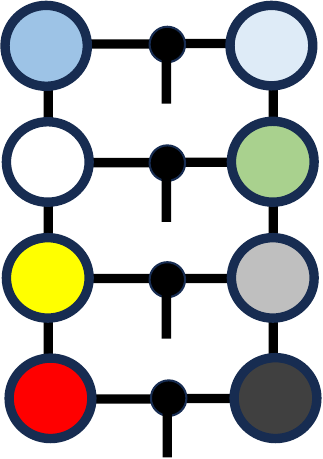}
\end{center}
Now, if we want to approximate this tensor with an MPS, we may use the tools that we have already seen: e.g.\ a combination of the zip-up algorithm, possibly followed by a few sweeps of the quantum circuit DMRG algorithm. An alternative is to use the TCI algorithm that we will see next. Note that these algorithms scale
as $\chi^4$. There exist more recent approaches that scale as $\chi^3$, but they
are not yet public at the time of this writing.
\subsection{Adding two MPS}
\label{sec:adding}
Adding two MPS $\Psi$ and $\Phi$ is also straightforward. If the tensors that make up
$\Psi$ are $M_a(i_a)$ and those of $\Phi$ are $N_a(i_a)$, then the tensors of the MPS
that describe $\Phi+\Psi$ are
\begin{equation}
P_a(i_a) =
\begin{pmatrix}
M_a(i_a) & 0 \\
0 & N_a(i_a)
\end{pmatrix},
\end{equation}
as one can verify directly. Hence the bond dimension of the sum equals the sum of the individual bond dimensions. It may be necessary to compress the resulting MPS. For instance, in the trivial case where the two MPS are the same, the true bond dimension is unchanged.

\subsection{Solving linear problems with MPO/MPS}
The last operation we would like to be able to perform on these ``gigantic matrices'' is to solve linear problems of the form $Ax=b$, where $A$ is an MPO and $x$ and $b$ are MPS. The simplest situation occurs when $A$ is positive definite, i.e.\ all its eigenvalues are positive. In that situation, the functional
\begin{equation}
\mathcal{C}[x] = x^\dagger A x - x^\dagger b
\end{equation}
is convex and has a unique minimum $x_*$, which is our solution. The strategy is exactly the same as in the DMRG algorithm: minimize the functional tensor by tensor, i.e.
\begin{center}
\includegraphics[scale=0.25]{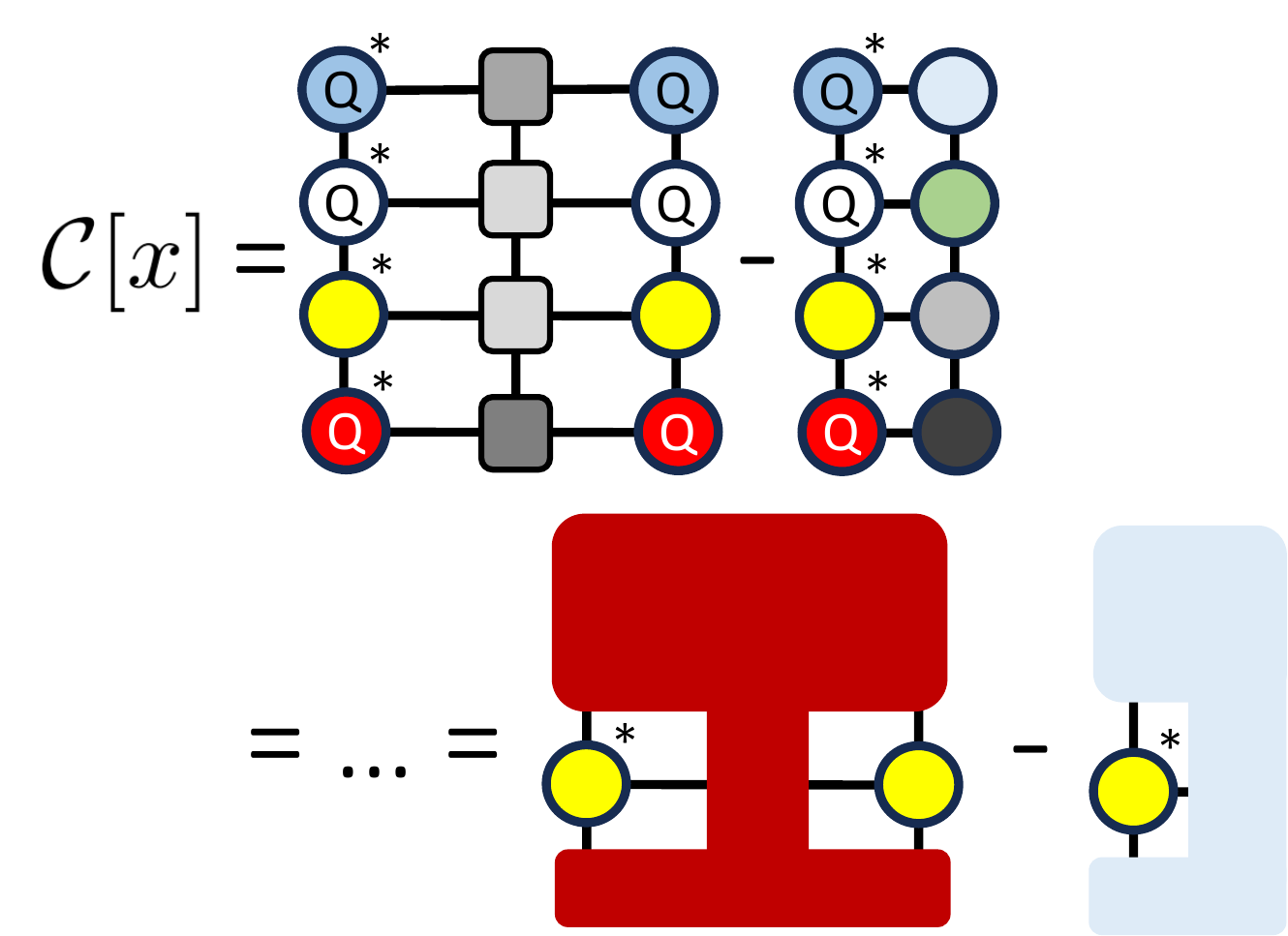}
\end{center}
For each tensor, we get an effective functional to minimize and we end up having to solve the following (small) linear problem:
\begin{center}
\includegraphics[scale=0.25]{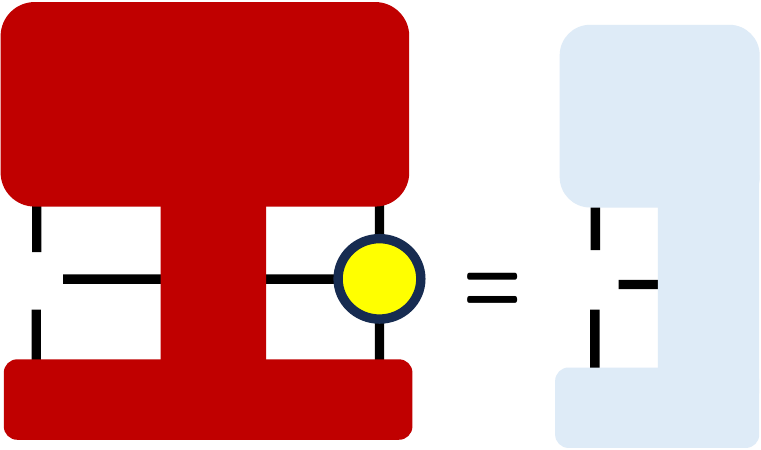}
\end{center}
Once again, this linear problem may be solved with any conventional linear algebra routine, including Krylov techniques. Once the problem is solved, we update the following tensor and optimize another one. We sweep over all the different tensors (or pairs of tensors in the case of a two-site algorithm) until convergence.

If the matrix $A$ is not positive definite, this approach has no guarantees of convergence. There are several ways to address this difficulty. The first is to ignore the problem and run the algorithm anyway. Indeed, although there are no guarantees of convergence, if the algorithm converges, it yields the correct solution. The second is to bring the problem back to the positive-definite case: if $x$ is a solution of $Ax=b$, then it is also a solution of $A^\dagger A x = A^\dagger b$. One may then construct $\tilde b = A^\dagger b$ and $\tilde A = A^\dagger A$, and run the algorithm with these inputs. The drawback is that the rank of $\tilde A$ may be significantly larger than that of $A$.

\section{Tensor Cross Interpolation for learning tensor networks}
\label{sec:tci}

 We will now discuss a novel and important algorithm -- Tensor Cross Interpolation (TCI) -- that has a very special place in the zoo of tensor network algorithms for several reasons.
This algorithm takes as an input a ``virtual'' tensor
$F_{\sigma_1,\sigma_2\cdots{}\sigma_{N} }$ and returns as an output an MPS that approximates $F_\mathbf{\sigma}$ in the best possible way:
\begin{equation}
F_{\sigma_1,\sigma_2\cdots{}\sigma_{N} } \approx \sum_{\{\alpha_i\}}
M^1_{\alpha_1}(\sigma_1)
M^2_{\alpha_1\alpha_2}(\sigma_2)\cdots{}
M^{N}_{\alpha_{N-1}}(\sigma_N).
\end{equation}
$F_\mathbf{\sigma}$ is virtual in the sense that the input of the algorithm is \emph{not} the actual tensor (which would be an exponentially large object with $d^N$ elements), as was used in the naive factorization. Rather, it is a function that takes
$\mathbf{\sigma} = (\sigma_1,\sigma_2\cdots{}\sigma_{N})$ as an input and
returns the corresponding value $F_\mathbf{\sigma}$. TCI is very different from many algorithms that we have seen so far: here $F_\mathbf{\sigma}$ is actually known by the user; what is not known is its MPS representation. Once one has this MPS (or MPO; TCI works equally well for them by just flattening the input and output indices together), one can start using all the other algorithms that we have seen already. In that sense, TCI is really the ``gateway'' that allows one to take a problem that is \emph{not} formulated in terms of a tensor network, and transform it into this framework. In that sense, TCI is pivotal in extending the scope of tensor networks to new kinds of problems beyond quantum many-body physics and computing. We will see examples of that in the context of solving partial differential equations.

Another peculiarity of TCI is that it is a \emph{learning} algorithm akin to what is done in machine learning. More precisely, it is an active learning algorithm, since TCI decides on the data $(\mathbf{\sigma},F_\mathbf{\sigma})$ that will be requested. As in machine learning, only a very tiny fraction of the possible configurations $\mathbf{\sigma}$ will be explored, and the fact that the resulting model interpolates correctly between the configurations can be spectacular. On the other hand, there are strong differences compared with deep neural networks: the optimization has nothing to do with gradient descent (and is much more effective), and the resulting function is much more structured (for instance, we can easily calculate e.g.\ integrals). The cost for these added features is a more restrictive set of applications: TCI is only effective for problems where the level of ``entanglement'' is limited.

A final peculiarity of TCI is algebraic. So far, most of what we did was associated with unitary matrices using e.g.\ the $QR$ or the SVD decomposition, with a central role played by the canonical form of an MPS. TCI will use another corner of linear algebra: Gaussian elimination with a key role devoted to the ``Cross Interpolation'' decomposition (discussed below), ``Schur complement'', and ``partial rank revealing LU'' decomposition.

The presentation below is mostly based on section III of \cite{nunez2022}, with a few
more advanced aspects borrowed from \cite{nunez2025}. Readers can also have a look at
the tensor4all open-source library that implements these algorithms
(https://tensor4all.org). While most developments in tensor networks emerged
in the theoretical physics community, this particular aspect has its roots in mathematics; see \cite{goreinov1997, bebendorf2000, goreinov2010} for matrix compression and \cite{oseledets2010, oseledets2011, savostyanov2011, savostyanov2014, dolgov2020} for the extension to tensor trains. We urge the reader to pay close attention to the notation, which is particularly important for TCI. Indeed, the biggest challenge in implementing these algorithms lies in bookkeeping various slices of $(\sigma, F_\mathbf{\sigma})$
that are held in memory.

\subsection{Another way to factorize matrices}
\begin{figure*}[htb]
\includegraphics[width=1.0\textwidth]{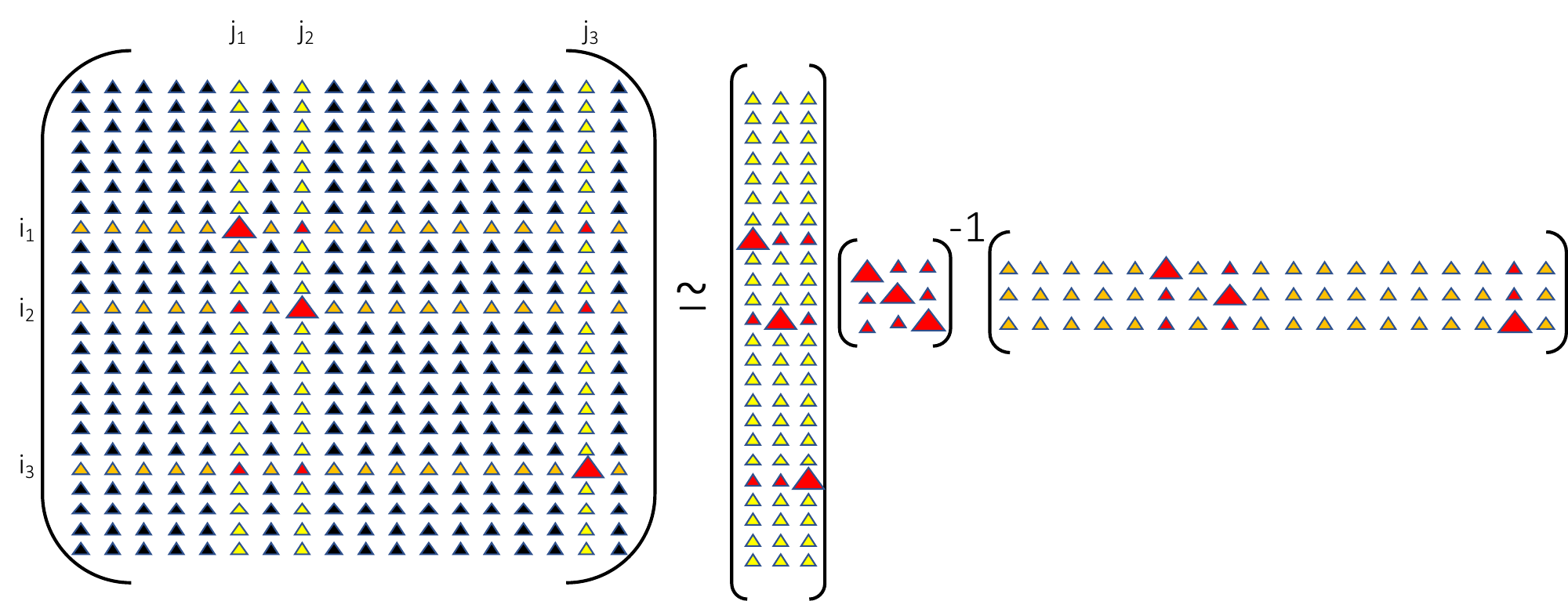}
\caption{\label{fig:ci}
   Illustration of the cross interpolation (CI) of a matrix.
   The large red triangles indicate real pivots and the smaller red
   triangles indicate automatically generated pivots. The right-hand side only contains small sub-parts of the matrix. Adapted from Nunez et al., PRX \textbf{12}, 041018 (2022).
}
\end{figure*}

Before we can get into TCI, we need a new matrix factorization formula for low rank
(or approximately low rank) matrices that is based on Gaussian elimination and the concept of
Schur complement \cite{golub1996}. This formula (the ``cross interpolation'') will be almost as good as SVD (SVD is the optimum), but with a key advantage: it can be performed without the need to access the full matrix $A$: only a set of $\chi$ rows and columns will be needed.

\subsubsection{Revisiting Gaussian elimination}
We consider an arbitrary matrix $A$ that we put in a $2\times 2$ block form:
\begin{equation}
A =
\begin{pmatrix}
A_{11} & A_{12} \\
A_{21} & A_{22}
\end{pmatrix}.
\end{equation}
Following the strategy of Gaussian elimination, we can put this matrix in triangular form (provided the $A_{11}$ block is invertible):
\begin{eqnarray}
\begin{pmatrix}
1 & 0 \\
- A_{21} A_{11}^{-1} & 1
\end{pmatrix}
\begin{pmatrix}
A_{11} & A_{12} \\
A_{21} & A_{22}
\end{pmatrix}
 \nonumber \\
=
\begin{pmatrix}
A_{11} & A_{12} \\
0 & A_{22} - A_{21}A_{11}^{-1} A_{12}
\end{pmatrix}.
\end{eqnarray}
Working in the same way from the other side, we can eliminate the other off-diagonal block to finally obtain
\begin{eqnarray}
\begin{pmatrix}
1 & 0 \\
- A_{21} A_{11}^{-1} & 1
\end{pmatrix}
\begin{pmatrix}
A_{11} & A_{12} \\
A_{21} & A_{22}
\end{pmatrix}
\begin{pmatrix}
1 & -A_{11}^{-1} A_{12} \\
0 & 1
\end{pmatrix} \nonumber \\
=
\begin{pmatrix}
A_{11} & 0 \\
0 & A_{22} - A_{21}A_{11}^{-1} A_{12}
\end{pmatrix}.
\end{eqnarray}
This equation will play a key role in multiple places. The quantity
$A_{22} - A_{21}A_{11}^{-1} A_{12}$ will appear over and over, and
we shall therefore use its official name. It is the ``Schur complement''
$[A / A_{11}]$ of $A$ with respect to block 11. The block-triangular matrices can be trivially inverted, and we arrive at a ``block $LDU$'' decomposition $A=LDU$ in terms of a block-lower-triangular matrix $L$, block-diagonal matrix $D$, and block-upper-triangular matrix $U$:
\begin{eqnarray}
\label{eq:lu2-repeat}
\begin{pmatrix}
A_{11} & A_{12} \\
A_{21} & A_{22}
\end{pmatrix} =
\nonumber \\
\begin{pmatrix}
1 & 0 \\
A_{21} A_{11}^{-1} & 1
\end{pmatrix}
\begin{pmatrix}
A_{11} & 0 \\
0 & [A / A_{11}]
\end{pmatrix}
\begin{pmatrix}
1 & A_{11}^{-1} A_{12} \\
0 & 1
\end{pmatrix}.
\end{eqnarray}
Among the various corollaries of this equation, it provides a closed form for the
determinant:
\begin{equation}
\label{eq:Schur}
{\rm det}\, A =  {\rm det}\,  [A_{11}]  \ {\rm det}\, [A / A_{11}].
\end{equation}
The Schur complement has many other nice properties; see \cite{nunez2025} for a discussion.
For instance, one does not need to take the Schur complement directly with respect to an entire block $A_{11}$. Instead, one may take it sub-block by sub-block, and if one does so, the order in which the Schur complements are taken does not matter.

\subsubsection{Cross Interpolation}
The cross interpolation formula approximates $A\approx A_{\mathrm{CI}}$, where $A_{\mathrm{CI}}$ is defined as
\begin{align}
A_{\mathrm{CI}}  =
\begin{pmatrix}
A_{11} \\ A_{21} \end{pmatrix}
(A_{11})^{-1}
\begin{pmatrix}
A_{11} & A_{12}
\end{pmatrix}.
\end{align}
In other words, the Schur complement is the \emph{error} of the cross interpolation:
\begin{align}
A = A_{\mathrm{CI}} +
 \begin{pmatrix}
0 & 0 \\ 0 & [A/A_{11}]
\end{pmatrix}.
\end{align}
An important remark is that to construct $A_{\mathrm{CI}}$, one does not need to
know anything about $A_{22}$. Indeed, as we have seen, when a matrix is of (low) rank $\chi$, we only need $\chi$ independent vectors (the first matrix in the definition of
$A_{\mathrm{CI}}$) and $\chi$ rows (which tells us how the other vectors decompose in terms of the independent ones). The cross interpolation formula has two important properties: (i) First, it is exact when evaluated on the blocks that have been used to construct it ($A_{11}$, $A_{21}$ and $A_{12}$), as evident in the above equation. We refer to this as the interpolation property. (ii) Second, it is exact if $A_{11}$ is a $\chi\times\chi$ matrix and $A$ is exactly of rank $\chi$. We have already proven this second assertion, but let us prove it again (the equation we will write will be useful later). We construct the sub-matrix of $A$ that contains the 11-block plus a single extra row $i_0$ and a single extra column $j_0$. Using the Schur complement, we have
\begin{eqnarray}
\label{eq:proof_maxvol}
&\left| {\rm det }
\begin{pmatrix}
A_{11} & A_{1j_0} \\
A_{i_01} & A_{i_0j_0}
\end{pmatrix}
\right| =  \\
&| {\rm det} A_{11} | \times |A_{i_0j_0} - A_{i_01} A_{11}^{-1} A_{1j_0}|\nonumber
\end{eqnarray}
(with a slight abuse of notation that mixes indexing with block indexing). The left-hand side is zero by definition (it is a $(\chi+1)\times (\chi+1)$ matrix); hence $A_{i_0j_0} - A_{i_01} A_{11}^{-1} A_{1j_0} = 0$, i.e.\ the cross interpolation is exact.

\subsubsection{Practical cross interpolation}
\begin{figure}
  \centerline{    \includegraphics[scale=0.47]{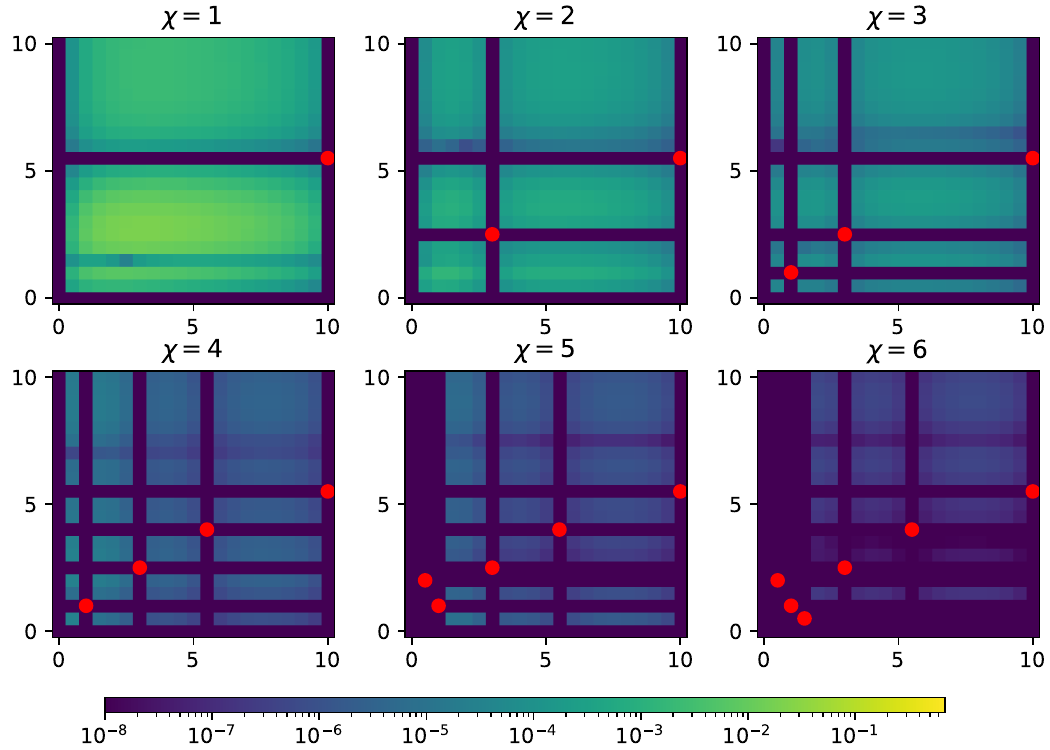}}
  \caption{Error $|A_{ij}-[A_{\mathrm{CI}}]_{ij}|$ versus $i$ and $j$ at different stages of the Cross-Interpolation for a $M\times M$ matrix with $M=20$.
    In this toy example, $A_{ij}=\left(\frac{i/M}{i/M+1}\right)^{4}(1+e^{-(j/M)^{2}})\left[1+(j/M)\cos(j/M)e^{-(j/M)\frac{i/M}{(i/M)+1}}\right]$.
    The red dots indicate the pivots. The $x$ and $y$ axes have been rescaled to be in
    $[0,10]$. Adapted from Jeannin et al.\ PRB \textbf{110}, 035124 (2024).}
  \label{fig:annex_crossi}
\end{figure}

In practice, to build up $A_{\mathrm{CI}}$, we need to choose the $A_{11}$ block properly.
Let us introduce the notation that we will use to refer to the chosen rows and columns.
Let $\mI=\{i_1,i_2,\ldots,i_\chi\}$ (respectively,
$\mJ=\{j_1,j_2,\ldots,j_\chi\}$) denote a list of the rows (columns)
of $A$ (that will form the $A_{11}$ block).
Indexing these sets gives the corresponding index: $\mI_a \equiv i_a$ is its $a^{\text{th}}$ element. The list of the indices of
all rows (columns) is denoted $\mathbb{I}=\{1,2,\ldots,M\}$ ($\mathbb{J}=\{1,2,\ldots,N\}$). Following usual programming convention (as in Python/MATLAB/Julia),
we denote by $A(\mI,\mJ)$ the submatrix
of $A$ comprised of the rows $\mI$ and columns $\mJ$; $A(\mI,\mJ)_{ab} \equiv A_{\mI_a,\mJ_b}$. We have:
\begin{align}
A &= A(\mathbb{I},\mathbb{J}),\\
A_{\mathrm{CI}} &= A(\mathbb{I},\mJ)
A(\mI,\mJ)^{-1}
A(\mI,\mathbb{J}).
\label{eq:matci}
\end{align}
Equation (\ref{eq:matci}) is illustrated graphically in Fig.~\ref{fig:ci}.
The rows and columns of $A(\mI,\mJ)$ are called the
{\it pivots}, and $A(\mI,\mJ)$ is the {\it pivot matrix}.
The pivots are chosen one by one iteratively in such a way as to maximize the
determinant of the matrix $A(\mI,\mJ)=A_{11}$ in order to guarantee that the chosen vectors are truly independent. This is known as the maximum volume (maxvol) principle.
Another way to look at the maxvol principle is that each new pivot is chosen to be the one where the current error of $A_{\mathrm{CI}}$ is maximum (maxerror), so that adding this pivot brings the largest amount of new information into the approximation. The proof of the equivalence between maxvol and maxerror is in Eq.~\eqref{eq:proof_maxvol}.
A practical example of how the error decreases for the cross interpolation of a (toy)
matrix is shown in Fig.~\ref{fig:annex_crossi}.

The important thing to remember about cross interpolation is that it is given in terms
of slices of the matrix $A$: it is entirely defined in terms of the two lists
$\mI$ and $\mJ$ of the rows and columns of the pivot matrices.

\subsubsection{Stable evaluation of the cross interpolation}
We need one final ingredient to use cross interpolation in practice. Since as one adds more pivots, the $A_{11}$ matrix becomes increasingly singular, we do not want to
calculate $A_{11}^{-1}$ explicitly because this becomes numerically unstable (even for moderate values of $\chi$). Several ways exist to stabilize the evaluation of
\begin{align}
\begin{pmatrix}
A_{11} \\ A_{21} \end{pmatrix}
(A_{11})^{-1}.
\end{align}

The first is to perform a $QR$ factorization of the first matrix, writing
\begin{align}
\begin{pmatrix}
A_{11} \\ A_{21} \end{pmatrix}
= \begin{pmatrix}
Q_{11} \\ Q_{21} \end{pmatrix} R.
\end{align}
Since $Q$ is an isometry, it is well-conditioned. All the (possibly very small)
singular values of $A_{11}$ are in the triangular matrix $R$, which disappears from the calculation.
Indeed, we have
\begin{align}
\begin{pmatrix}
A_{11} \\ A_{21} \end{pmatrix}
(A_{11})^{-1} =
\begin{pmatrix}
1 \\ Q_{21}Q_{11}^{-1} \end{pmatrix}.
\end{align}

The second way to stabilize this calculation (now our preferred way) is to realize
that Eq.~\eqref{eq:lu2-repeat} can be rewritten as
\begin{eqnarray}
\label{eq:prrlu}
\begin{pmatrix}
A_{11} & A_{12} \\
A_{21} & A_{22}
\end{pmatrix} =
\nonumber \\
\begin{pmatrix}
A_{11} & A_{12} \\
A_{21} & A_{21}A_{11}^{-1} A_{12}
\end{pmatrix}
+
\begin{pmatrix}
0 & 0 \\
0 & [A / A_{11}]
\end{pmatrix}.
\end{eqnarray}
In other words, if one ignores the Schur complement in Eq.~\eqref{eq:lu2-repeat}, one is left with the cross interpolation. We can use Eq.~\eqref{eq:lu2-repeat} iteratively,
performing the decomposition pivot after pivot (as stated above, this is legit, the proof can be found in \cite{nunez2025}), building a decomposition in the form
$A_{\mathrm{CI}} = LDU$, where $L$ is lower triangular, $D$ is diagonal, and $U$ is upper triangular. This is nothing but the well-known $LU$ decomposition, which can be used to invert matrices for example. The only caveat is that here it is partial (we stop it after getting the $\chi$ pivots, we don't go all the way through) and rank-revealing (we use the maxvol criterion to select the pivots). The procedure is called prrLU (partial rank-revealing LU) decomposition. But again, it is just a neat way to obtain the cross interpolation in a stable way.

\subsection{TCI: Extension of CI to n-dimensional tensors}
Now that we have everything we need to factorize matrices, we may extend cross interpolation to tensors. This is what TCI does.

\subsubsection{TCI: naive approach}
In the same way in which, as we proved, any tensor can be turned into an MPS using e.g.\ SVD,
any tensor can be iteratively decomposed using cross interpolation. We first flatten all indices except $\sigma_1$, apply cross interpolation to the resulting matrix, and repeat the procedure until the tensor has been entirely factorized. Graphically, this (very naive) algorithm has the following form:
\begin{center}
  \includegraphics[width=0.45\textwidth]{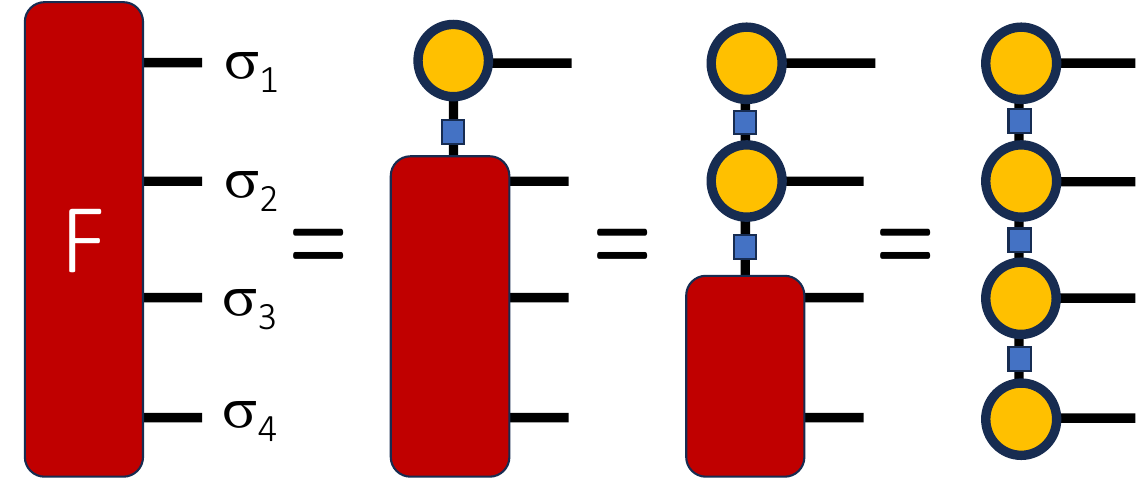}
\end{center}
Here, the small blue squares stand for the inverses of the pivot matrices. This algorithm is not practical, since applying cross interpolation to an exponentially large matrix requires an exponentially large amount of memory and computing time. However, it has
the merit of showing that such a decomposition exists. More interestingly, it shows the structure of the ``pivots'' of a TCI representation. Indeed, the cross interpolation is defined in terms of the lists $\mI$ and $\mJ$. Now we
have one such list on each side of the pivot matrices (the blue squares above). Each element of this list is itself a list
that contains the values of the corresponding indices. We call such a list a \emph{multi-index}. More explicitly, we have for our example
\begin{center}
  \includegraphics[width=0.4\textwidth]{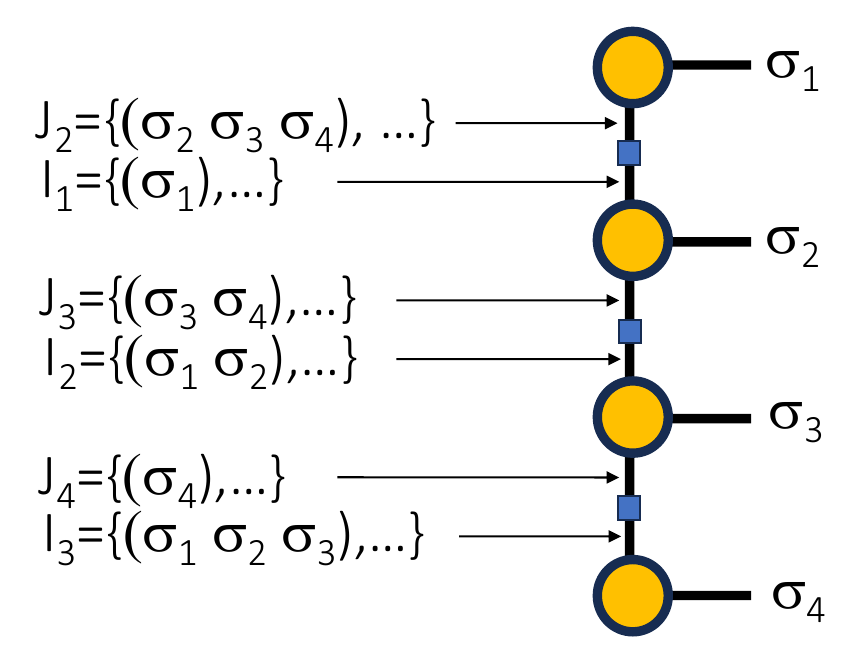}
\end{center}
The trickiest part of writing a TCI code is the correct bookkeeping of these lists of lists.

\subsubsection{TCI: formal form}
Let us introduce the above notation more formally. A TCI representation is essentially an MPS but we keep the pivot matrices explicit so that the TCI is entirely made of ``slices'' of the original tensor. For  any $\alpha$ such that $1 \leq \alpha \leq N$,
we consider ``row'' multi-indices $(\sigma_1,\sigma_2, \ldots, \sigma_{\alpha})$ and
``column'' multi-indices $(\sigma_{\alpha},\sigma_{\alpha +1},
\ldots,\sigma_N)$.
The pivot lists are defined as $\mI_\alpha = \{i_1,i_2, \ldots,i_\chi\}$
for the ``rows'' (the multi-indices have size $\alpha$) and
$\mJ_\alpha = \{j_1,j_2,\ldots,j_\chi\}$ for the ``columns''
(the multi-indices have size $N-\alpha+1$).
For notational convenience, we define $\mI_0$ and $\mJ_{N+1}$ as
singleton sets each comprised of an empty multi-index. Last, we use the symbol $\oplus$ to denote the concatenation of multi-indices:
\begin{equation}
  (\sigma_1,\sigma_2, \ldots,\sigma_{\alpha -1}) \oplus (\sigma_\alpha) \oplus (\sigma_{\alpha
  +1}, \ldots,\sigma_N) \equiv (\sigma_1, \ldots,\sigma_N).
\end{equation}

We are now ready to define the TCI representation formally. The definitions look a bit scary, but they are nothing else than what we obtained above using the naive algorithm.
The pivot matrices $P_\alpha$ (blue squares) are defined as
\begin{equation}
[P_\alpha]_{ij} \equiv F_{[\mI_\alpha]_i \oplus [\mJ_{\alpha+1}]_j},
\end{equation}
with $P_N=1$ for notational convenience. Likewise, the orange three-legged tensors $T_\alpha$ are defined as
\begin{equation}
[T_\alpha]_{i\sigma j} \equiv F_{[\mI_{\alpha-1}]_i
\oplus \sigma \oplus [\mJ_{\alpha+1}]_j}.
\end{equation}
We also introduce the matrix $T_\alpha(\sigma)$ defined as
\begin{equation}
T_\alpha(\sigma)_{ij} \equiv [T_\alpha]_{i\sigma j}.
\end{equation}
to make contact with the standard MPS form.
Using this notation, we have
\begin{equation}
   \label{eq:defTCI}
  F_\mathbf{\sigma} \approx
  [F_\text{TCI}]_{\mathbf{\sigma}} \equiv
   \prod_{\alpha=1}^N T_\alpha(\sigma_\alpha) P_\alpha^{-1},
\end{equation}
or graphically
\begin{center}
  \includegraphics[width=0.4\textwidth]{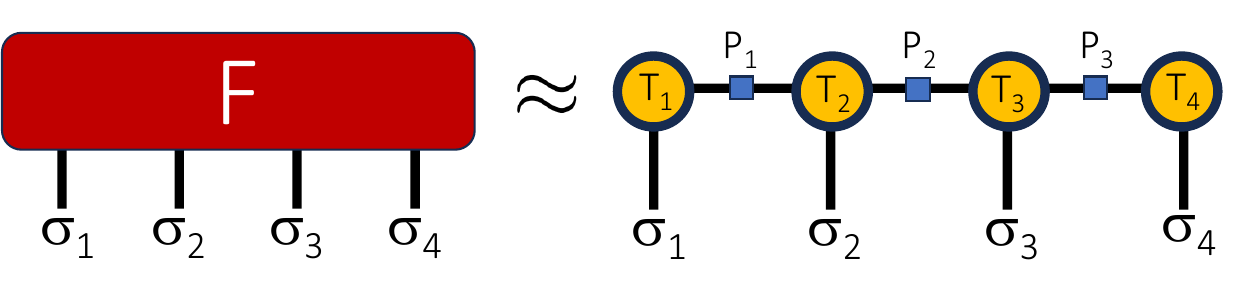}
\end{center}
The TCI representation is defined entirely by the selected sets of
``rows'' $\mI_\alpha$ and ``columns'' $\mJ_\alpha$, so that
constructing an accurate representation of $F_\mathbf{\sigma}$ amounts to optimizing
the selection of $\mI_\alpha$ and $\mJ_\alpha$ for $1 \leq \alpha \leq
N$. Only $O(N d\chi^2) \ll d^N$ entries of $F_\mathbf{\sigma}$ are used in the approximation.

\subsubsection{Practical TCI algorithm}
\label{sec:TCIalgo}
We start with an initial point $(\sigma_1, \ldots,\sigma_N)$ that we split in the $N-1$ different ways
$$(\sigma_1, \ldots,\sigma_N) = (\sigma_1, \ldots,\sigma_\alpha) \oplus (\sigma_{\alpha+1}, \ldots,\sigma_N)$$
to obtain one element for each of the sets $\mI_\alpha$ and $\mJ_\alpha$.
This yields the initial $\chi=1$ TCI, which is exact if the tensor $F_\mathbf{\sigma}$ factorizes as a product of tensors of a single variable.
To improve on this TCI, we are going to sweep over pairs of tensors
$(T_\alpha, T_{\alpha+1})$, as is done in two-site DMRG. The sweeping is performed until convergence. For each pair, we use the following procedure:
First, we introduce yet another tensor $\Pi_\alpha$ as
\begin{equation}
[\Pi_\alpha]_{i\sigma\sigma' j} \equiv F_{[\mI_{\alpha-1}]_i
\oplus \sigma \oplus \sigma' \oplus [\mJ_{\alpha+2}]_j}.
\end{equation}
Second, we replace $[T_\alpha(\sigma_\alpha) P_\alpha^{-1} T_{\alpha+1}(\sigma_{\alpha+1})]_{ij}$ inside the TCI by $[\Pi_\alpha]_{i\sigma_\alpha\sigma_{\alpha+1} j}$ because the former is a cross interpolation of the latter, hence we might use the more precise form just as well. Next, we continue the cross interpolation of $\Pi_\alpha$
(seen as a matrix $[\Pi_\alpha]_{(i\sigma),(\sigma'j)}$)
by adding a new pivot, i.e.\ one new entry to the lists $\mI_\alpha$ and $\mJ_{\alpha+1}$. The procedure can be represented graphically as
\begin{center}
  \includegraphics[width=0.4\textwidth]{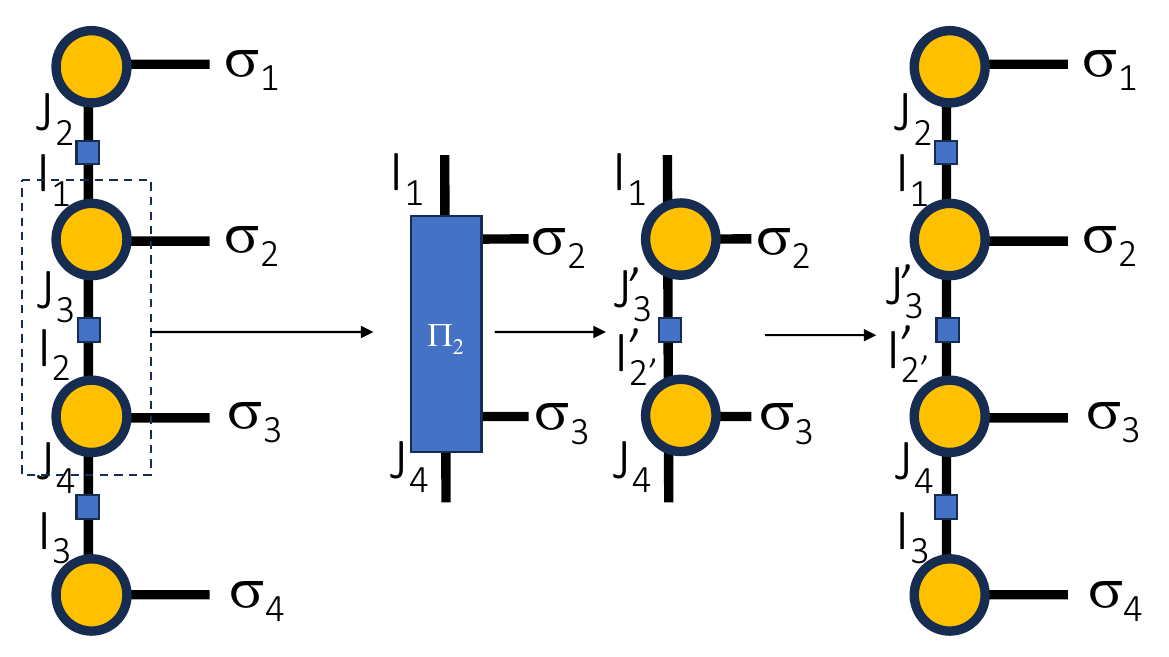}
\end{center}
And that's it. This is a fully functional TCI algorithm (although variants exist that are more suitable for specific purposes).

During the sweeping, we monitor the so-called pivot error
\begin{equation}
\epsilon_\Pi = \max_{i\sigma\sigma'j} \left|
[\Pi_\alpha]_{i\sigma\sigma' j} - [T_\alpha(\sigma) P_\alpha^{-1} T_{\alpha+1}(\sigma')]_{ij} \right|
\end{equation}
between the $\Pi_\alpha$ tensor and its cross interpolation.
We stop iterating when this error falls below a certain threshold during an entire sweep.

A subtle point remains which we have swept under the rug so far: the error
$\epsilon_\Pi$ turns out to be equal to
\begin{align}
\epsilon'_\Pi = \max_{i\sigma\sigma'j} \left|
F_{[\mI_{\alpha-1}]_i\oplus\sigma\oplus \sigma' \oplus[\mJ_{\alpha+2}]_j} \right. \nonumber \\ \left.
- [F_\text{TCI}]_{[\mI_{\alpha-1}]_i\oplus\sigma\oplus \sigma' \oplus[\mJ_{\alpha+2}]_j}
  \right|,
\end{align}
the error of the TCI approximation for the corresponding pivots.
Therefore, improving the cross interpolation of $\Pi_\alpha$ does indeed improve the TCI approximation itself (at least for these pivots). To prove this point, we need to remember that the cross interpolation is exact on the pivots. We also need to realize that there is a form of ``nesting condition'' that connects the different pivot lists: a pivot $i_\alpha\in\mI_\alpha$ takes the form $i_\alpha = i_{\alpha-1} \oplus \sigma_\alpha$ with $i_{\alpha-1} \in\mI_{\alpha-1}$ (a similar condition applies for the $\mJ_\alpha$). Using these two ingredients, one easily sees that there is a telescopic condition for the restriction of the TCI on these pivots.

Let's see how this works concretely.
We start by restricting $\sigma_1$ to values that belong to $\mI_1$. For these values, $T_1$ and $P_1$ cancel due to the interpolation property. Schematically, this reads
\begin{center}
  \includegraphics[width=0.3\textwidth]{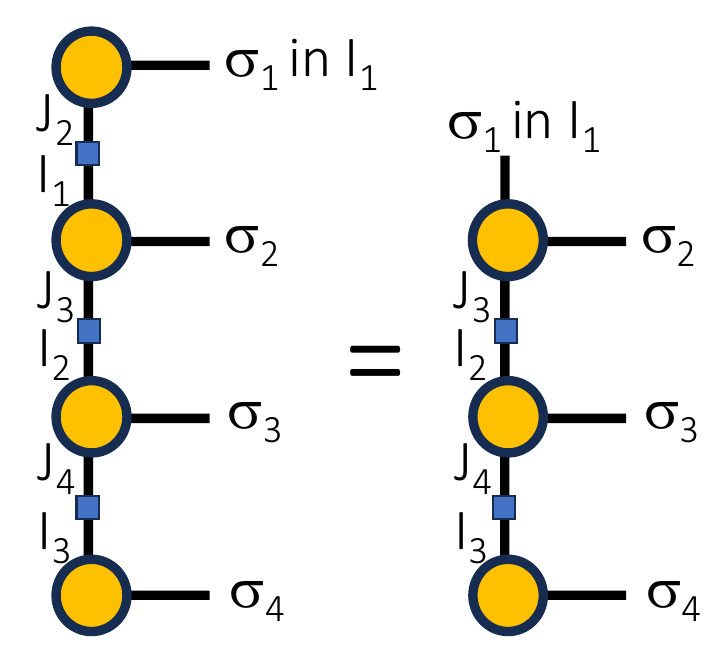}
\end{center}
We continue by requesting that $\sigma_1 \oplus \sigma_2 \in \mI_2$, which we can do because of the nested condition. The interpolation property implies that:
\begin{center}
  \includegraphics[width=0.3\textwidth]{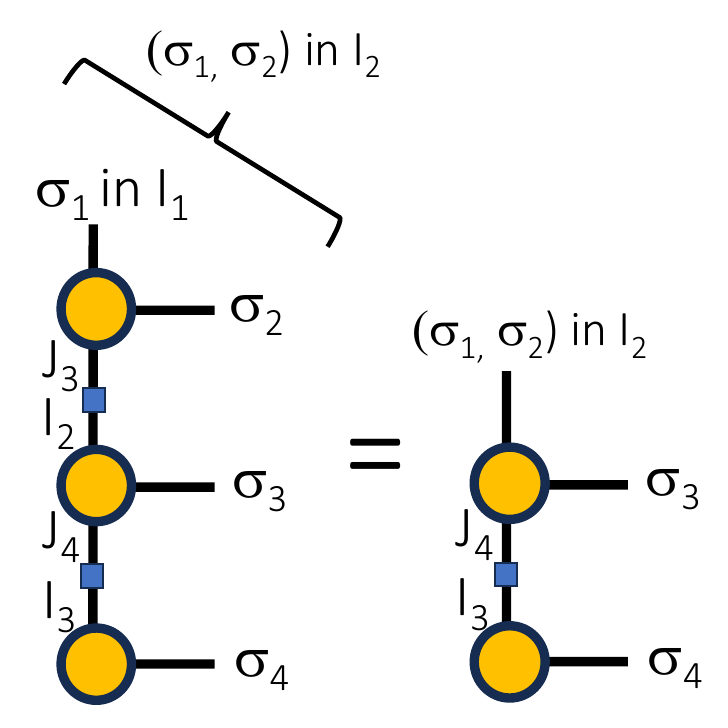}
\end{center}
and we can continue like that down the TCI representation. Since the same thing can be done with the $\mJ_\alpha$, we can also go up from the bottom of the TCI. See \cite{nunez2022} or \cite{nunez2025} for a more formal proof of the statement.

\subsubsection{Application to integrals}
\label{sec:integrationTCI}
The TCI representation has numerous uses.  As mentioned before, it unlocks for other fields the multitude of algorithms that were originally developed for many-body physics.

One very straightforward application is multi-dimensional integration. It is an alternative to the Monte Carlo approach to which we will compare it below. The convergence behavior of TCI-powered integration as $\chi$ is increased depends on the integrand, but when the method works, it compares very favorably to Monte Carlo in two respects: its convergence is much faster than allowed by the law of large numbers, and it is immune to the sign problem that plagues Monte Carlo whenever the integrand oscillates.

In its plainest version, multi-dimensional integration is quite straightforward.
Let us consider a function $f(x_1,\ldots,x_N)$. We discretize it using a plain quadrature rule
with $d$ points per dimension $a_1, \ldots, a_d$ and the corresponding weights $w_1, \ldots, w_d$.
For instance, we could use the Gauss--Kronrod-21 rule (with $d=21$) or even the trapezoidal rule. We write
\begin{equation}
\int dx_1\cdots{}dx_N f(x_1\cdots{}x_N) \approx \sum_{\sigma_1\cdots{}\sigma_N} F_\mathbf{\sigma}
w_{\sigma_1}\cdots{}w_{\sigma_N},
\end{equation}
with
\begin{equation}
F_\mathbf{\sigma} \equiv f(a_{\sigma_1},\ldots,a_{\sigma_N}).
\end{equation}
The problem is, of course, that the sum runs over $d^N$ different configurations, which is impractical. This is known as the curse of dimensionality. If, however, we can factorize
$F_\mathbf{\sigma}$ using TCI, then calculating this sum is reduced to $N$ matrix-vector multiplications that can be typically computed much faster:
\begin{equation}
\sum_{\sigma_1\cdots{}\sigma_N} F_\mathbf{\sigma}
w_{\sigma_1}\cdots{}w_{\sigma_N} \approx
 \prod_{\alpha=1}^N \left[\sum_\sigma w_{\sigma} T_\alpha(\sigma) P_\alpha^{-1}\right].
\end{equation}

\begin{figure}
  \centering
  \includegraphics[width=0.6\linewidth]{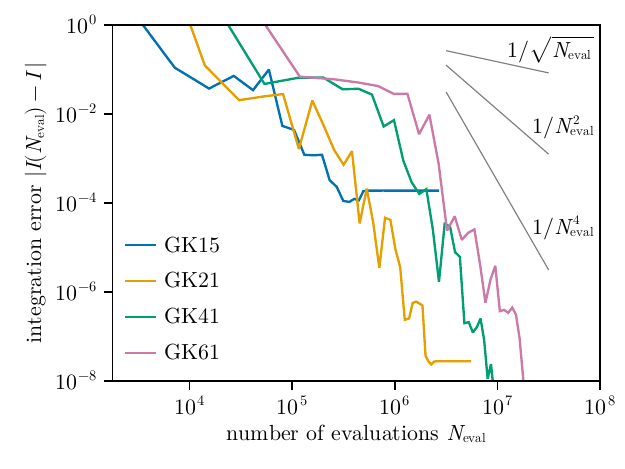}
  \caption{Convergence of the estimate $I(N_\mathrm{eval})$
   versus the number of evaluations of the integrand
  $N_\mathrm{eval}$ requested by TCI for the integral defined in Eq.~\eqref{eq:10dIntegral}. $I(N_\mathrm{eval})$ is computed using TCI with 15-, 21-, 41- and 61-point Gauss--Kronrod quadrature in each dimension. With 41- and 61-point quadrature, the value converges to $I =-5.4960415218049$. Adapted from \cite{nunez2025}.}
  \label{fig:TCI_integral_convergence}
\end{figure}

To illustrate the power of TCI, let us use it to compute a 10-dimensional integral that would be extremely hard (if not impossible) to compute with Markov Chain Monte Carlo because it contains a highly oscillatory argument. First, we apply TCI to the integrand on a Gauss--Kronrod grid. Second, we compute the integral trivially by contracting the MPS with the weights. This second step can be viewed as a scalar product between the integrand MPS and the rank-$1$ weight MPS. Our example reads:
\begin{equation}
\label{eq:10dIntegral}
    I =
    10^3 \int_{-1}^{1} dx_1\int_{-1}^{1}dx_2\cdots{}\int_{-1}^{1}dx_{10}
    \cos\!\left(10 [x^2_1+x^2_2+\cdots{}+x^2_{10}] \right)
    \exp\!\left[-10^{-3}\left( x_1+x_2+\cdots{}+x_{10} \right)^4\right].
\end{equation}
The result is shown in Fig.~\ref{fig:TCI_integral_convergence}.

Here, TCI converges approximately as $\sim 1/N_{\mathrm{eval}}^4$, where
$N_\mathrm{eval}$ is the number of evaluations of the integrand.
For comparison, Monte Carlo integration would converge as $\sim 1/\sqrt{N_{\mathrm{eval}}}$ and encounter a huge sign problem due to the cosine term in the integrand; the prefactor is probably so large that even obtaining one digit would be tough for this integral. The saturation that one observes for the blue and orange curves corresponds to regimes where the error is no longer limited by the factorization introduced by TCI, but by the grid not being dense enough.

For instance, for the orange curve, we get a precision of around 8 digits for a few million calls to the integrand, while the direct approach would require
$21^{10}\sim 1.6 \times 10^{13}$ calls. This is a speed-up by seven orders of magnitude.

\section{The Quantics representation of functions}
\label{sec:quantics}

\begin{figure}
\begin{center}
\includegraphics[width=1\textwidth]{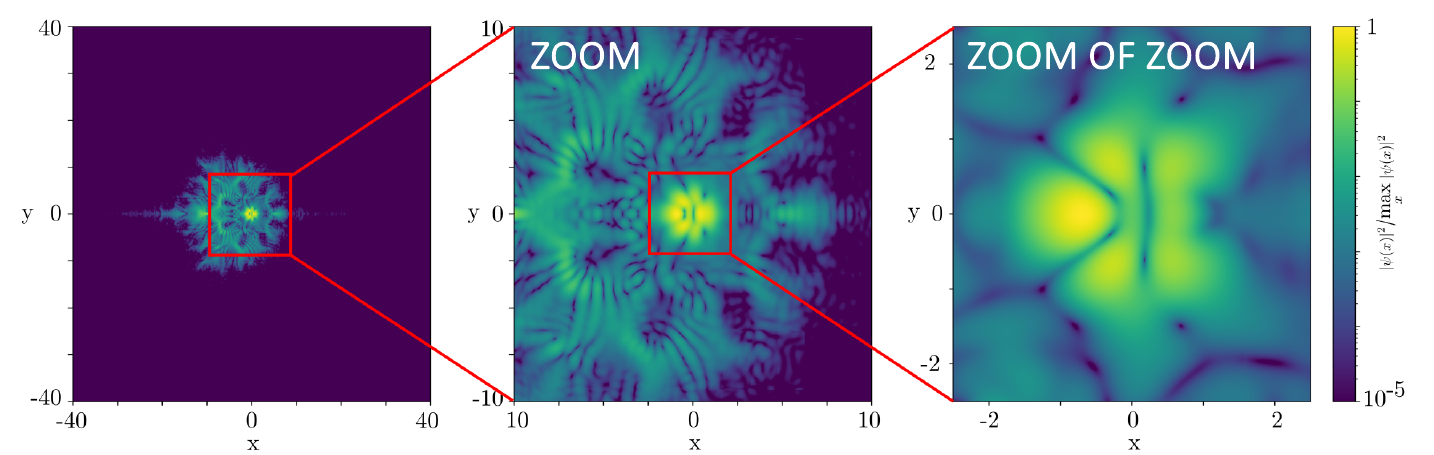}
\end{center}
\caption{
\label{fig:gross}
Snapshot of a simulation of the Gross--Pitaevskii equation using the quantics representation (in a quasiperiodic potential). In this simulation, performed on a simple workstation, 20 bits were used per dimension, hence $10^{12}$ grid points in total, far beyond what can be reached in brute-force simulations. Adapted from \cite{niedermeier2025}.}
\end{figure}
We're now in possession of a full stack of algorithms for manipulating MPO and MPS.
Before finishing our tour of tensor networks, let us discuss an entirely new set of applications: partial differential equations (PDEs). This is an emerging and very active field. Indeed, PDEs are everywhere -- in every field of physics, but also in finance, biology, etc.

The methods of this section apply to generic PDEs, but for concreteness let’s consider the Gross--Pitaevskii equation
\begin{equation}
i \partial_t \Psi(\vec r,t) = \Delta\Psi (\vec r,t) + V(\vec r)\Psi(\vec r,t) + g |\Psi(\vec r,t)|^2
\Psi(\vec r,t)
\end{equation}
with the initial condition $\Psi(\vec r,t=0) = \Psi_0(\vec r)$. To integrate such an equation with tensor networks, we require several ingredients that will be introduced in turn:
\begin{itemize}
\item We need a way to discretize $\Psi(\vec r,t)$ into a tensor that admits a low-rank MPS representation. This is the ``quantics'' representation.
\item We need to construct differential operators (e.g.\ the Laplacian) as an MPO in this representation.
\item We need to transform the inputs of the problem, $\Psi_0(\vec r)$ and $V(\vec r)$, into MPS. This will be the role of TCI.
\item We need an algorithm for solving the dynamics. For that, we have many choices: we may use most of the existing techniques for solving these equations. For instance, we may use an explicit integration scheme such as Runge--Kutta, an implicit scheme such as Crank--Nicolson, a spectral approach, or something in-between as in \cite{niedermeier2025}. We only have to replace the usual vectors and matrices with their corresponding MPS and MPO counterparts.
\item To calculate the right-hand side of the equation, we need MPO-MPS multiplication as well as element-wise multiplication for the potential and non-linear terms.
\end{itemize}
The result should be, for suitable problems, an exponentially fine grid at the cost of a regular grid. For problems with vastly different length scales, this may be an important breakthrough. See Fig.~\ref{fig:gross} for an illustration of the simulation of the Gross--Pitaevskii equation.

\subsection{The basics of quantics}
Let us start with a single dimension. We will discuss higher dimensionalities later, but that is not
very different. We consider a function $\Psi(x)$ with $x\in [-b,b]$. There are many different ways to discretize this function. The one we describe now is deceptively simple yet very powerful. We discretize the input interval into $2^N$ equally spaced points
\begin{equation}
x_n = -b + 2b \frac{n}{2^N},
\end{equation}
with the integer $n\in \{0,\ldots,2^N-1\}$. We also define the discretized function $\Psi_n$
as $\Psi_n = \Psi(x_n)$. The interesting step comes now: we write this integer $n$ in
\emph{binary} form $n = n_{N}n_{N-1}\cdots{}n_2n_1$, where the $n_i\in \{0,1\}$ are the different bits or, explicitly,
\begin{equation}
n = \sum_{a=1}^{N}  n_a 2^{a-1}.
\end{equation}
The quantics representation consists of writing $\Psi_n = \Psi_{n_1\cdots{}n_N}$
as an MPS. What makes it interesting is that the different bits are associated with different scales: changing $n_1$ changes $n$
by just one unit while changing $n_N$ corresponds to an exponentially large change of magnitude $2^{N-1}$.
\begin{center}
\includegraphics[width=0.3\textwidth]{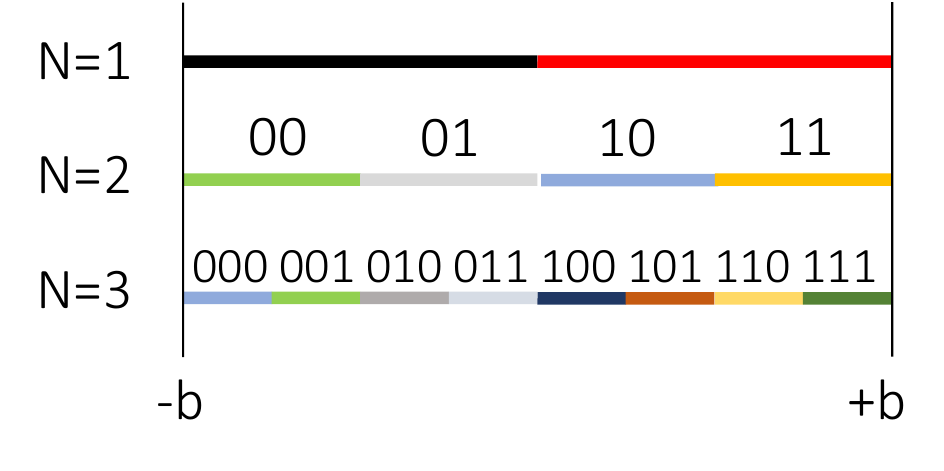}
\end{center}
Hence, the success of the quantics representation depends on the level of ``entanglement'' of different scales. A growing body of evidence indicates that many interesting problems have very limited entanglement in the quantics representation.

Let us look at a few examples. The exponential, obviously, has a rank-1 quantics representation, since it can be put in the form of the simple product
\begin{equation}
e^{a x} = e^{-ba +ba\sum_{\alpha=1}^N n_\alpha 2^{\alpha - N}}
= e^{-ba} \prod_{\alpha=1}^N e^{ba n_\alpha 2^{\alpha - N}}.
\end{equation}
Likewise, the functions $\cos(x)$, $\sin(x)$, $\cosh(x)$ and $\sinh(x)$ have rank 2 because they are each sums of two exponentials. Interestingly, the sum of two cosines, $\cos(k_1 x) + \cos(k_2 x)$, has rank 4 even if
$k_1$ and $k_2$ differ by orders of magnitude. This makes quantics very appealing when vastly different length scales play a role in the problem. The function $f(x)=x$ also has rank 2, since it can be expressed as the sum of local terms
\begin{equation}
f(x)= -b + \frac{2b}{2^N}\sum_{a=1}^{N}  n_a 2^{a-1},
\end{equation}
and can therefore be obtained with the same technique we used to construct the MPO of the
TFI Hamiltonian. Likewise, the rank of $x^n$ is $n+1$.
In fact any polynomial of degree $n$ can be represented exactly by an MPS of rank $n+1$, as we show next.

\subsection{Explicit quantics representation of polynomials}
Polynomials are an important class of functions that can be represented by quantics at low cost. Since smooth functions can be well approximated by polynomials \cite{trefethen2019}, this is an important result, because it means that smooth functions will admit a low-rank approximate quantics representation.

Let $P(x) = \sum_{k=0}^Q a_k x^k$ be a polynomial of degree $Q$ with $x\in[0,1]$. We will prove that it admits a quantics representation of rank $Q+1$ by constructing it explicitly using a technique very close to the one we used to construct the MPO of the TFI Hamiltonian. Let us define
\begin{equation}
K_N^k \equiv \left( \sum_{\alpha=1}^N \frac{n_\alpha}{2^{\alpha}} \right)^k,
\end{equation}
so that $P(x) = \sum_{k=0}^Q a_k K_N^k$. We seek to construct a set of matrices $M^\alpha(n_\alpha)$ such that
\begin{equation}
\begin{pmatrix}
K^0_N \\ K^1_N \\ \vdots \\ K^Q_N
\end{pmatrix}
= M^N(n_N)
\begin{pmatrix}
K^0_{N-1} \\ K^1_{N-1} \\ \vdots \\ K^Q_{N-1}
\end{pmatrix}.
\end{equation}
If we can do that, then we automatically have (proof by iterating the formula)
\begin{equation}
P(x) =
\begin{pmatrix}
a_0 & a_1 & \cdots & a_Q
\end{pmatrix}
M_N(n_N) M_{N-1}(n_{N-1}) \cdots M_2(n_2)
\begin{pmatrix}
1 \\ n_1/2 \\ \vdots \\ (n_1/2)^Q
\end{pmatrix}.
\end{equation}
The next step is to write a recursion relation that expresses $K_N^k$ in terms of
$K_{N-1}^k$ and the variable $n_N$, i.e.\ we isolate $n_N$ in the definition of
$K_N^k$. Using the binomial law, we get
\begin{align}
K_N^k &= \left( \frac{n_N}{2^{N}} + \sum_{\alpha=1}^{N-1} \frac{n_\alpha}{2^{\alpha}} \right)^k \nonumber \\
&= \sum_{a=0}^k
\begin{pmatrix} k \\ a \end{pmatrix}
\left( \frac{n_N}{2^{N}}\right)^{k-a} K_{N-1}^a,
\end{align}
from which we obtain
\begin{equation}
M_N(n_N) =
\begin{pmatrix}
1 & 0 & 0 & \cdots & 0 \\
\left( \frac{n_N}{2^{N}}\right) & 1 & 0 & \cdots & 0 \\
\left( \frac{n_N}{2^{N}}\right)^2 & 2\left( \frac{n_N}{2^{N}}\right) & 1  & \cdots & 0 \\
\vdots & \vdots & \vdots & \ddots & \vdots\\
\left( \frac{n_N}{2^{N}}\right)^{Q} &
\begin{pmatrix} Q\\ 1 \end{pmatrix} \left( \frac{n_N}{2^{N}}\right)^{Q-1} &
\begin{pmatrix} Q \\ 2 \end{pmatrix} \left( \frac{n_N}{2^{N}}\right)^{Q-2} &
\cdots &
1
\end{pmatrix}.
\end{equation}
This completes the proof.

A significant body of mathematical literature deals with the rank of quantics representations of functions. We will not attempt to explore it here. Qualitatively, a smooth function (which looks locally like a polynomial) is approximately low-rank. It is important to observe that quantics can efficiently represent non-smooth functions as well. For example the $\delta$ function
\begin{equation}
\delta(x-y) = \prod_\alpha \delta_{x_\alpha,y_\alpha}
\end{equation}
is of rank 1, while not being smooth at all.

\subsection{The magic quantics tensor}
To work with quantics, we will need many different MPOs, for instance the MPO $D$ to calculate the (finite difference) derivative of a function $(D\Psi)_n = \Psi_{n+1}-\Psi_n$, or the discrete Laplacian $(\Delta\Psi)_n = \Psi_{n+1}-2\Psi_n+\Psi_{n-1}$. In this section, we introduce a single tensor $M_{xycx'y'c'}$ which allows us to construct many of these objects in a straightforward way. This tensor is intimately linked to the way we learn to do additions in elementary school. It is so handy that we call it the ``magic quantics tensor''. The original mathematical literature for such constructions can be found in
\cite{kazeev2012,kazeev2013,kazeev2013b,roberts2014}.

We want to construct an MPO for the (exponentially large) matrix
\begin{equation}
\label{eq:def_magic_quantics}
\Theta_{nm,n'm'} = \delta_{n,n'}\delta_{m,n'+m'}.
\end{equation}
Applying $\Theta$ to a two-variable quantics function $\Psi_{n'm'}$ corresponds to
the change of variables
$$(\Theta\Psi)_{nm} = \Psi_{n,m-n}.$$
Likewise, applying its transpose gives
$$(\Theta^T\Psi)_{nm} = \Psi_{n,m+n}.$$

Let's construct the MPO of $\Theta$ bit by bit starting from the least important ones $n_1$ and $m_1$. The construction follows the algorithm used to perform additions in elementary school (except that it is in base 2, not 10). In base 2, the elementary addition of
$11 + 19 = 30$ takes the form:
\begin{center}
\includegraphics[width=0.3\textwidth]{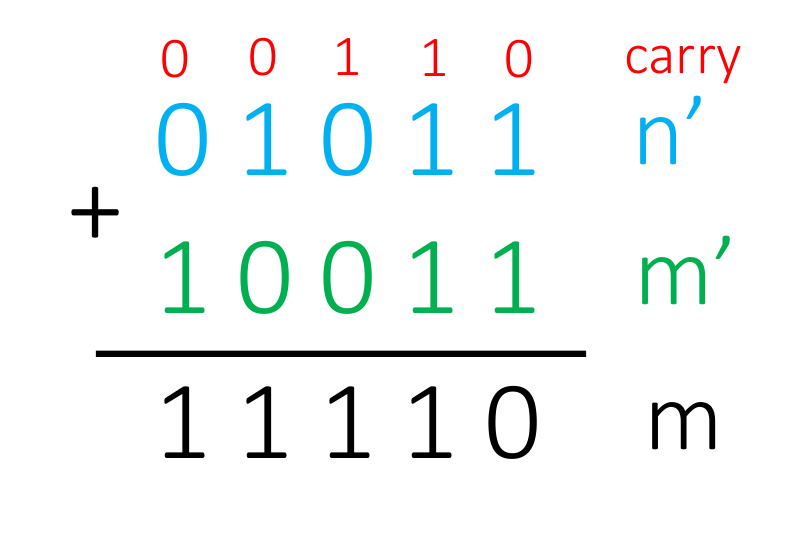}
\end{center}
Now, if we examine the above bit by bit, we find that Eq.~\eqref{eq:def_magic_quantics} implies for the first bits that
\begin{equation}
\label{eq:magic_1}
n_1 = n_1',\quad\text{and}\quad m_1 = m_1'+n_1'\ [2].
\end{equation}
To get the conditions for the next bit, we also need the ``carry'' $c_1$, which is one if both $m'_1=1$ and $n'_1=1$, and zero otherwise. Now, the condition Eq.~\eqref{eq:def_magic_quantics} for the second set of bits reads
\begin{equation}
\label{eq:magic_2}
n_2 = n_2',\quad\text{and}\quad m_2 = m_2'+n_2'+ c_1 \ [2].
\end{equation}
The carry is given by $c_2 = \lfloor (m_2'+n_2'+ c_1) / 2\rfloor$ (the quotient of the Euclidean division). The $M$ tensor essentially captures these constraints automatically.
It is defined as
\begin{itemize}
\item $M_{xycx'y'c'}= 1$, if $x=x'$, $y = x'+y'+c'[2]$ and $c = \lfloor (x'+y'+ c') / 2\rfloor$,
\item $M_{xycx'y'c'}= 0$ otherwise.
\end{itemize}
\begin{center}
\includegraphics[scale=0.3]{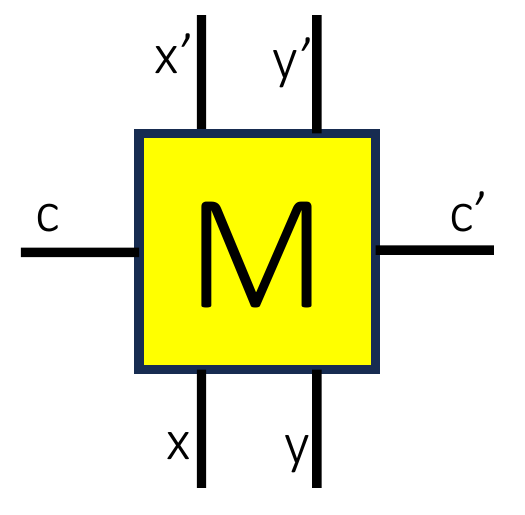}
\end{center}
The construction of the MPO of $\Theta$ is now straightforward: we only need to make sure that the output carry at one stage (the $c$) is equal to the input carry at the next one (the $c'$).
This is exactly what the contraction of two tensors does. Graphically, the MPO is
\begin{center}
\includegraphics[scale=0.3]{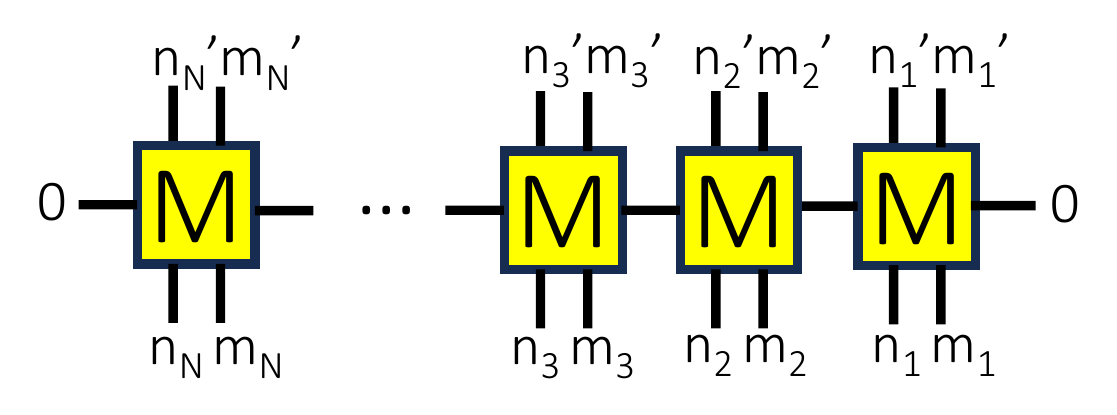}
\end{center}
The zero on the right comes from the fact that the addition starts with no carry. If one replaces the zero on the left by the vector $(1,1)$, the definition of $\Theta$ is modified such that the addition is performed \emph{modulo} $2^N$, i.e. $\Theta_{nm,n'm'} = \delta_{n,n'}\delta_{m,n'+m' [2^N]}$.

In many situations, we do not need the $n$ output, and we can simply trace over it, defining
\begin{equation}
\Theta_{m,n'm'} = \sum_n \Theta_{nm,n'm'} = \delta_{m,n'+m'}.
\end{equation}

We can do all sorts of things with this MPO. The first is to perform convolutions. If we have two MPS $\Psi_n$ and $\Phi_n$, then
\begin{equation}
\Lambda_m \equiv \sum_{n'm'} \Theta_{m,n'm'} \Psi_{n'} \Phi_{m'} =
\sum_n \Psi_{n}\Phi_{m-n}.
\end{equation}
or in graphical form
\begin{center}
\includegraphics[scale=0.3]{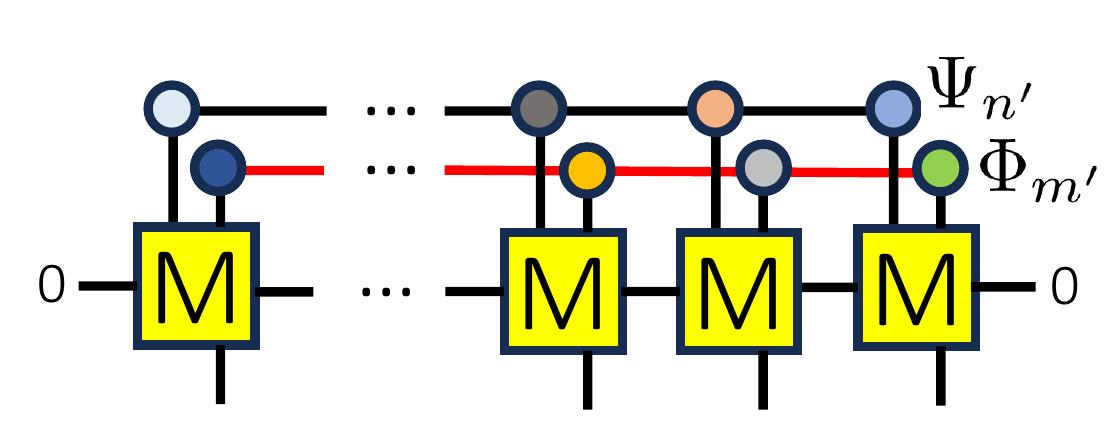}
\end{center}
Contracting the above network is done following the zip-up algorithm and/or DMRG (quantum circuit version). We can also use it to perform translations $T(n)$ by a vector $n =
n_Nn_{N-1}\cdots{}n_1$:
\begin{equation}
[T(n)\Psi]_m = \Psi_{m+n}.
\end{equation}
This is simply given by
\begin{center}
\includegraphics[scale=0.3]{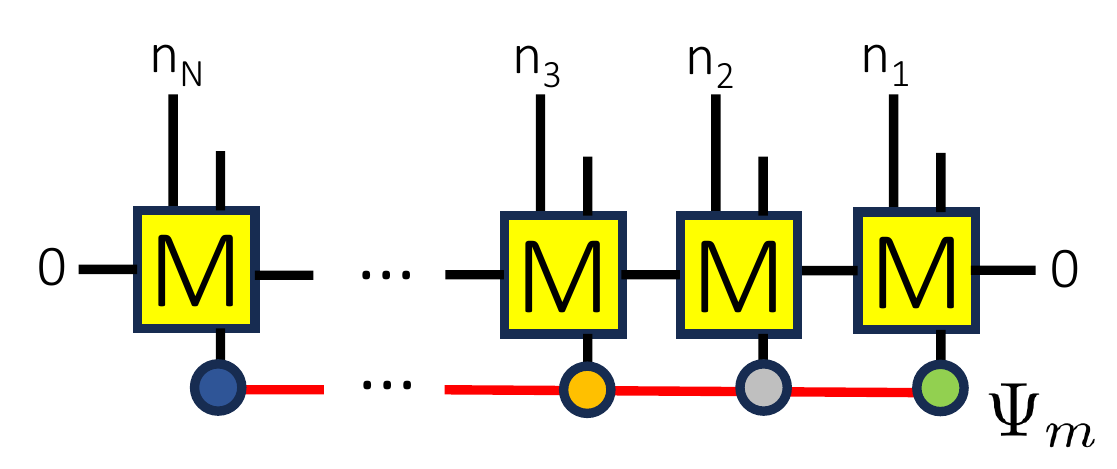}
\end{center}
while $T(-n)\Psi$ is simply obtained by placing $\Psi$ on the other side
\begin{center}
\includegraphics[scale=0.3]{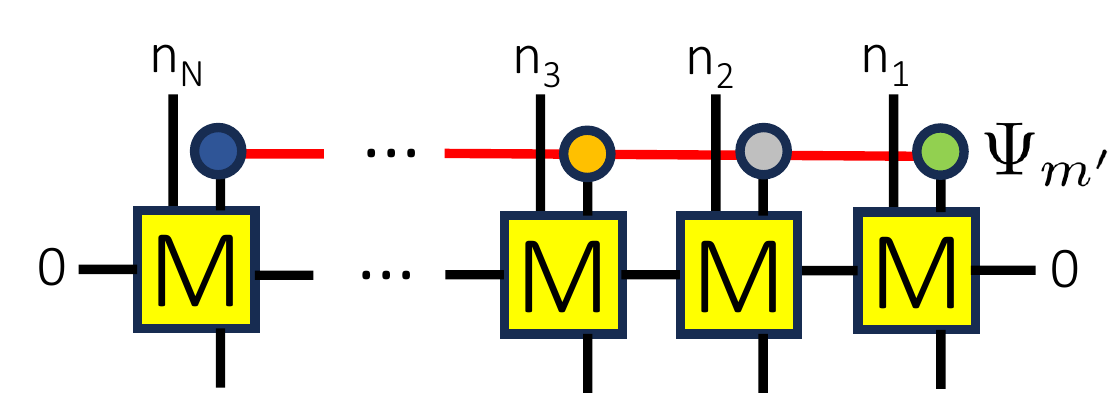}
\end{center}
To obtain, for instance, the discrete Laplacian, we simply add the two MPOs above with
$n = 0\cdots{}001$.

\subsection{Indefinite integral}
When we learn calculus, one of the first things that we realize is that calculating the derivative of a function is easy; we have a fixed set of rules to apply (it can even be automated with symbolic calculus or automatic differentiation). However, calculating integrals is hard, and it is seldom possible to do it explicitly. In quantics, this asymmetry is no longer present: calculating a derivative is easy (as shown above), and obtaining the indefinite integral is equally simple; the corresponding MPO has rank $\chi=2$, as we show now.

Suppose we have a function $\Psi_n$ (in quantics representation), and we want to calculate
\begin{equation}
F_n = \sum_{m=0}^n \Psi_m =\sum_m \Theta(n-m)\Psi_m.
\end{equation}
In other words, we are looking for an MPO that represents the Heaviside function
$\Theta(n-m)$ in quantics (beware of the notation conflict with the previous section). We will follow the same approach as for the magic tensor and
construct a local tensor $\text{II}$ that implements  $\Theta(n-m)$ bit by bit, starting from the most relevant bits $n_{N-1}$ and $m_{N-1}$ and proceeding to the least relevant ones. Let's do it on an example:
\begin{eqnarray}
n &= 110\textbf{0}10111 \nonumber \\
m &= 110\textbf{1}10010 \nonumber
\end{eqnarray}
To compare these two numbers, we start from the left and compare their bits. The first three are equal, so the comparison is undecided. When we reach the fourth bits (in bold), we know that $m>n$, regardless of the value of the remaining bits. So, all we need is a tensor that performs the bit-by-bit comparison (this is local), and the only ``communication'' between tensors is a flag that indicates when the value of $\Theta(n-m)$ has been decided.

To do this, we define the tensor $\text{II}_{xy,ab}$ such that the MPS has the form
\begin{eqnarray}
\Theta(n-m) &=& \sum_{i_{N-1}\cdots{}i_0} \text{II}_{n_{N-1}m_{N-1},0i_{N-1}}
\nonumber \\
&\cdots{}&\text{II}_{n_{2}m_{2},i_{3}i_2}
\text{II}_{n_{1}m_{1},i_{2}i_1}  \text{II}'_{n_{0}m_{0},i_{1} }
\end{eqnarray}
i.e.\ $x$ corresponds to a bit of $n$, $y$ corresponds to a bit of $m$, $a$ to a
``message'' received from the left, and $b$ to a ``message'' sent to the right. By convention, this message is 0 if the value of $\Theta(n-m)$ is undecided yet (i.e.\ given the bits seen so far), and the message is 1 if the value of $\Theta(n-m)$ is fully decided (meaning that its value is independent of the remaining bits). The definition of $\text{II}$ reads
\begin{eqnarray}
\text{II}_{xy,ab} = \delta_{a,1} \delta_{a,b} +
\delta_{a,0} [ \delta_{x,1}\delta_{y,0} \delta_{b,1}
+ \delta_{x,y} \delta_{b,0} ],
\end{eqnarray}
which can be interpreted as follows:
\begin{itemize}
\item If the input message is 1, then return 1 and send 1 as the input message to the next tensor (the value of the MPO is already decided, regardless of the remaining bits).
\item If the input message is 0, then
  \begin{itemize}
  \item if $x>y$, return 1 and send message 1 to the next tensor;
  \item if $x=y$, send message 0 to the next tensor;
  \item in the remaining case ($x<y$), return 0.
  \end{itemize}
\end{itemize}
One can check iteratively that $\text{II}$ indeed does the job. 
The last tensor on the right (which corresponds to $n_0$ and $m_0$) is slightly
different because it decides on the value of $\mu=\Theta(0)$.
Let us call this last tensor $\text{II}'_{xya}$. Then
\begin{equation}
\text{II}'_{xya} = \delta_{a,1}  +
\delta_{a,0} [ \mu \delta_{x,y} + \delta_{x,1} \delta_{y,0} ].
\end{equation}

\subsection{Generalization to higher dimensions}
To generalize quantics to two or more dimensions, we have several strategies at our
disposal. Let's consider a function $\Psi_{x,y,z}$, already discretized, with
$x = x_N\cdots{}x_1$, $y = y_N\cdots{}y_1$, and $z = z_N\cdots{}z_1$. The first strategy is to put all the variables one after the other, as follows:
\begin{center}
\includegraphics[scale=0.25]{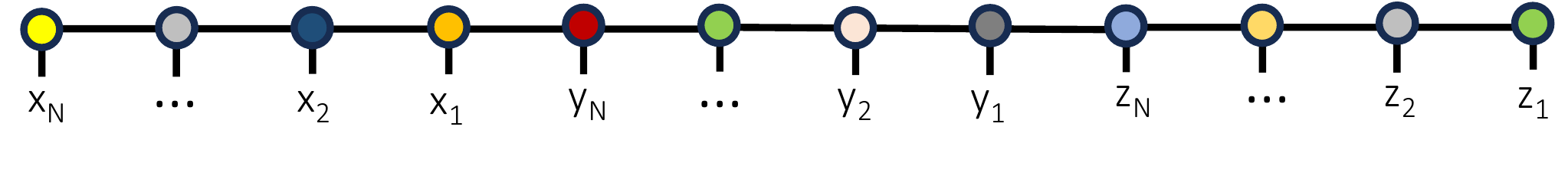}
\end{center}
But we can also put the indices that correspond to the same scale together (this is called interleaved):
\begin{center}
\includegraphics[scale=0.25]{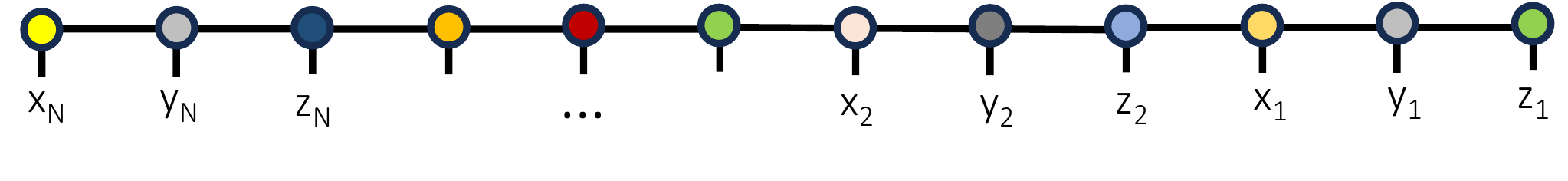}
\end{center}
Choosing between the above two choices is not straightforward and will depend on the application. In two dimensions, we have often observed that the representation with the
lowest bond dimension is the ``mirror'' configuration, which combines some of the advantages of the above two:
\begin{center}
\includegraphics[scale=0.25]{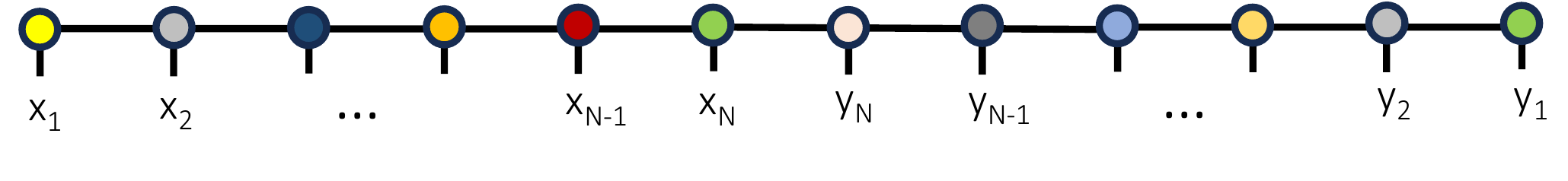}
\end{center}
However, in order to generalize the mirror to three dimensions, we would need to
introduce tree product states and tree product operators:
\begin{center}
\includegraphics[scale=0.3]{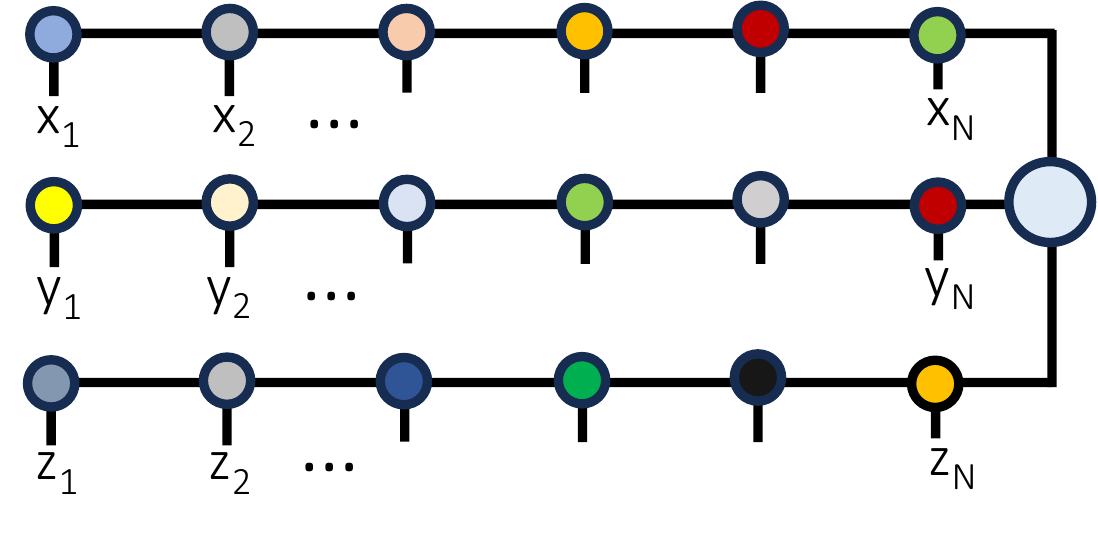}
\end{center}
Those are not particularly difficult and literature is currently emerging on the subject. Essentially all the algorithms that have been defined for MPO/MPS can be extended to trees.

\subsection{Application to the Poisson equation}
We finished going through the ``must-have'' list for quantics. We now have a complete toolbox at our disposal that we can start to use to compute various things with functions. Let’s first discuss a straightforward example, the Poisson equation
\begin{equation}
\Delta U = n(\vec r).
\end{equation}
It belongs to an important class of (elliptic) PDEs. It appears in electrostatics, obviously, but also in fluid dynamics, heat transfer, elastic theory, etc. The most direct algorithm to solve it is the following sequence:
\begin{itemize}
\item Choose a representation, say mirror or interleaved.
\item Feed $n(\vec r)$ to TCI to obtain the corresponding MPS.
\item Construct the MPO of the Laplacian (analytically).
\item Give this MPO and the MPS of $n(\vec r)$ to your favorite tensor network linear solver.
\end{itemize}
Once again, the unknown here is the bond dimension needed to describe $n(\vec r)$ (if it is too high, the method will not be advantageous), as well as that needed for the solution $U(\vec r)$. Very often, this bond dimension is small and does not depend on $N$. In that case, one can reach an exponentially fine grid at a polynomial cost, fulfilling the type of promises typically expected of quantum computers.

More tricky is how to implement boundary conditions. For instance, one might want to include a metallic conductor $\mathcal{M}$ inside the simulation volume, keeping it at a constant potential $U(\vec r)=V_g$ for $\vec r\in\mathcal{M}$ (Dirichlet boundary condition). A possibility is to turn to the Helmholtz equation
\begin{equation}
\Delta U - \rho(\vec r)\,U = n(\vec r),
\end{equation}
where $\rho(\vec r)$ is the density of states in the material. A very large $\rho(\vec r)$ (formally infinite) corresponds to a perfect metal (Dirichlet). We then proceed as previously, adding to the MPO of $\Delta$ the diagonal MPO obtained by feeding $\rho(\vec r)$ to TCI and using the copy tensor:
\begin{center}
\includegraphics[scale=0.3]{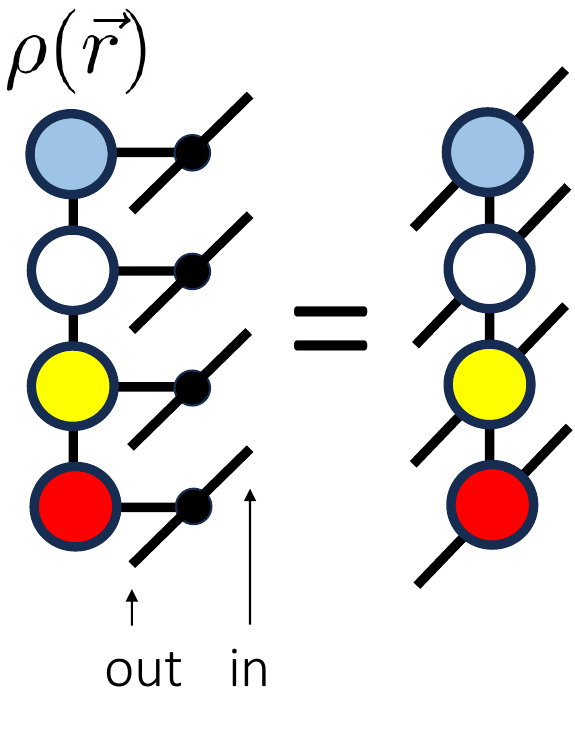}
\end{center}
Once again, the efficiency of the solver will be tightly linked to the rank of $\rho(\vec r)$. How to find a representation where $\rho(\vec r)$ keeps a low rank (in other words: how to describe geometries in quantics) is currently one of the toughest questions in the field. There exist many simple geometries, for instance a sphere, where the simple representations introduced above will \emph{not} be of low rank.

\subsection{Application to the Schrödinger equation}
The Schrödinger equation is very similar to the Poisson equation; we just need to exchange the linear solver for an eigenvector solver. Here, we want to solve
\begin{equation}
- \Delta \Psi +  U(\vec r) \Psi = E \Psi,
\end{equation}
where the input potential $U(\vec r)$ can be obtained from TCI or directly from the solution of a Poisson problem as discussed in the preceding section. Once all ingredients have been put in MPO form, we end up with an eigenvalue problem. This is a job for the original DMRG algorithm. It is quite paradoxical that DMRG, which was initially designed to solve complex many-body problems, could be used off the shelf to solve a much more mundane problem, a mere Schrödinger equation. The exponential complexity here translates into an exponentially fine grid.

\subsection{The quantum Fourier transform as a low-rank MPO}
Let us discuss one last important point: the Fourier transform. The discrete Fourier transform $\hat \Psi_\omega$ of a vector $\Psi_t$ is defined as
\begin{equation}
\hat \Psi_\omega = \sum_t F_{\omega t} \Psi_t,\quad\text{with}\quad F_{\omega t} = \frac{1}{\sqrt{2^N}} e^{-i2\pi \omega t/2^N}.
\end{equation}
The Fourier operator $F_{\omega t}$ has a very nice property: it is a low-rank MPO with a rank around $\chi=15$ for machine precision ($\chi=10$ is typically sufficient in practice). Interestingly, this discrete Fourier transform is nothing but the quantum Fourier transform (QFT) central to many quantum computing algorithms (e.g.\ Shor's algorithm or quantum phase estimation). Hence, the existence of a low-rank MPO has deep consequences for the entanglement generation in the corresponding part of the quantum algorithm.

An algorithm was constructed to perform ``superfast'' Fourier transforms as early as 2012 \cite{dolgov2012}, but the existence of a low-rank MPO was recognized only in \cite{woolfe2014}, where the low rank was observed in some numerical experiments. The actual proof of the statement was done in \cite{chen2023}. The presentation below leans heavily on \cite{chen2024}, which has the advantage of being very transparent as well as providing an explicit construction of the MPO through yet another magic tensor.

\subsubsection{Explicit construction of the Quantum Fourier Transform MPO}
In this paragraph, we will construct the QFT-MPO explicitly following \cite{chen2024}. Note that this part was not present in the oral lectures. There are several alternative ways to construct this MPO, but none, to our knowledge, are as elegant. The least elegant (yet effective) way is to feed the definition of $F_{\omega t}$ to TCI.

To proceed, we will need slightly more precise notation. Let us introduce the partial integers
\begin{eqnarray}
t^{k:q} &=& t_k + 2t_{k+1} +\cdots + 2^{q-k} t_{q}, \\
\omega^{k:q} &=& 2^{q-k} \omega_k + 2^{q-k-1}\omega_{k+1} +\cdots +  \omega_{q},
\end{eqnarray}
so that $t = t^{1:N}$ and $\omega = \omega^{1:N}$. We also introduce the QFT with the corresponding bits as
\begin{equation}
F^{k:q}= \frac{1}{\sqrt{2^{q-k+1}}} e^{-i2\pi \omega^{k:q} t^{k:q}/2^{q-k+1}},
\end{equation}
so that $F_{\omega t} = F^{1:N}$. The important point about this notation is that the ordering of the variables in time and frequency is \emph{reversed}. Crucially, the MPO is \emph{not} of low rank if we keep the same ordering. This makes sense intuitively: what happens on small time scales corresponds to large frequencies, so that we want to keep the corresponding bits close to each other.

Now, to understand why the QFT MPO is low rank, let us split the variables into the first $k$ bits and the last $N-k$ ones (the value of $k$ is arbitrary), writing
\begin{eqnarray}
t &=& t^{1:k} + 2^{k} t^{k+1:N}, \\
\omega &=& 2^{N-k}\omega^{1:k} + \omega^{k+1:N},
\end{eqnarray}
so that $F_{\omega t}$ contains the product of four terms. Two of these terms correspond to the QFT with, respectively, the first bits $1:k$ and the last bits $k+1:N$. The term $e^{-i2\pi t^{k+1:N}\omega^{1:k}}=1$ is irrelevant, so that we get
\begin{equation}
F_{\omega t} = F^{1:k} F^{k+1:N} e^{-i2\pi A}
\end{equation}
with
\begin{equation}
A = \frac{t^{1:k}\omega^{k+1:N}}{2^N}.
\end{equation}
The term $e^{-i2\pi A}$ is the only factor that links the first set of bits with the second one. Now, the key to the argument is that $A\in [0,1]$ belongs to a small compact interval. The factorization now arises from $e^{-i2\pi A}$ being \emph{smooth} in that interval, so that it can be approximated (with exponential accuracy) by a polynomial:
\begin{equation}
e^{-i2\pi A}\approx\sum_{\alpha=1}^\chi a_\alpha A^{\alpha-1},
\end{equation}
i.e.\ as a sum of functions that factorize. So we get
\begin{equation}
F_{\omega t} \approx \sum_{\alpha=1}^\chi
F^{1:k} a_\alpha \left(\frac{t^{1:k}}{2^{k-1}}\right)^{\alpha-1}
\times
\left(\frac{\omega^{k+1:N}}{2^{N-k+1}}\right)^{\alpha-1} F^{k+1:N}.
\end{equation}
That is, a sum of $\chi$ terms, each of which is a product of a factor involving the first $k$ qubits and another factor involving the remaining qubits. This concludes the proof that the low-rank factorization of the QFT exists.

Now, with the same ideas, we can do even better and construct the corresponding MPO explicitly. The key, as always, is to identify the information that needs to be transferred from one tensor to the next in the iterative construction. Suppose that we have managed to build the vector
\begin{equation}
\hat C_N = F^{1:N}
\begin{pmatrix}
 \exp(-i2\pi t^{1:N} c_1/2^N) \\
 \exp(-i2\pi t^{1:N} c_2/2^N) \\
\vdots \\
 \exp(-i2\pi t^{1:N} c_\chi/2^N)
\end{pmatrix},
\end{equation}
where the $c_\alpha \in [0,1]$ are a set of real numbers to be defined below. By convention, $c_1=0$, so that the first entry of this vector is $F^{1:N}$, i.e.\ we only need to multiply this vector by $(1\ 0\cdots{}0)^T$ to obtain our MPO. The task is to construct a $\chi\times\chi$ local matrix $M^N(t_N,\omega_N)$ such that $\hat C_N = M^N(t_N,\omega_N) \hat C_{N-1}$.  We proceed as before by splitting $t^{1:N} = t^{1:N-1} + 2^{N-1} t_N$ and $\omega^{1:N} = 2 \omega^{1:N-1} + \omega_N$ to obtain
\begin{equation}
\label{eq:qft1}
F^{1:N}\exp(-i2\pi t^{1:N} c_\alpha/2^N) =
e^{-i\pi (t_N\omega_N+c_\alpha t_N)} F^{1:N-1}
\exp\left(-\frac{i2\pi t^{1:N-1}}{2^{N-1}} \frac{c_\alpha+\omega_N}{2}\right).
\end{equation}
Now, the ``smoothness'' argument will be applied to the function
\begin{equation}
f(A) = \exp\left(-\frac{i2\pi t^{1:N-1}}{2^{N-1}} A \right),
\end{equation}
which we expand as a sum of polynomials of order $\chi$. A common choice is to take the $c_\alpha$ as the Chebyshev grid points $c_\alpha=[1-\cos(\pi[\alpha-1]/\chi)]/2$, since this leads to rapid convergence \cite{trefethen2019}. Noting that $P_\alpha(A) = \prod_{\beta\ne\alpha} (A-c_\beta)/(c_\alpha -c_\beta)$ is the Lagrange interpolating polynomial, our factorization formula reads
\begin{equation}
f(A) \approx \sum_\alpha f(c_\alpha) P_\alpha(A).
\end{equation}
Inserting this interpolative decomposition into Eq.~\eqref{eq:qft1} gives
\begin{equation}
F^{1:N}\exp(-i2\pi t^{1:N} c_\alpha/2^N) \approx \sum_\beta
e^{-i\pi (t_N\omega_N+c_\alpha t_N)}
P_\beta\left(\frac{c_\alpha+\omega_N}{2}\right)
F^{1:N-1}\exp\left(-\frac{i2\pi t^{1:N-1} c_\beta}{2^{N-1}} \right).
\end{equation}
From this, we can directly read the matrix elements of $M^N$
\begin{equation}
M^N_{\alpha\beta} = e^{-i\pi (t_N\omega_N+c_\alpha t_N)}
P_\beta\left(\frac{c_\alpha+\omega_N}{2}\right).
\end{equation}
We need only to specify the initial vector $\hat C_0 = (1\ 1\cdots{}1)$ to complete the construction. This MPO is slightly different from the ones we have seen before: its construction is not purely algebraic, but involves some input from the theory of approximate interpolants.

\subsubsection{Application to the heat equation}
\begin{figure}
  \begin{center}
    \includegraphics[width=10cm]{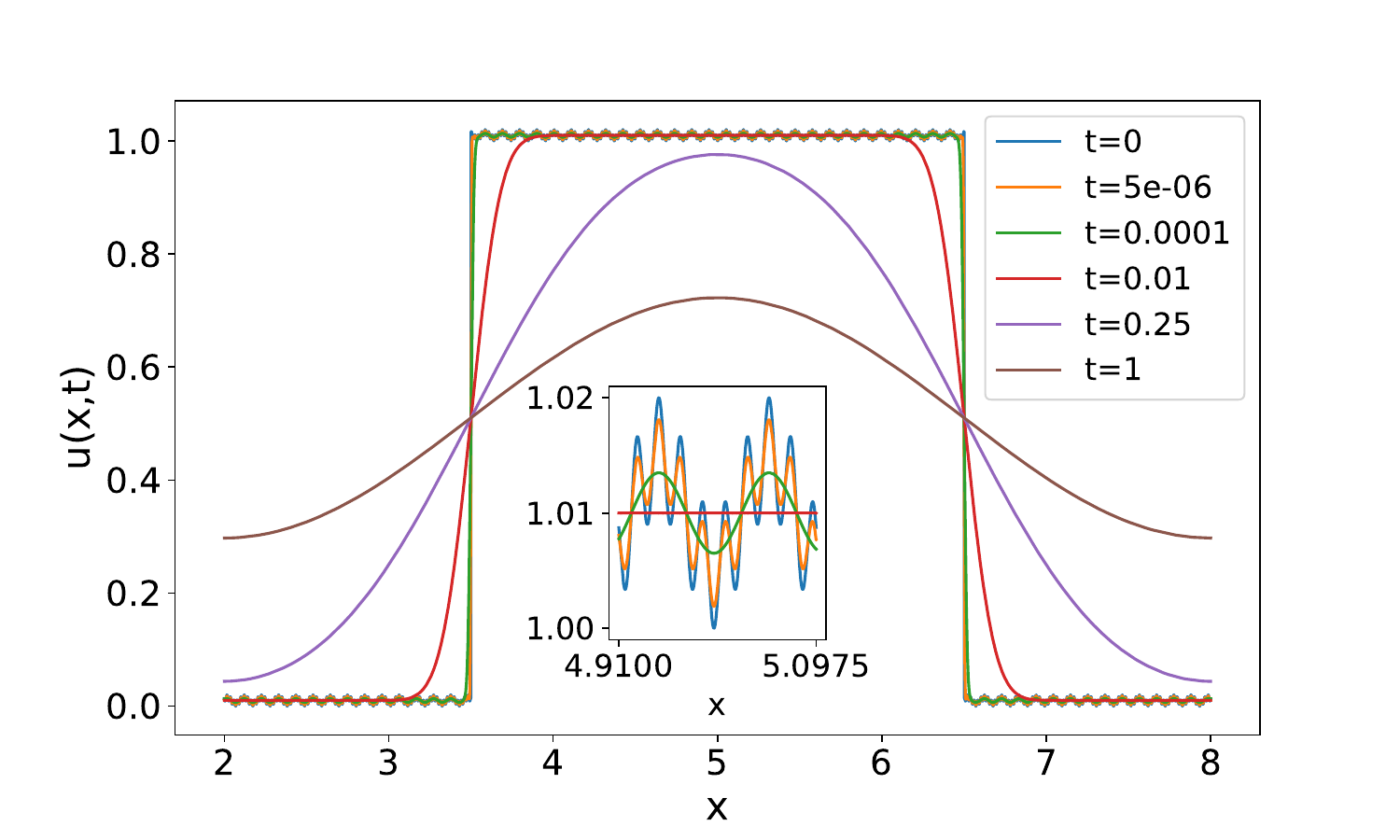}
    \caption{Illustration of using QFT to solve the heat equation~\eqref{eq:heat} using quantics. The plot shows $u(x,t)$ versus $x$ for different times, using an initial condition that contains Heaviside functions and rapidly oscillating terms:
      $u_0(x)=[1+\cos(120x)\sin(180x)]/100 + \theta(x-7/2)[1 -\theta(x-13/2)]$. We used a 1D grid with $2^{30}$ points. The inset shows a zoom close to $x=5$. \label{fig:heat}
      (Adapted from \cite{nunez2025}).}
  \end{center}
\end{figure}

The Fourier transform can have many applications, so the idea that it may be performed
exponentially faster on a function that has been put in quantics form is rather appealing. Let us quickly discuss one application, as an illustration of these possibilities. Suppose that we want to solve the heat equation
\begin{equation}
\label{eq:heat}
\partial_t u = \partial_{xx} u,
\end{equation}
with $u(x,t=0)=u_0(x)$ in one dimension. This equation admits a simple solution
in $k$ space (Fourier transform with respect to the spatial dimension): $\hat u(k,t)= e^{-k^2 t} \hat u_0(k)$. An algorithm to solve it for a given time $t$ is therefore the following:
\begin{itemize}
  \item Feed $u_0(x)$ to TCI.
  \item Feed the kernel $e^{-k^2 t}$ to TCI. Construct the corresponding diagonal MPO as we did for the Helmholtz equation.
  \item Apply the Fourier transform to the initial condition $\hat u_0(k)=F u_0(x)$.
  \item Apply the MPO of $e^{-k^2 t}$ to the result $\hat u(k,t)= e^{-k^2 t} \hat u_0(k)$.
  \item Apply the inverse Fourier transform $\hat u(x,t)=F^{-1}\hat u(k,t)$.
\end{itemize}
And that's it. Essentially, we obtain the solution at any given time by performing
two calls to TCI and three calls to an MPO-MPS multiplication routine. (In practice, we can even do better using specific algorithms tailored for element-wise products between two MPS, but that's for another time.) What's more, there is nothing specific to one dimension in the above algorithm. An illustration of a practical calculation following this scheme is shown in Fig.~\ref{fig:heat}. These calculations are almost instantaneous on a laptop, even for a grid containing a billion points. This is one of the cases where quantics seems to perform particularly well.

\section{Conclusion}
\label{sec:conclusion}

So this is it. We have come a long way and learned about many algorithms that you may now use for various purposes, both related and unrelated to the quantum many-body problem. 
We hope to have conveyed how versatile and powerful tensor network representations can be. Yet, despite the flexibility of the approach and the impressive results, readers should be warned not to view everything as a tensor network (although this can be very tempting!). Other representations exist that can succeed when tensor networks fail. Ultimately, to address an exponentially large problem one must take advantage of one structure or another, and this structure is not necessarily related to entanglement.
To highlight this point, an earlier version of these notes included a light discussion of quantum Monte Carlo in the context of variational Monte Carlo (a technique that is experiencing an important revival in the context of deep-neural-network-based variational ansatz) and Green function Monte Carlo (an ``exact'' technique that has no sign problem for the transverse-field Ising model). The discussion was too superficial to be of any real use, so we have removed it, but the point remains: there are many different approaches that can be used to address exponentially large problems.

With respect to tensor networks, many other aspects of the field had to be left out. In particular, we made an effort to present the MPO/MPS toolbox independently from its traditional application domain of many-body physics, where the MPS is almost always the many-body ground state and the MPO the Hamiltonian. These problems possess additional structure, and good algorithms take advantage of this to be faster or more accurate. For instance, variants of DMRG exist for cases when the system is invariant under translation (infinite DMRG or iDMRG) or in the presence of a local symmetry (e.g.\ to account for particle conservation or even SU(2) local symmetries).
Furthermore, many more tensor network types are used beyond MPO and MPS, although these two are by far the most popular, largely because of how easy and efficient it is to write algorithms for them.

We highlighted the special importance of the TCI algorithm. This is a bet rather than a certainty. We believe that the field is at a crossroads where many new applications that have nothing to do with many-body physics (such as quantics for PDEs) emerge and become very impactful. TCI opens the door to these new developments by making it possible to map problems onto the tensor network formalism. If quantum computers one day become a real thing (in the sense of a useful machine rather than the beautiful experiments that they are today), then tensor networks and TCI will very likely play a key role in mapping the input of a problem into a tensor network from which one can build a quantum circuit.

\section*{Acknowledgments}
X.W. and C.W.G. thank Tero Heikkilä and Stephan Ilic for the invitation to give these lectures and the warm hospitality at the University of Jyväskylä.
We also thank the students of the school as well as of our own group, who by their questions and feedback helped to improve this text. A special mention goes to Adrien Moulinas, who found a number of embarrassing typos in the equations.  X.W. would like to thank all his tensor network collaborators, with a special mention for Miles Stoudenmire, with whom he started to simulate quantum computers, and Yuriel Nuñez Fernandez, with whom he explored the tensor cross interpolation aspects.

\paragraph{Author contributions}
X.W. wrote the text and gave the lectures. C.W.G. prepared the hands-on sessions, which he supervised together with the help of X.W. C.W.G. thoroughly edited and improved the original text. C.-H. H. attended the lectures and performed the numerical computations used in the illustrations Fig.~\ref{fig:ghz} and Fig.~\ref{fig:tebd}. All authors reviewed the final text.

\appendix

\section{Code repositories}
\label{app:code}

Two small code repositories accompany these lectures:
\begin{itemize}
\item \url{https://gricad-gitlab.univ-grenoble-alpes.fr/theorypheliqs/whocancompete-tutorial} contains the pedagogical Python examples prepared for the hands-on sessions. They implement elementary gates, a full-state quantum computer emulator, MPS routines, and an MPS-based quantum computer emulator.
\item \url{https://gricad-gitlab.univ-grenoble-alpes.fr/theorypheliqs/whocancompete-example} contains participant-contributed code by Chen-How Huang. It generates the data for Fig.~\ref{fig:ghz} and Fig.~\ref{fig:tebd}: the GHZ-state timing comparison and the imaginary-time TEBD calculation for the transverse-field Ising model.
\end{itemize}

\bibliography{whocancompete}

\end{document}